\documentclass[10pt,a4paper]{article}

\usepackage[english]{babel}
\usepackage[T1]{fontenc}
\usepackage[utf8]{inputenc}
\usepackage[babel=true]{microtype}
\usepackage{a4wide}

\usepackage{graphicx,amsfonts,amssymb,amsmath,hyperref}

\newif\ifhyper
\hypertrue
\ifhyper
\hypersetup{
   citecolor = {green},
   colorlinks = {true}, % false
   linkcolor = {red},
   urlcolor = {blue} % magenta
}
\fi 

\usepackage{setspace}

\usepackage[smaller]{acronym}

\usepackage{calc}
\usepackage{array}
\usepackage{adjustbox}
\usepackage{graphbox}
\usepackage{lastpage}

\PassOptionsToPackage{square,numbers}{natbib}
 \usepackage{natbib}

\usepackage{ifthen}

\usepackage{textgreek}

\usepackage{multicol}
\usepackage{multirow}

\usepackage{verbatim}

\usepackage{textcase}
\usepackage{xspace}
\usepackage{authblk}
\usepackage{multirow}
\usepackage{blkarray}

\usepackage{braket}

\usepackage{tikz}

\definecolor{TensorGreen}{HTML}{9AEE68}
\definecolor{TensorGreenOne}{HTML}{7BE93A}
\definecolor{TensorGreenTwo}{HTML}{B9F396}
\definecolor{TensorAwesomeGreen}{HTML}{53C061}

\definecolor{TensorPink}{HTML}{EE689A}
\definecolor{TensorPinkOne}{HTML}{E93A7B}
\definecolor{TensorPinkTwo}{HTML}{F396B9}

\numberwithin{equation}{section}
 
\newcommand{\Eref}[1]{Eq.~(\ref{#1})}

\newcommand{\fref}[1]{Fig.~(\ref{#1})}
\newcommand{\tref}[1]{Table~\ref{#1}}

\newcommand{\sref}[1]{Sec.~\ref{#1}}

\newcommand{\beq}{\begin{equation}}
\newcommand{\eeq}{\end{equation}}
\newcommand{\beqa}{\begin{eqnarray}}
\newcommand{\eeqa}{\end{eqnarray}}

\newcommand{\ps}[1]{{\color{black}#1}}

\newcommand{\mr}[1]{{\color{black}#1}}

\begin{document}

\title{A programming guide for tensor networks \\ with global $SU(2)$ symmetry}
\author[1,2]{Philipp Schmoll}
\author[3]{Sukhbinder Singh}
\author[1,4,5]{Matteo Rizzi}
\author[1,6,7]{Rom\'an Or\'us}
\affil[1]{Institute of Physics, Johannes Gutenberg University, 55099 Mainz, Germany}
\affil[2]{Graduate School Materials Science in Mainz, Staudingerweg 9, 55128 Mainz, Germany}
\affil[3]{Max Planck Institute for Gravitational Physics (Albert Einstein Institute), 14476 Potsdam, Germany}
\affil[4]{\mr{Institute of Quantum Control (PGI-8), Forschungszentrum J\"ulich, D-52425 J\"ulich, Germany}}
\affil[5]{\mr{Institute for Theoretical Physics, University of Cologne, D-50937 K\"oln, Germany}}
\affil[6]{Donostia International Physics Center, Paseo Manuel de Lardizabal 4, E-20018 San Sebasti\'an, Spain}
\affil[7]{Ikerbasque Foundation for Science, Maria Diaz de Haro 3, E-48013 Bilbao, Spain}
\renewcommand\Affilfont{\small}

\maketitle

\begin{abstract}

This paper is a manual with tips and tricks for programming tensor network algorithms with global $SU(2)$ symmetry. We focus on practical details that are many times overlooked when it comes to implementing the basic building blocks of codes, such as useful data structures to store the tensors, practical ways of manipulating them, and \ps{adapting typical functions for symmetric tensors}. Here we do not restrict ourselves to any specific tensor network method, but keep always in mind that the implementation should scale well for simulations of higher-dimensional systems using, e.g., Projected Entangled Pair States, where tensors with many indices may show up. To this end, the structural tensors (or intertwiners) that arise in the usual decomposition of $SU(2)$-symmetric tensors are never explicitly stored throughout the simulation. Instead, we store and manipulate the corresponding fusion trees -- an algebraic specification of the symmetry constraints on the tensor -- in order to implement basic $SU(2)$-symmetric tensor operations. \ps{This fusion tree approach is readily extensible to anyonic systems, as we demonstrate for a chain of Fibonacci anyons.}

\end{abstract}

\newpage 
\enlargethispage{1cm}
{
\hypersetup{linkcolor=black}
\tableofcontents
}
\clearpage

\section{Introduction} % (fold)
\label{sec:introduction}

In recent years, the study of quantum entanglement in strongly correlated systems has fostered the applicability of so-called \emph{tensor networks} (TN)~\cite{tn}. These are representations of quantum many-body states based on their entanglement structure. Individual tensors play the role of the DNA of the wave function, and the overall state for the whole system emerges by gluing the single tensors with quantum entanglement. As such, TNs have been rediscovered several times, and have found applications in many fields, e.g., in condensed matter and statistical physics~\cite{cmapp}, quantum chemistry~\cite{qchemapp}, quantum gravity~\cite{qgapp}, tensor calculus~\cite{tcapp}, and more recently even in machine learning~\cite{mlapp} and linguistics~\cite{liapp}. 

It has also become clear that TNs are the basis of not only new theoretical conceptions, but also of new numerical simulation algorithms. This is especially true for strongly correlated quantum many-body systems. Such systems were conquered in one spatial dimension during the 90s by the Density Matrix Renormalization Group Algorithm (DMRG)~\cite{dmrg}, which can be understood as a variational method over some TNs called Matrix Product States (MPS)~\cite{mps}. Recently there has been a burst of new numerical methods for other types of systems, including higher-dimensional systems~\cite{PEPS}, critical systems~\cite{MERA}, and much more. If compared to the development of MPS-based methods, such new algorithms are evolving at a slower rate because they are intrinsically more complex. However, they hold great promise for the future of numerical simulations, and their development is a must. 

In this context, a key ingredient in many numerical simulations is the implementation of the symmetries of the system. This is particularly true for global on-site symmetries, i.e., those symmetries that leave the quantum state invariant once they act on the whole physical system. Examples are in fact ubiquitous, e.g., the global $\mathbb{Z}_2$ symmetry of the quantum Ising model (spin up/down), the $\mathbb{Z}_q$ symmetry of the quantum $q$-Potts model, the $\mathbb{Z}_2$ parity symmetry of many-body fermionic systems, the $U(1)$ symmetry of particle-conserving Hamiltonians, the $SU(2)$ symmetry of magnetic interactions invariant under rotations, and the $SU(N)$ symmetry of the multiband Hubbard model. As such, imposing that physical states must be symmetric is a huge constraint on the many-body Hilbert space. One can expect, then, that these constraints play an important role in the development of numerical algorithms. In particular, it sounds like a good idea to come up with methods that target directly the physical (symmetric) set of states, which may be quite small, instead of messing around with the full Hilbert space of the system, which is insanely huge. Even addressing symmetries cannot avoid the exponential growth of the Hilbert space though it can slower it down by a factor, which could even be exponential itself in the best case scenario.

The implementation of symmetries in TN methods is thus a top priority. We are in no way the first ones to discuss this, though~\cite{symgen}. The topic has a long history dating back to the implementation of symmetries in DMRG~\cite{symdmrg}. Symmetries in other TN methods have also been successfully implemented, including algorithms based on Multiscale Entanglement Renormalization Ansatz (MERA)~\cite{symmera, sukhi} and Projected Entangled Pair States (PEPS)~\cite{PEPS,iPEPSSU2Hubig}. Still, when comparing the story for different TN methods, the implementation of non-abelian symmetries in PEPS algorithms is quite recent. While abelian symmetries were already implemented in early versions of PEPS numerical codes~\cite{abelPEPS}, implementing non-abelian symmetries for PEPS algorithms was a very different story altogether given the technicalities involved in the manipulation of tensors with many indices. Nevertheless, recent works have implemented $SU(2)$ symmetry for some simple versions of the infinite-PEPS method~\cite{iPEPS} in order to study frustrated quantum antiferromagnets~\cite{su2peps}. Still, there are ways to improve how such non-abelian symmetries can be implemented in PEPS-based methods, which is particularly important if the plan is to develop simulations that are ``as accurate as possible'', based on, e.g., variational tensor updates~\cite{varPEPS}.

Considering the above, our goal with this paper is to offer a detailed explanation of some tips and tricks involved in the programming of $SU(2)$-invariant TNs, focusing on the building blocks of algorithms, and keeping always in mind that we would like to have a scheme that scales well for higher-dimensional algorithms based on PEPS. We do not focus on any TN method in particular, so the formalism is general in this respect. We focus, however, on the case of $SU(2)$ symmetry, since it is a non-trivial example of a non-abelian symmetry that is quite common in nature. While the group $SU(2)$ lacks some ingredients present in more complex symmetries (e.g., multiplicities, as in $SU(3)$), the formalism that we unfold can be extended also to those cases without too much complication. The same is true for the implementation of multiple symmetries such as, e.g. $SU(2) \times U(1)$, which is not within the scope of this paper. \ps{Finally, since our formalism is based on fusion trees it is readily generalized to quantum symmetries e.g. $SU(2)_k$, which underlie anyonic systems. (The Hilbert space of anyonic systems has a natural description in terms of fusion trees.) We demonstrate this by accurately simulating the ground state of a chain of Fibonacci anyons (described by $SU(2)_3$) with a nearest neighbour antiferromagentic interaction.} Additionally, we stress that the purpose of this paper is eminently practical, and that it has been conceived as a manual for those interested in the programming part of TNs. We thus leave the applications of this formalism for forthcoming works. Finally, in this paper we also assume that the reader has some basic knowledge about TNs and associated numerical methods. For more basic background about TNs, we recommend the reader to have a look at the many good reviews and introductions written on the topic~\cite{tn}. The implementation of (multiple) abelian symmetries in tensor networks is well explained in, e.g., Ref.~\cite{matteoanthology}.

Tensor networks are widely used as numerical tools for ground state searches in quantum many-body systems. Once the ground state is found they can be used to efficiently compute physical observables such as magnetization or correlation functions. These algorithms rely on the manipulation of tensors that include a number of standard operations. The most striking one is certainly the contraction, which is the operation that contracts different tensors over their common indices to produce a new tensor. Other relevant operations include reshaping, i.e. the combination or separation of indices, permutations of indices, but also the singular value decomposition and truncation. This set of standard operations needs to be translated for the case of symmetric tensors, thus posing a large part of the symmetric low-level implementation. In this paper we aim to describe the required functions in a comprehensive way that can be followed for an implementation of symmetries in tensor networks. 

The structure of this paper is as follows. In \sref{sec:appetizer_the_spin_1_2_heisenberg_chain} we show, as an appetizer and in order to motivate the rest of the paper, a practical example of the application of our formalism for the spin-1/2 Heisenberg quantum chain. In \sref{sec:fundamentals} and \sref{sec:fusion_trees} we will introduce symmetric tensors and explain how the structure of the imposed symmetry group shows up. We will discuss different approaches to handle the symmetry and give \emph{a} way to store all the necessary information in the tensors {in \sref{sec:blue_data_structure_for_an_su_2_symmetric_tensor}}. The following sections are essentially describing the set of operations that we will rely on when writing tensor network algorithms, where we will come back often to the basic sections about symmetric tensors and their handling. {A general template that applies to all of these operations is presented in \sref{sec:a_general_template_for_symmetric_tensor_operations}. The first operations, transforming the structural part of the symmetric tensor, are described in \sref{sec:transforming_the_fusion_tree}.} \sref{sec:reversing_an_index} introduces and explains reversals. This feature, usually not needed for generic tensors, becomes important for the manipulation of symmetric ones. In \sref{sec:permutation_of_indices} we will explain the permutation of indices, e.g., the swapping of two tensor indices. In \sref{sec:reshaping_a_tensor} we will generalize the combination and separation of indices to the case of symmetric tensors. Using all these necessary structures and functions we will then explain contractions in \sref{sec:tensor_contraction}. Even though the main functionality is the same for non-symmetric and symmetric tensors, we will see that the imposed symmetry leads to some important differences. In \sref{sec:symmetric_singular_value_decomposition} we will deal with the singular value decomposition and truncation of tensors, two operations that are indispensable for many tensor network algorithms. Some final remarks about our implementation with respect to generalizations and efficiency are to be found in \sref{sec:generalizations_of_fusion_trees_efficiency_and_prospect}, and we will conclude the paper with some conclusions in \sref{sec:conclusions}. 

As a disclaimer, let us stress that our implementation may not be optimal depending on the specific task at hand. However, we believe that the general approach that we pursue here is, with minor extensions to be described at the end of the paper, one of the best possible implementations in order to program tensor network algorithms for higher-dimensional systems, where one may encounter tensors  with many indices.

% section introduction (end)

\section{Appetizer: the spin-1/2 Heisenberg chain} % (fold)
\label{sec:appetizer_the_spin_1_2_heisenberg_chain}

\begin{figure}
 	\centering
	\includegraphics[width = 0.65\textwidth]{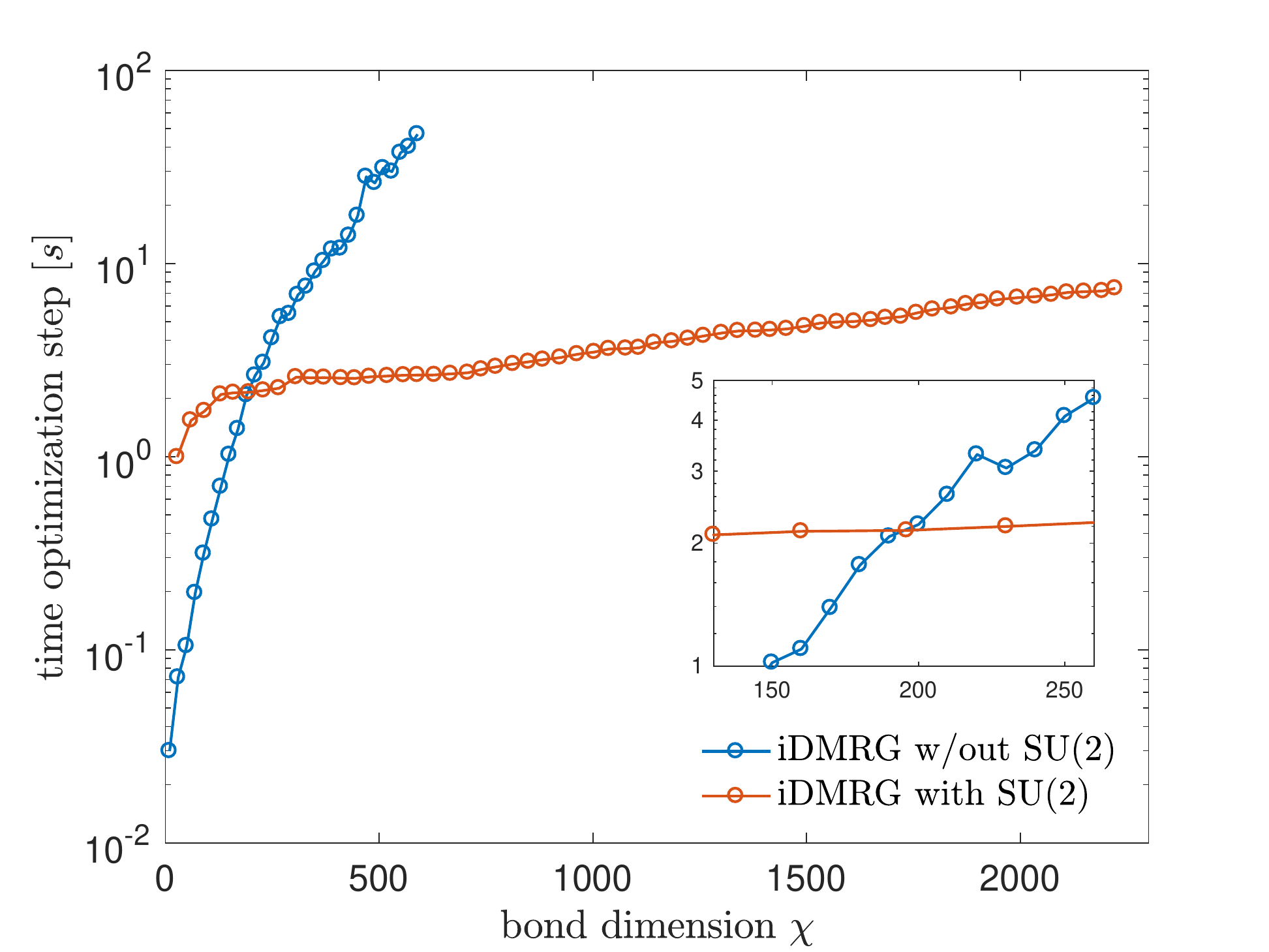}
	\caption{Time (in seconds) required to perform one optimization step in the two-site infinite-DMRG algorithm for non-symmetric and $SU(2)$-symmetric tensors. The symmetric algorithm outperforms the non-symmetric one for large bond dimensions. For the symmetric algorithm, the bond dimension shown here is the total bond dimension, see \Eref{eq:TotalBondDimension}. The crossover for other algorithms, in particular for higher-dimensional systems, is expected at lower bond dimensions due to the different overall complexity and scaling of the computational cost.} 
	\label{fig:plots_HeisenbergModel_timingAnalysis_iDMRG}
\end{figure}
\begin{figure}
 	\centering
	\includegraphics[width = 0.65\textwidth]{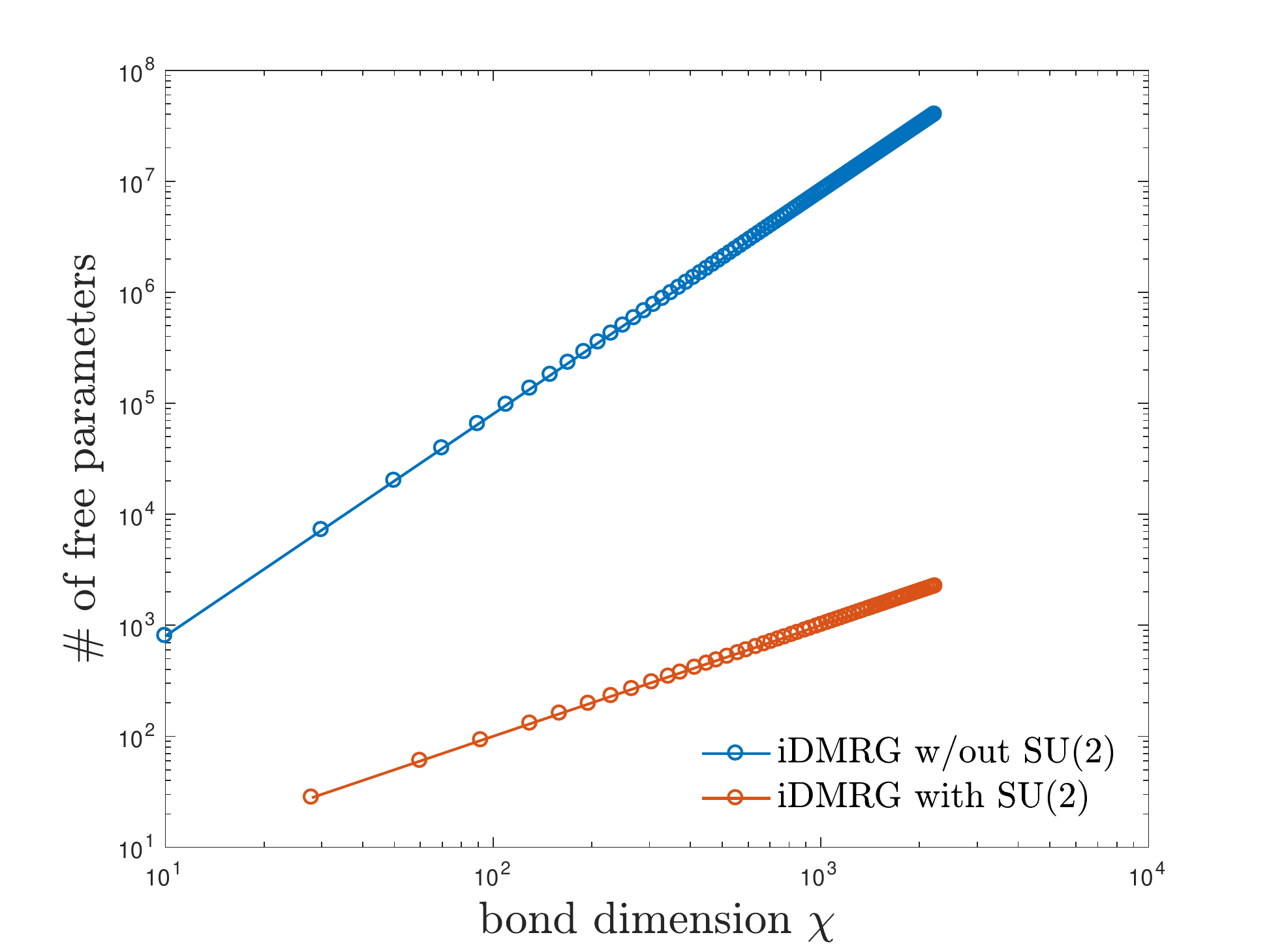}
	\caption{Number of free parameters in the infinite MPS as a function of the bond dimension, both for non-symmetric and $SU(2)$-symmetric tensors. For the symmetric algorithm, the bond dimension shown here is the total bond dimension, see \Eref{eq:TotalBondDimension}.} 
	\label{fig:plots_HeisenbergModel_parameters_iDMRG}
\end{figure}

Let us start by directly showing some results obtained with the implementation that we shall describe. We computed ground-state properties of the spin-1/2 Heisenberg chain with nearest-neighbour interactions, described by the Hamiltonian
\begin{align}
	H = J \sum_i \vec S_{i} \cdot \vec S_{i+1} \ .
\end{align}
We simulated this model with our own implementation of a $SU(2)$-invariant (i.e., targeting total spin 0 for the whole chain) infinite-DMRG algorithm with a two-site unit cell. An important step in this algorithm, apart from the whole formalism that we shall describe in this paper, is the construction of a symmetric  {matrix product operator (MPO)} for the Hamiltonian, which is described in more generality in Ref.~\cite{ourLadderPaper}. For small bond dimension, our approach has a computational overhead with respect to the non-symmetric algorithm due to the manipulation of tensors before certain operations. There is, however, a crossing in the computational cost, and the symmetric algorithm quickly becomes faster for large bond dimensions, clearly outperforming its non-symmetric counterpart. This can be seen in \fref{fig:plots_HeisenbergModel_timingAnalysis_iDMRG}, where one can see clearly that the symmetric algorithm is the only plausible option for large total bond dimension. In \fref{fig:plots_HeisenbergModel_parameters_iDMRG} one can also see that, for a fixed bond dimension, the total number of free parameters in the MPS is much lower for the symmetric algorithm than for the non-symmetric one. For instance, for bond dimension $\chi = 1500$ we go from $\sim 10^7$ variational parameters in the non-symmetric algorithm, to $\sim 10^3$ in the symmetric version. This reduction of parameters is at the hearth of the speed-up for large bond dimensions in  \fref{fig:plots_HeisenbergModel_timingAnalysis_iDMRG}, and it also implies significant memory savings in this regime. 

\begin{figure}
 	\centering
	\includegraphics[width = 0.75\textwidth]{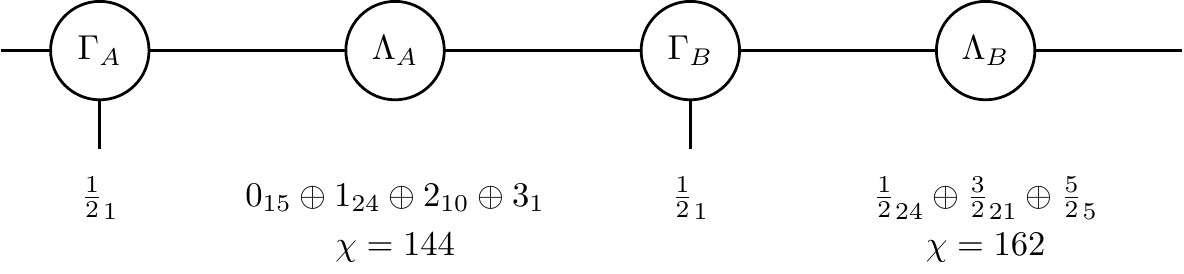}
	\caption{$SU(2)$ representations $j_t$ (where $j$ are the quantum numbers and $t$ their degeneracies) on the bond indices of the two-site MPS ansatz for a symmetric bond dimension of $\chi_{\rm sym} = 50$. The total bond dimension $\chi$ is higher due to the projections of the quantum numbers (refer to \sref{sub:symmetric_tensors}).}
	\label{fig:TikZ_Files_iDMRG_IrrepsDegeneracies}
\end{figure}

\begin{figure}
 	\centering
	\includegraphics[width = 0.65\textwidth]{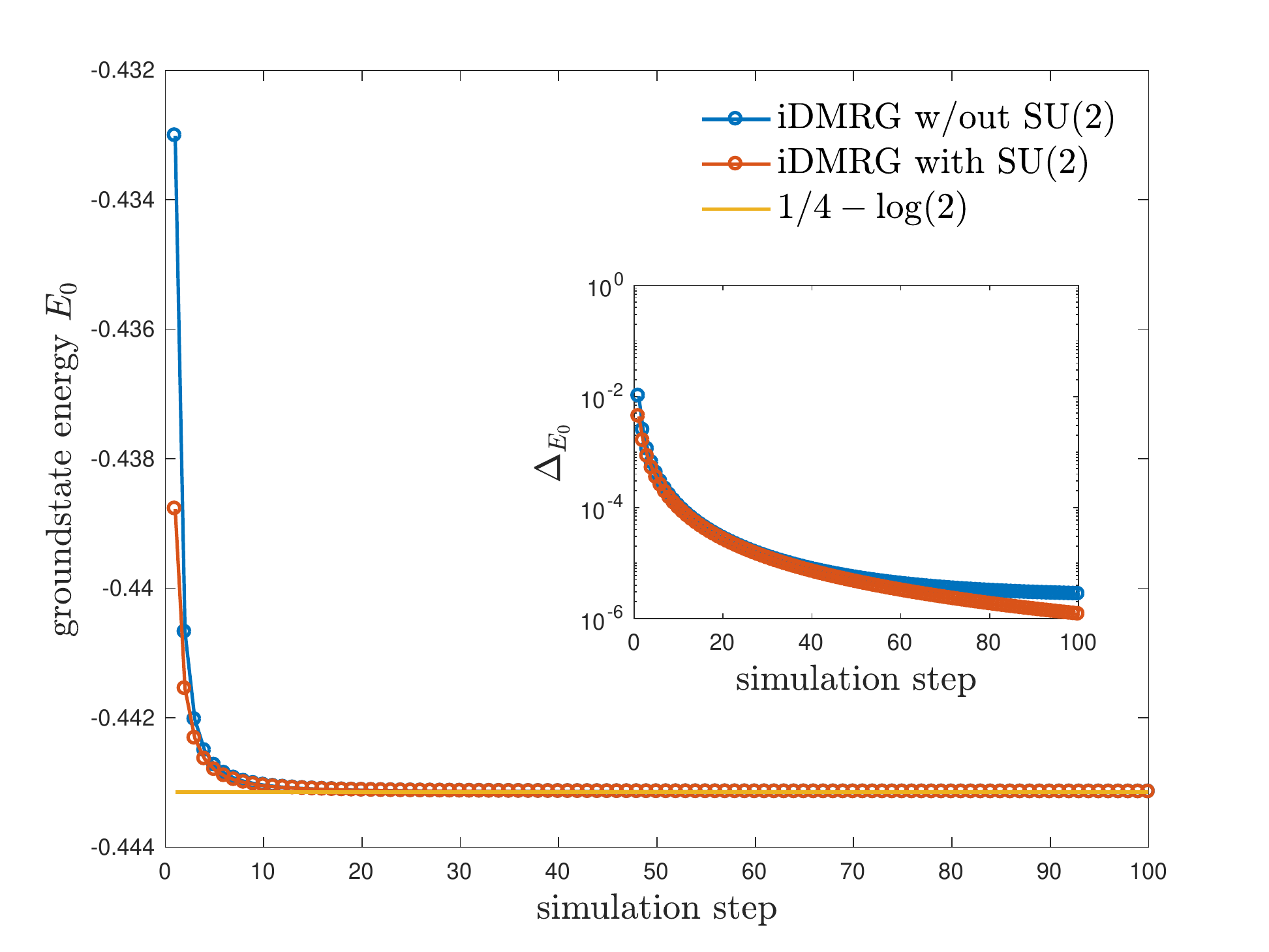}
	\caption{Convergence of the ground state energy for a fixed bond dimension $\chi = 50$ for the symmetric and non-symmetric infinite-DMRG algorithm. For comparison, the exact value of the energy is also shown as a yellow line. The inset shows the error with respect to the exact value.}
	\label{fig:plots_HeisenbergModel_convergence_iDMRG_SU2}
\end{figure}

We also show a comparison of the convergence of the ground state energy for a fixed (symmetric) bond dimension of $\chi_{\rm sym} = 50$. In terms of $SU(2)$ quantum numbers the variational parameters are distributed in the different sectors as shown in \fref{fig:TikZ_Files_iDMRG_IrrepsDegeneracies}.
The two different sets of quantum numbers for $\Lambda_A$ and $\Lambda_B$ are due to the spin-1/2 representation on the physical legs. The results are shown in \fref{fig:plots_HeisenbergModel_convergence_iDMRG_SU2}. As seen in the inset, one can clearly see that the symmetric algorithm converges faster to the exact ground state energy: while the non-symmetric algorithm reaches an accuracy plateau, the symmetric algorithm continues to decrease the error with respect to the exact ground state energy (yellow line in the main plot). Thus, according to the previous plots, for the same accuracy of the simulations a symmetric algorithm requires less free parameters than the non-symmetric one and, for relatively large bond dimension, also much less computational and memory cost. 
These results are an example of the type of simulations that one can perform with the formalism that we explain in detail in this paper. As a matter of fact, these results are just a proof of principle. Much more elaborated results are presented in Ref.~\cite{ourLadderPaper}, where we analyze the properties of $SU(2)$-symmetric ladders with chiral interactions in the thermodynamic limit. More results on the application of this formalism to two-dimensional systems with PEPS will also appear soon. 

% section appetizer_the_spin_1_2_heisenberg_chain (end)

\section{Fundamentals} % (fold)
\label{sec:fundamentals}

The basic building blocks of symmetric TN algorithms are symmetric tensors. By preserving the symmetry explicitly in the tensors at every algorithmic step, {we ensure that the optimization remains in the physical subspace of the overall Hilbert space of the many-body system}. In this section we develop some of the fundamental concepts about tensors and tensor networks with symmetries that shall be used throughout the whole paper.

\subsection{Symmetric quantum states} % (fold)
\label{sub:symmetric_quantum_states}

We start by considering a lattice $\mathcal L$ made of $N$ sites, where each site is described by a Hilbert space $\mathbb V$ of finite dimension $d$. A pure state in the total Hilbert space, $\ket{\psi}\, \in\, \mathbb V^{\otimes N}$, can always be written as
\begin{align}
	\ket{\psi} = \sum_{i_1,i_2,\hdots,i_N} c_{i_1,i_2,\hdots,i_N} \ket{i_1,i_2,\hdots,i_N}\ ,
\end{align}
where $\ket{i_k}$ denotes a single-site basis at site $k$, $i_k = 1, \ldots, d$. A TN decomposition of the state $\ket{\psi}$ consists of a set of tensors and a graph or network of directed edges\footnote{As we will see, for symmetric TNs, it is important that edges are directed.}, which determines how the tensors can be contracted together to recover the probability amplitudes $c_{i_1,i_2,\hdots,i_N}$. 

Next, we consider a compact, completely reducible group $\mathcal G$ with $U: \mathcal G \rightarrow L(\mathbb V)$ a unitary matrix representation of $\mathcal G$ on the space $\mathbb V$ of one site. (A large part of the formalism that we will describe in this paper applies to more general groups, however, from henceforth we will always have $\mathcal G=SU(2)$ in mind.) For each element $g$ of the group, $U_g: \mathbb V \rightarrow \mathbb V$ denotes a unitary matrix and $U_{g_1g_2} = U_{g_1} U_{g_2}$\footnote{Thus, we do not consider projective representations.}. 

Under the action of the symmetry, the space $\mathbb V$ of each site decomposes as the direct sum of irreducible representations (from now on ``irreps'') of the group $\mathcal G$ as
\begin{align}\label{eq:spacedecompose}
	\mathbb V \cong \bigoplus_j d_j \mathbb V_j \cong \bigoplus_j \left( \mathbb D_j \otimes \mathbb V_j \right)\ ,
\end{align}
where $\mathbb V_j$ denotes a subspace for the irrep labeled by the charge $j$, and $d_j$ is the number of times $\mathbb V_j$ appears in the decomposition of $\mathbb V$. Conveniently, this can also be written in terms of a $d_j$-dimensional degeneracy vector space $\mathbb D_j$. From now on, it is a good idea to work with the single-site basis $\ket{j,t_j,m_j}$ in $\mathbb V$, where $t_j = 1,\hdots,d_j$ denotes states within the degeneracy space $\mathbb D_j$ and $m_j$ denotes states within the irrep space $\mathbb V_j$. For $\mathcal G = SU(2)$, the charge $j$ corresponds to the total spin and $m_j$ corresponds to the spin projection along the quantization axis.

We are interested in quantum states $\ket{\psi}\, \in\, \mathbb V^{\otimes N}$ that are \emph{invariant} under the global on-site symmetry, 
\begin{align}
	\left( U_g \right)^{\otimes N}\ket{\psi} = \ket{\psi} \hspace{1.0cm} \forall\, g\, \in\, \mathcal G \ .
\end{align} 
{For the cases} of $\mathcal G = U(1)$ and $\mathcal G = SU(2)$ symmetry, {$\ket{\psi}$ would {then} correspond to a state with zero particles and zero total spin respectively.}

% subsection symmetric_quantum_states (end)

\subsection{Symmetric tensors} % (fold)
\label{sub:symmetric_tensors}

\begin{figure}
	\centering
	\includegraphics[width = 1\textwidth]{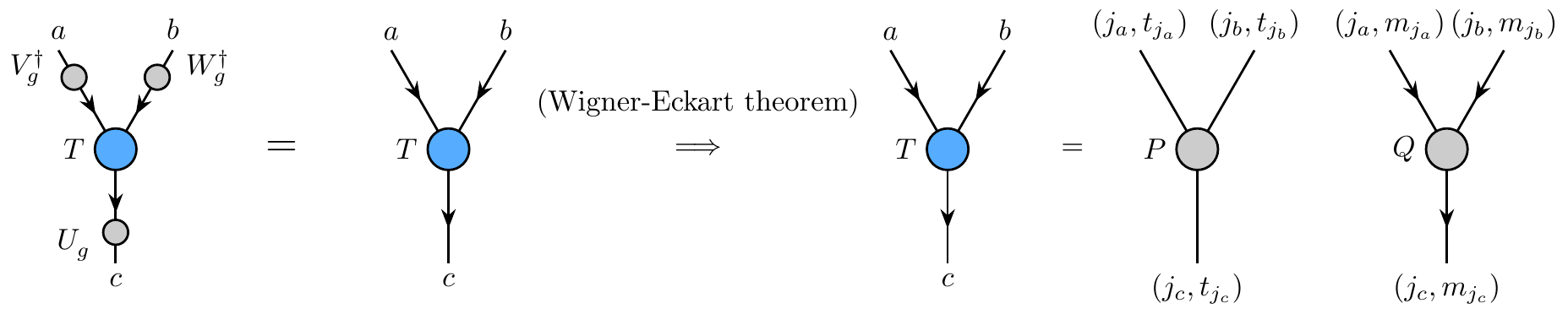}
	\caption{A symmetric tensor remains invariant under the simultaneous action of $\mathcal G$ on all its indices. The tensor can then be decomposed into a \emph{degeneracy} part $P$ and a \emph{structural} part $Q$, where $P$ holds the unconstrained parameters and $Q$ is entirely determined by the symmetry (it is an \emph{intertwiners} of the group). For a 3-index tensor, the structural tensors $Q$ are the Clebsch-Gordan coefficients. Note that the variables $j_a,j_b,j_c$ only appear as labels for the tensors $P$ and $Q$, their dimensions are however specified by $t_{j_a},t_{j_b},t_{j_c}$ and $m_{j_a},m_{j_b},m_{j_c}$ respectively.}
	\label{fig:TikZ_Files/SymmetricTensor_Rank3_1}
\end{figure}

A tensor is a multi-linear map between tensor product vector spaces, and each index of the tensor labels a basis on one of these spaces. In order to introduce the action of a symmetry, we will also assign a direction $w_i \in \{0,1\}$ to each index $i$. We use the following convention for incoming and outgoing indices: 
\begin{align}
	w_i = \left\{ \begin{array}{l l}
		-1 		& \text{if $i$ is an incoming index}\\
		+1 		& \text{if $i$ is an outgoing index}	
	\end{array}
	\right\}
\end{align}
A tensor is said to be \textit{symmetric} if it is invariant under the simultaneous action of $\mathcal G$ on all its indices (that is, on the vector spaces associated with the indices). This is shown in \fref{fig:TikZ_Files/SymmetricTensor_Rank3_1} using the usual graphical representation for tensors and TNs (see, e.g., Ref.~\cite{tn}). The direction of an index determines how the group acts on the index. We follow the convention that the symmetry operators act as the adjoint on incoming indices.

{From here on we fix the basis of symmetric tensors to the one corresponding to the irrep basis $\{\ket{j,t_j,m_j}\}$ for each index of the tensor. This implies that each index $i$ of a symmetric tensor corresponds to a triple
\begin{align}
i \equiv (j_i,t_{j_i},m_{j_i}).
\label{eq:indexTripleSym}
\end{align}
The total effective dimension $|i|$ for index $i = (j_i,t_{j_i},m_{j_i})$ is then given by 
\begin{align}
	\chi = \sum_{j_i} t_{j_i} \times | m_{j_i} | = \sum_{j_i} t_{j_i} \times (2 j_i + 1) ,
	\label{eq:TotalBondDimension}
\end{align}
where $| m_{j_i} |$ is the size of index $m_{j_i} = \{ -j_i,\hdots,+j_i \}$, and therefore $| m_{j_i} | = 2j_i + 1$.}

In this basis, a symmetric tensor with $n$ indices can be decomposed in such a way that the degrees of freedom that are not fixed by the symmetry can be isolated in a set of \emph{degeneracy} tensors $\{P_{j_1,j_2,\ldots,j_n}\}$, whereas the rest of the degrees of freedom correspond to \emph{structural} tensors $\{Q_{j_1,j_2,\ldots,j_n}\}$ which are entirely determined by the symmetry. (This follows from the Wigner-Eckart theorem~\cite{wignereckart}.) In practice, this means that only the degeneracy tensors $P$ encode the variational parameters in a symmetric tensor network optimization, which implies a reduction in memory cost as well as a computational speed-up in TN algorithms.

Let us elaborate this decomposition for a 3-index symmetric tensor $T_{abc}$. By fixing a particular value of total spin labels $j_a,j_b,j_c$ we select a ``block'' of components from the tensor. We denote this block by $T_{j_aj_bj_c}$. Then the Wigner-Eckart theorem implies that this block factorizes into a tensor product of a degeneracy tensor $P_{j_aj_bj_c}$ and a structural tensor $Q_{j_aj_bj_c}$, namely,
\begin{align}\label{eq:degblocks}
T_{j_aj_bj_c} = P_{j_aj_bj_c} \otimes Q_{j_aj_bj_c}.
\end{align}
Here $P_{j_aj_bj_c}$ contains all the degrees of freedom of the block $T_{j_a j_b j_c}$ that are not fixed by the symmetry, whereas $Q_{j_aj_bj_c}$ is completely determined by the symmetry. We will sometimes refer to a degeneracy tensor $P_{j_aj_bj_c}$ as a ``degeneracy block'' labeled by charges $(j_a,j_b,j_c)$. In terms of components as in \Eref{eq:degblocks}, $T_{abc}$ can be written as
\begin{align}
	T_{abc} = \bigoplus_{j_a j_b j_c} \left( P_{j_aj_bj_c} \right)_{t_{j_a}t_{j_b}t_{j_c}} \otimes \left(Q_{j_aj_bj_c} \right)_{m_{j_a}m_{j_b}m_{j_c}}
	\label{cano}
\end{align}
for compatible combinations $(j_a,j_b,j_c)$.

Such decompositions can be exploited to store symmetric tensors compactly in a numerical optimization. For example, the 3-index tensor $T$ considered above can be stored compactly as the following list of data (instead of storing all its components):
\begin{align}\label{eq:tensordata}
	T_{abc} = \left\lbrace  \{j_a,t_{j_a}\},  \{j_b,t_{j_b}\} , \{j_c,t_{j_c}\} , \vec W , \left\lbrace P_{j_aj_bj_c} \right\rbrace \right\rbrace,
\end{align}
which contains the charges, degeneracies and directions $\vec{W} \equiv [w_a, w_b, w_c]$ for every index of the tensor, and a list of all the degeneracy blocks along with their respective charge labels. The number of parameters to store for the symmetric tensor is now generally given by 
\begin{align}
	\sum_{\rm{compatible}\ j_a j_b j_c} t_{j_a} (2j_a + 1) \cdot t_{j_b} (2j_b + 1) \cdot t_{j_c} (2j_c + 1)
\end{align}
which can be further reduced due to the fixed parameters of the symmetry in all tensors $Q_{j_aj_bj_c}$. Therefore, the cheapest way to store this 3-index $SU(2)$-invariant tensor only needs a number of
\begin{align}
	\sum_{\rm{compatible}\ j_a j_b j_c} t_{j_a} t_{j_b} t_{j_c}
\end{align}
components. This corresponds exactly to the product of degeneracies for every block in the tensor labeled by a compatible set of the quantum numbers $(j_a,j_b,j_c)$. As a concrete example let us consider the 3-index tensor with vector spaces $\mathbb V = \mathbb V_0 + 3 \mathbb V_1$, which corresponds to choosing $i = ([0,1],[1,3],\{[0],[-1,0,+1]\}$ for all three indices using the notation of \Eref{eq:indexTripleSym}. The number of parameters for the full tensor would now be
\begin{align}
	\prod_{i = a,b,c} \sum_{j_i} t_{j_i} \times (2j_i + 1) = 10^3 = 1000\ .
\end{align}
The $SU(2)$-symmetric tensor (consisting of the five blocks $T_{000}, T_{011} T_{101}, T_{110}$ and $T_{111}$) however only has a number of $1 + 9 + 9 + 9 + 27 = 55$ free parameters.\\

Though the structural tensors $\left\lbrace Q_{j_a j_b j_c} \right\rbrace$ are completely determined by the symmetry group, for tensors with more than three indices they can be a complicated function of the index data. Furthermore, the index data does not generally specify a structural tensor uniquely\footnote{The reason for this is an internal freedom for the construction of structural tensors for more than three indices. This will be explained in detail in \sref{sec:fusion_trees}.}. Thus, it might seem surprising that we do not plan to store a list of structural tensors $\left\lbrace Q_{j_aj_bj_c} \right\rbrace$ in the data (\ref{eq:tensordata}), even for more general tensors. There is an important reason behind this. As we will describe later in the paper, we will work with (and store as part of the tensor data) so-called fusion trees -- a kind of a tree graph decorated with charge labels -- which completely specify a structural tensor. Furthermore, manipulations of structural tensors correspond to \textit{algebraic} (group-theoretic) manipulations of the corresponding fusion tree. Thus, we will \emph{never} deal with a structural tensor explicitly in our code. 

There are several advantages of working directly with fusion trees. First, of course, we save memory since have we do not have to store the structural tensors. Second, manipulations of fusion trees (thus of the corresponding structural tensors) are algebraic and therefore virtually free of the potential floating-point errors that could occur when manipulating structural tensors numerically (that is, by storing them as numerical arrays in memory), thus leading to more accurate simulations. (Accumulated errors in the components of the structural tensors prevent the exact preservation of the symmetry constraints.) Third, a formalism based on fusion trees can be readily adapted to simulate \emph{anyonic systems}, since anyon models have a natural description in terms of fusion trees. (More specifically, the data that specifies an anyon model -- such as fusion rules, braidings, $F$-moves, and so on -- can be conveniently described as elementary manipulations or moves of fusion trees)~\cite{anyons}. And finally, we also find that the fusion-tree approach is more convenient to scale up the formalism to tensors with many indices, as is required for e.g., 2d algorithms with Projected Entangled Pair States.

On the other hand, manipulations of fusion trees have to be implemented as a sequence of elementary algebraic moves, which often have to be carried out skillfully. In contrast, structural tensors can be manipulated (e.g. contracted or permuted) just like regular tensors and their algebraic (group-theoretic) properties can be disregarded. Thus, the use of structural tensors, in place of fusion trees, may lead to a simpler implementation of the symmetry at higher memory costs, possible loss of accuracy when contracting large tensors, and lack of extensibility to anyon systems.

Let us now turn to tensors with more than three indices. We will discuss general symmetric tensors in more detail in \sref{sec:fusion_trees}. Here we just describe how the decomposition \Eref{eq:degblocks} generalizes to tensors with a greater number of indices. A symmetric tensor with $k$ indices can be generally decomposed in terms of blocks $T_{j_{1} \hdots j_{k}}^{j^\text{int}_1,\hdots,j^\text{int}_l}$ which are labeled by the charges given by $j_{1} \hdots j_{k}$ associated with the $k$ indices, and also a set of $l=k-3$ \textit{internal} charges $j^\text{int}_{1} \hdots j^\text{int}_{l}$ (described in \sref{sec:fusion_trees}). Each of these blocks decompose as
\begin{align}\label{eq:gentensor}
	T_{j_{1} \hdots j_{k}}^{j^\text{int}_1,\hdots,j^\text{int}_l} = P_{j_{1} \hdots j_{k}}^{j^\text{int}_1,\hdots,j^\text{int}_l}  \otimes Q_{j_{1} \hdots j_{k}}^{j^\text{int}_1,\hdots,j^\text{int}_l},
\end{align}
which generalizes \Eref{eq:degblocks}. It is important to note that though the $P$ (and $Q$) tensors are labeled by the charges $\{j^\text{int}_1,\hdots,j^\text{int}_l , j_1 \hdots j_k\}$, they nonetheless have only $k$ indices each (that is, the same number of indices as the total tensor $T$). The size of tensor $P_{j_{1} \hdots j_{k}}^{j^\text{int}_1,\hdots,j^\text{int}_l}$ is equal to $t_{j_1}\times t_{j_2}\hdots \times t_{j_k}$. Analogously, the size of tensor $Q_{j_{1} \hdots j_{k}}^{j^\text{int}_1,\hdots,j^\text{int}_l}$ is equal to $|m_{j_1}|\times|m_{j_2}|\hdots \times |m_{j_k}|$.

% subsection symmetric_tensors (end)

\subsection{Clebsch-Gordan tensors} % (fold)
\label{sub:clebsch_gordan_tensors}

\begin{figure}
	\centering
	\includegraphics[scale=.9]{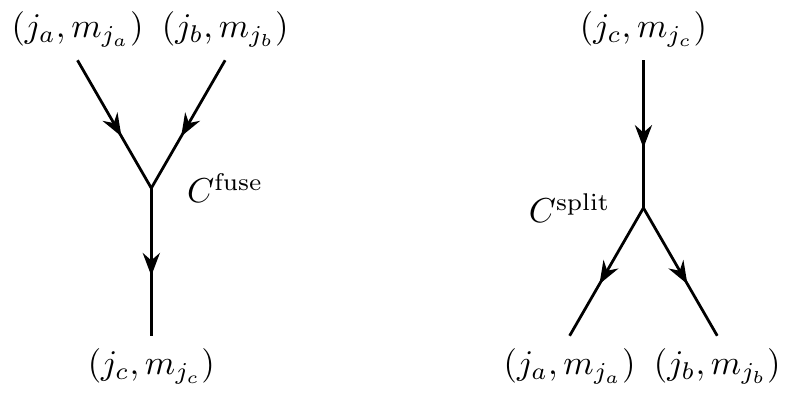}
	\caption{Graphical representation of the fusion tensor $C^\text{fuse}$ and the splitting tensor $C^\text{split}$. The graphical representation of $C^\text{split}$ is obtained by a vertical reflection of $C^\text{fuse}$.}
	\label{fig:TikZ_Files_ClebschGordan_1}
\end{figure}

\begin{figure}
	\centering
	\includegraphics[width=.9\textwidth]{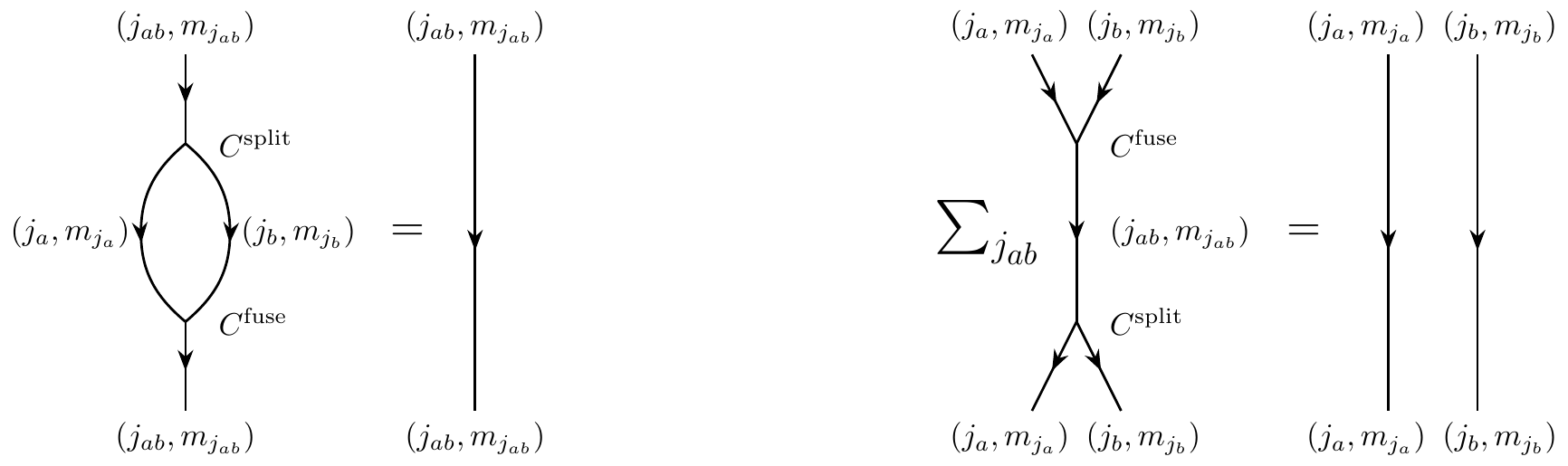}
	\caption{The tensors $C^\text{fuse}$ and $C^\text{split}$ satisfy the useful identities depicted here. The sums over all spin projections $m_{j_a}$, $m_{j_b}$ and $m_{j_{ab}}$ are implicit.}
	\label{fig:TikZ_Files_ClebschGordan_Identities_1}
\end{figure}

The structural tensors are nothing but the intertwiners of the symmetry group~\cite{hammermesh}. A 3-index structural tensor, in particular, is simply a tensor whose components are the Clebsch-Gordan coefficients of the group. We refer to a 3-index structural tensor as a \textit{Clebsch-Gordan tensor}. Structural tensors (or intertwiners) with three or more indices can always be decomposed in terms of Clebsch-Gordan tensors, which will therefore appear frequently in our discussion. A Clebsch-Gordan tensor $C^\text{fuse}$ describes the (unitary) change of basis from the tensor product of irreps $a$ and $b$ to a total irrep $c$. We will also say that irreps $a$ and $b$ \textit{fuse} to $c$ and denoted this as $a \times b \rightarrow c$. We have,
\begin{align}\label{eq:clebsch}
\ket{j_c,m_{j_c}} = \sum_{m_{j_a}m_{j_b}} (C^\text{fuse})_{(j_a,m_{j_a}),(j_b,m_{j_b})}^{(j_c,m_{j_c})} \ket{j_a,m_{j_a}} \otimes \ket{j_b,m_{j_b}}.
\end{align}
The Clebsch-Gordan coefficients that appear above vanish unless 
\begin{align}\label{eq:fusionrules}
j_c \in \{ \vert j_a - j_b \vert\, ,\, \hdots\, ,\, j_a + j_b\}.
\end{align}
We will say that irreps $j_a, j_b, j_c$ are \textit{compatible} if $j_c$ belongs to the above set. The set of all compatible triples $\{(j_a, j_b, j_c)\}$ are called the \textit{fusion rules} of the symmetry. The inverse Clebsch-Gordan tensor $C^\text{split}$ describes how irrep $c$ \textit{splits} into a tensor product of irreps $a$ and $b$, denoted as $c \rightarrow a \times b$. The graphical representation of the tensors $C^\text{fuse}$ and $C^\text{split}$ is shown is \fref{fig:TikZ_Files_ClebschGordan_1}. Since these tensors are isometric, they satisfy the relations depicted in \fref{fig:TikZ_Files_ClebschGordan_Identities_1}.

Next, we turn to structural tensors with more than three indices. As described previously, we will specify these by means of fusion trees.

% subsection clebsch_gordan_tensors (end)

% section fundamentals (end)

\section{Fusion trees} % (fold)
\label{sec:fusion_trees}

\begin{figure}
	\centering
	\includegraphics[scale=.9]{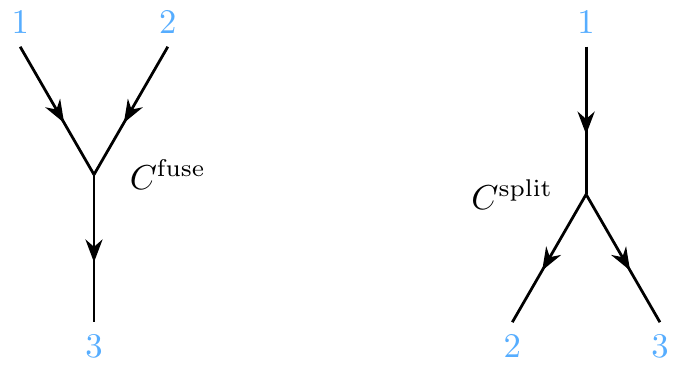}
	\caption{{Graphical representation of a fusion and a splitting node. Each node has an ``orientation'' (order of the incident edges) which is indicated by the numbers in blue. Elementary fusion nodes (left) are labeled in a clockwise order whereas elementary splitting nodes (right) are labeled in an anticlockwise order.}}
	\label{fig:TikZ_Files_ClebschGordan_2}
\end{figure}

A \emph{fusion tree} is a directed trivalent tree graph that describes how a set of irreps \textit{pairwise} fuse to a total irrep. In our context, fusion trees will be used to specify the structural part of a symmetric tensor. Thus, the irreps being fused are those carried by the indices of the tensor. The fusion tree for both $C^\text{fuse}$ and $C^\text{split}$ consists of only one node and is graphically represented in \fref{fig:TikZ_Files_ClebschGordan_2}. Note that this graphical representation is (intentionally) quite similar to the one introduced for actual tensors $C^\text{fuse}$ and $C^\text{split}$ in \fref{fig:TikZ_Files_ClebschGordan_1}, except that we no longer specify any $j$ and $m$ labels in the corresponding fusion tree.

\subsection{Data structure for a fusion tree} % (fold)
\begin{figure}[ht]
	\centering
	\includegraphics[width=\textwidth]{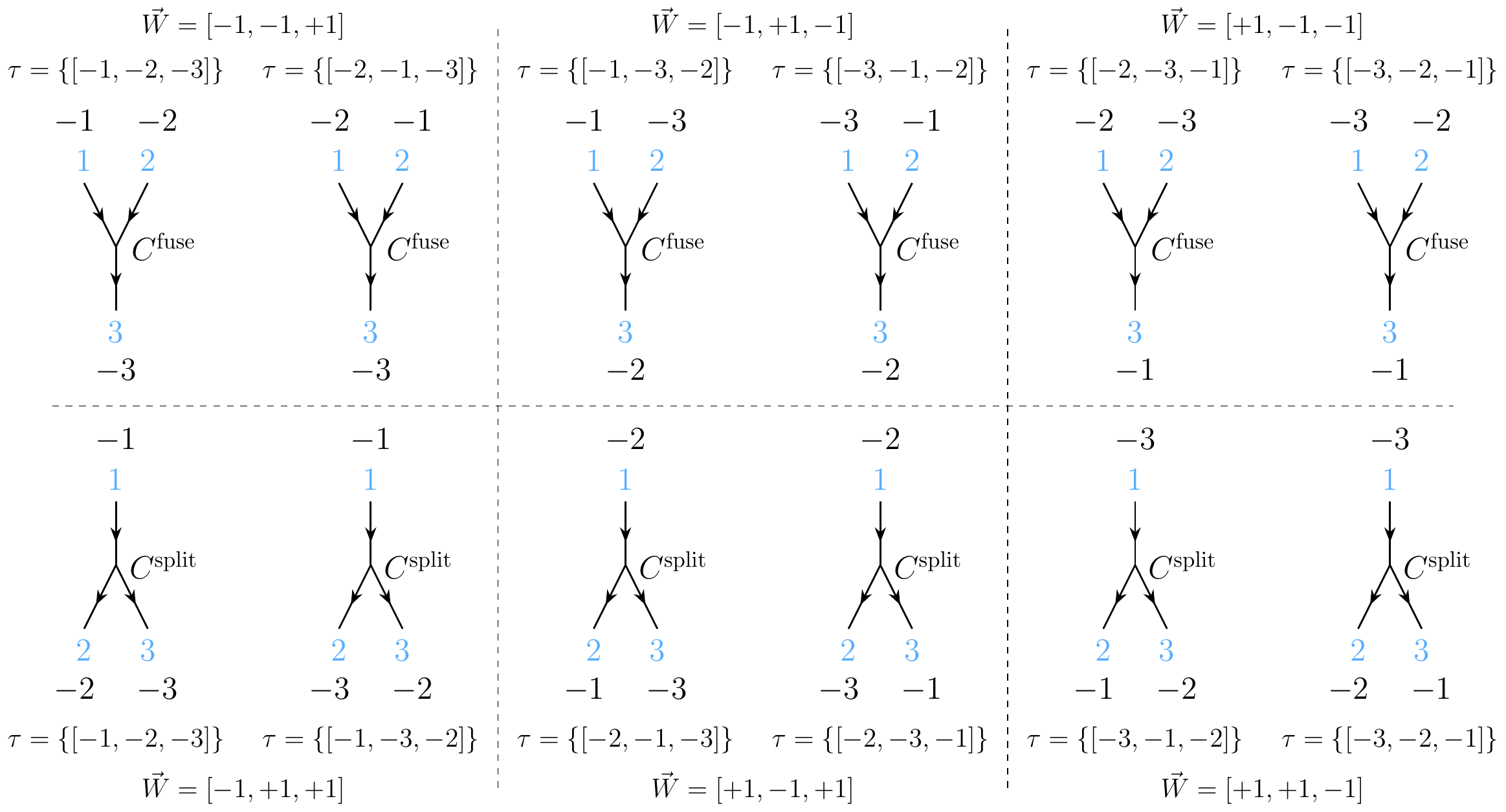}
	\caption{{Different ways of representing the Clebsch-Gordan coefficients for $C^\text{fuse}$ (fusion) and $C^\text{split}$ (splitting) as numerical arrays. Blue labels indicate the position of an index in the $\tau$ vector, whereas black labels indicate the order of the actual {tensor}.}}
	\label{fig:TikZ_Files_ClebschGordan_General_3}
\end{figure}

\label{sub:data_structure_for_a_fusion_tree}
From our description of fusion trees above, it follows that in order to store a general fusion tree on a computer one has to store the following data: (i) a tree graph, and (ii) the orientation of each node, namely, whether the node is a fusion or splitting node. Let us start with the tree graph. To this end, we label edges carrying incoming and outgoing edges by negative numbers $-1, -2, \hdots$, whereas edges carrying the intermediate irreps (the various pairwise fusion outcomes) are labeled by positive numbers $1, 2, \hdots$. Edges that carry irrep/spin 0 are labeled by $0$ and are referred to as \textit{dummy} edges or indices. Each node of the tree is specified by a three-component vector $\tau$, whose entries correspond to the labels of the three edges that intersect at the node. A fusion or splitting node has a ``fixed orientation'', namely, an order in which the three edge labels appear in the vector $\tau$ associated with the node. We will assume the node orientation specified by the blue labels in \fref{fig:TikZ_Files_ClebschGordan_2} and refer to it as the \emph{conventional orientation} of the fusion and splitting nodes. The edges labeled by negative integers correspond to the indices of the corresponding structural tensors. The indices of the structural tensors are ordered in the same way as those of the total symmetric tensor (and the order of indices of the total tensor is supplied when the tensor is specified). See \fref{fig:TikZ_Files_ClebschGordan_General_3} for an illustration of the $\tau$ vectors corresponding to various edge label assignments. Note again that the edge labels always appear according to the fixed node orientation described above.

\begin{figure}
	\centering
	\includegraphics[scale = 0.65]{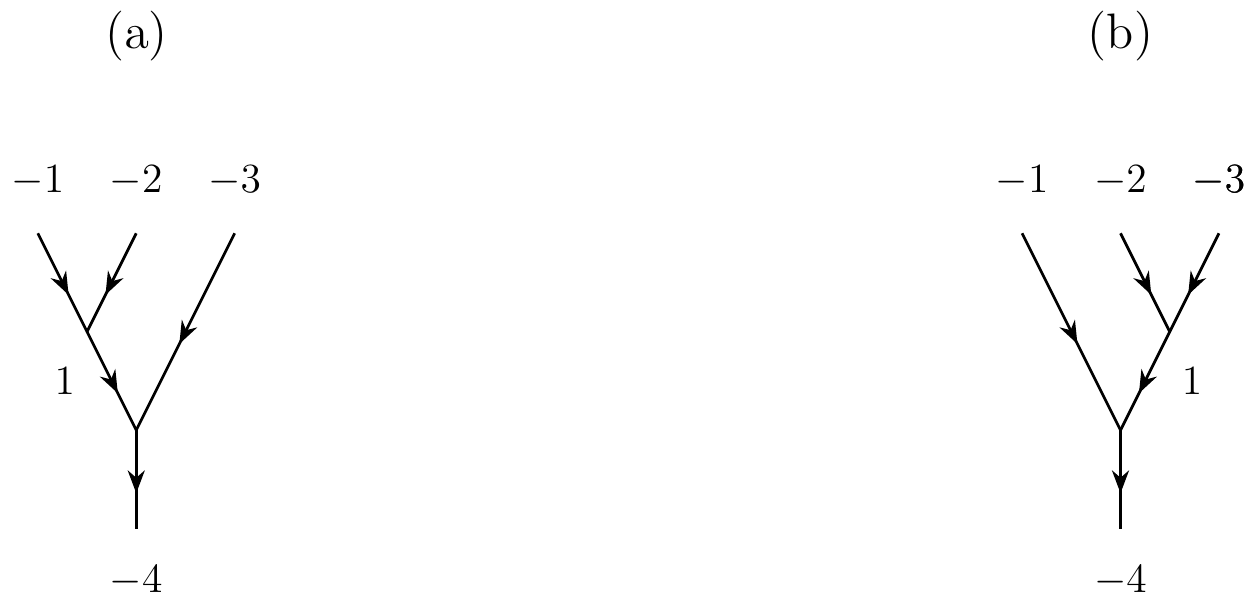}
	\caption{Different ways to pairwise fuse three spins into a single spin, described by different fusion trees. The 4-index structural tensors corresponding to these two trees are not the same, but they are related to each other by a change of basis that is implemented by an $F$-move {(see \sref{sub:f_moves})}.}
	\label{fig:TikZ_Files_Rank_4_TensorSpinNetwork}
\end{figure}

A general fusion tree -- corresponding to a $k$-index symmetric tensor -- has $k-2$ nodes, $k$ open edges, and $k-3$ internal edges. A general fusion tree is specified by a list of $\tau$ vectors, one for each node in the tree. As mentioned above one also needs to store a label to distinguish fusion nodes from splitting nodes, such that the tree can be uniquely constructed. A tree without this list is incomplete and the same list of $\tau$ vectors can lead to different configurations, see for instance \fref{fig:TikZ_Files_FusionTreeDirections_3}. To this end, we append to the data structure of a fusion tree a vector $\sigma$ such that for each node $i$:
\begin{align}
	\sigma_i = \left\{ \begin{array}{l l}
		-1 		& \text{if node $i$ is a fusion node}\\
		+1 		& \text{if node $i$ is a splitting node}	.
	\end{array}
	\right. 
\end{align}
Using (i) the tree graph and (ii) the list of the node orientations it is possible to construct arbitrary trees and also graphs which can contain loops\footnote{In our implementation we will only work with trees as internal structures for tensors and we will make further restrictions in the next section. Although graphs are covered by our fusion tree implementation they provide a less efficient description of tensors.}. Examples of simple fusion trees for 4-index tensors are given in \fref{fig:TikZ_Files_Rank_4_TensorSpinNetwork}, amounting to the different ways to fuse three irreps into one. Notice that, as opposed to a fusion tree with just one node, now the fusion tree \emph{is not unique}, since one can recouple the intermediate irreps in different ways. In practice, one chooses a particular fusion tree, which essentially corresponds to fixing a particular basis for the tensor. The different fusion trees, corresponding to different bases for the tensors, are related to each other by the so-called $F$-moves~\cite{ftensors}. We will talk extensively about $F$-moves in \sref{sub:f_moves}. 

\begin{figure}
	\centering
	\includegraphics[scale = 0.65]{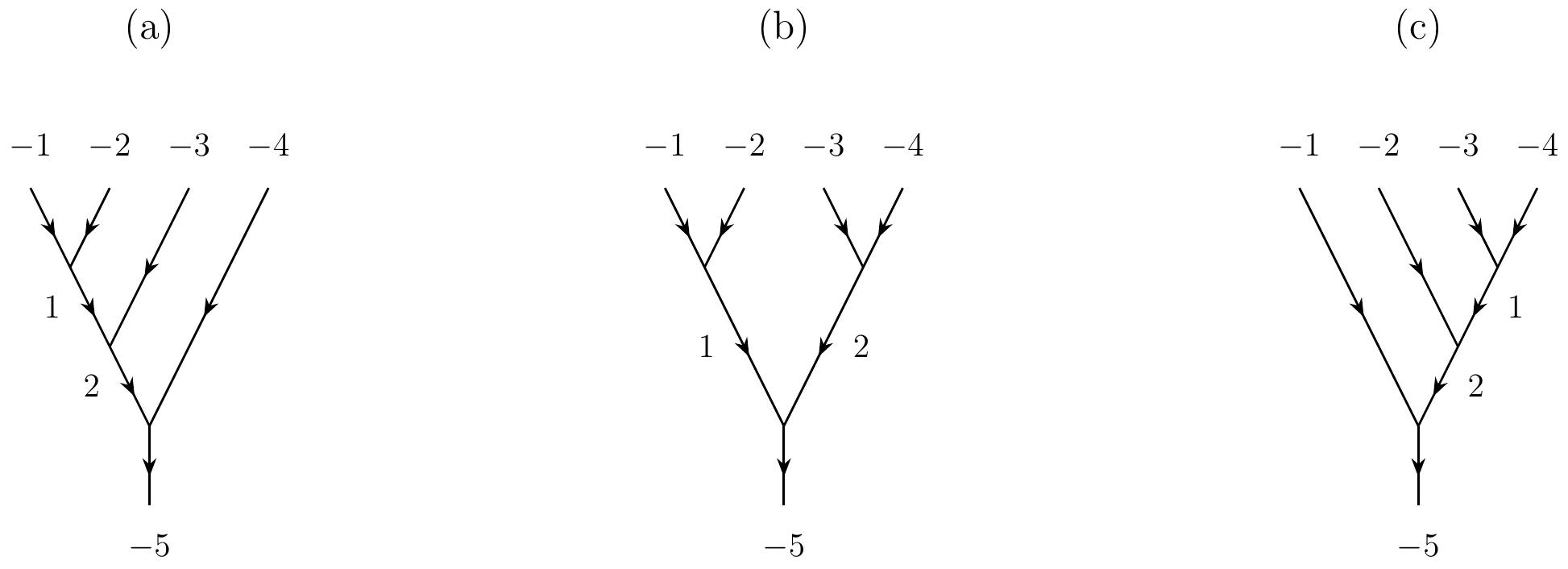}
	\caption{Some of the different ways of pairwise fusing four spins into a single spin, corresponding to different fusion trees.}
	\label{fig:TikZ_Files_Rank_5_TensorSpinNetwork}
\end{figure}

\fref{fig:TikZ_Files_Rank_5_TensorSpinNetwork} shows some fusion trees for five irreps (corresponding to the structural part of 5-index symmetric tensors with four incoming edges and one outgoing edge). Using our data structure these trees are stored as the following list of $\tau$ vectors:

\begin{align}\label{eq:example5irreps}
	\tau^{(a)} &= \lbrace \lbrack -1,-2,1 \rbrack , \lbrack 1,-3,2 \rbrack , \lbrack 2,-4,-5 \rbrack \rbrace \\
	\tau^{(b)} &= \lbrace \lbrack -1,-2,1 \rbrack , \lbrack -3,-4,2 \rbrack , \lbrack 1,2,-5 \rbrack \rbrace \\
	\tau^{(c)} &= \lbrace \lbrack -3,-4,1 \rbrack , \lbrack -2,1,2 \rbrack , \lbrack -1,2,-5 \rbrack \rbrace.
\end{align}

All three trees in this example would have the same list of node orientations $\sigma = [-1,-1,-1]$. Note that the list of the $\tau$ vectors itself does not follow any ordering and can be arbitrary as long as it matches with the ordering of the $\sigma$ vector. This has some implication in the implementation of certain functions and this point will reappear in later discussions. However, for a better readability, we will follow the convention that the internal edges appear in increasing order in $\tau$.

\begin{figure}
	\centering
	\includegraphics[width=.9\textwidth]{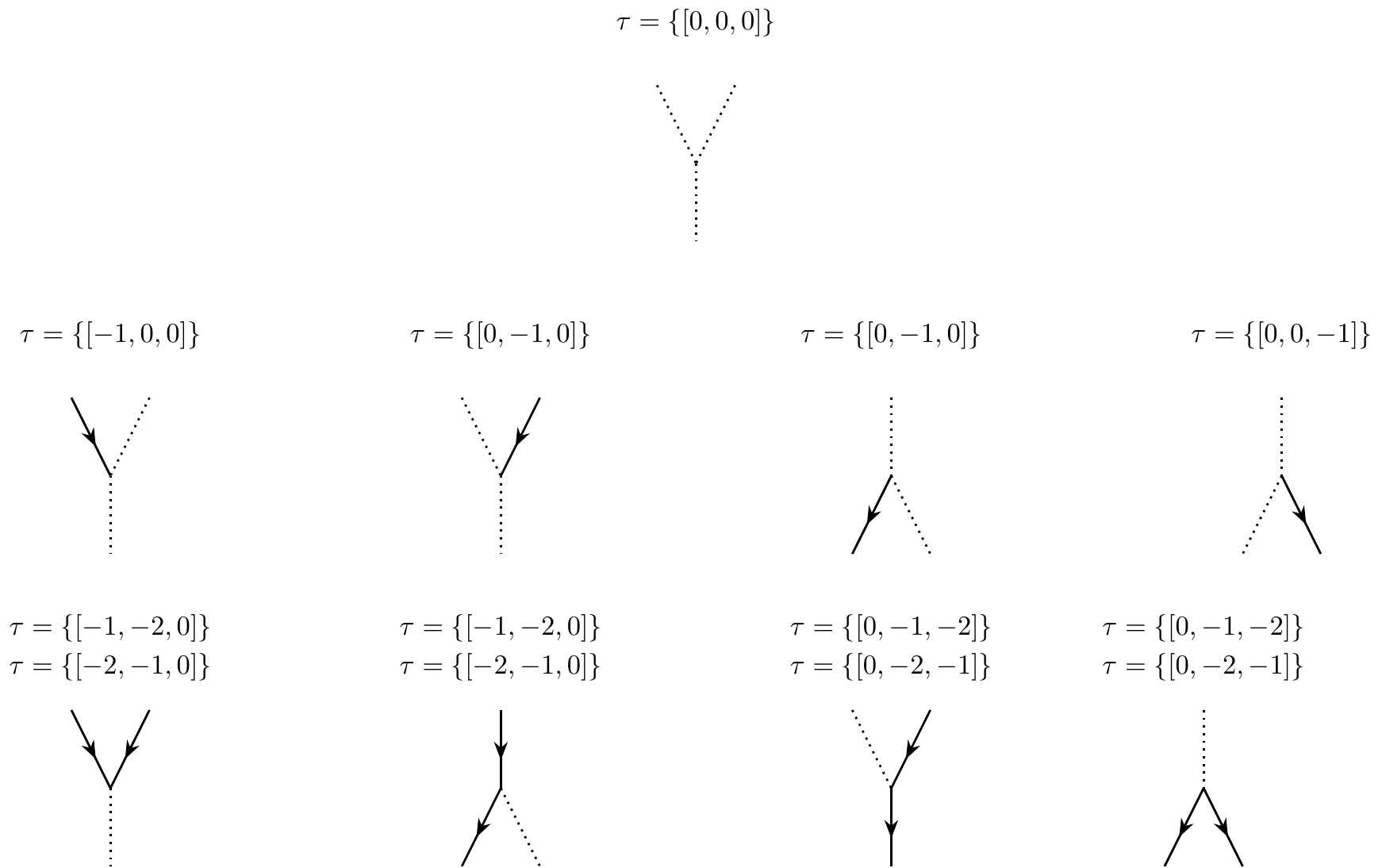}
	\caption{Fusion trees for tensors with 0 (top row),1 (middle row) and 2 (bottom row) indices, described by single Clebsch-Gordan tensors with 3, 2, or 1 ``dummy'' indices (dotted lines) respectively.}
	\label{fig:TikZ_Files_ClebschGordan_LowRank}
\end{figure}

It may seem that we have neglected structural tensors with two indices (matrices), one index (vectors), and no index (scalars) so far. It turns out that structural tensors with two, one, and zero indices can be seen as special instances of 3-index Clebsch-Gordan tensors, namely, Clebsch-Gordan tensors with one, two, and three dummy indices respectively, as shown in \fref{fig:TikZ_Files_ClebschGordan_LowRank}. This allows us to indeed focus only on Clebsch-Gordan tensors as the basic building blocks.

In the next section we will introduce more general forms of fusion trees that can be handled with our approach and which may appear in tensor network calculations.

% subsection data_structure_for_a_fusion_tree (end)

\subsection{Simple, Yoga, and Monster fusion trees} % (fold)
\label{sub:simple_yoga_and_monster_fusion_trees}

\begin{figure}
	\centering
	\includegraphics[width=.6\textwidth]{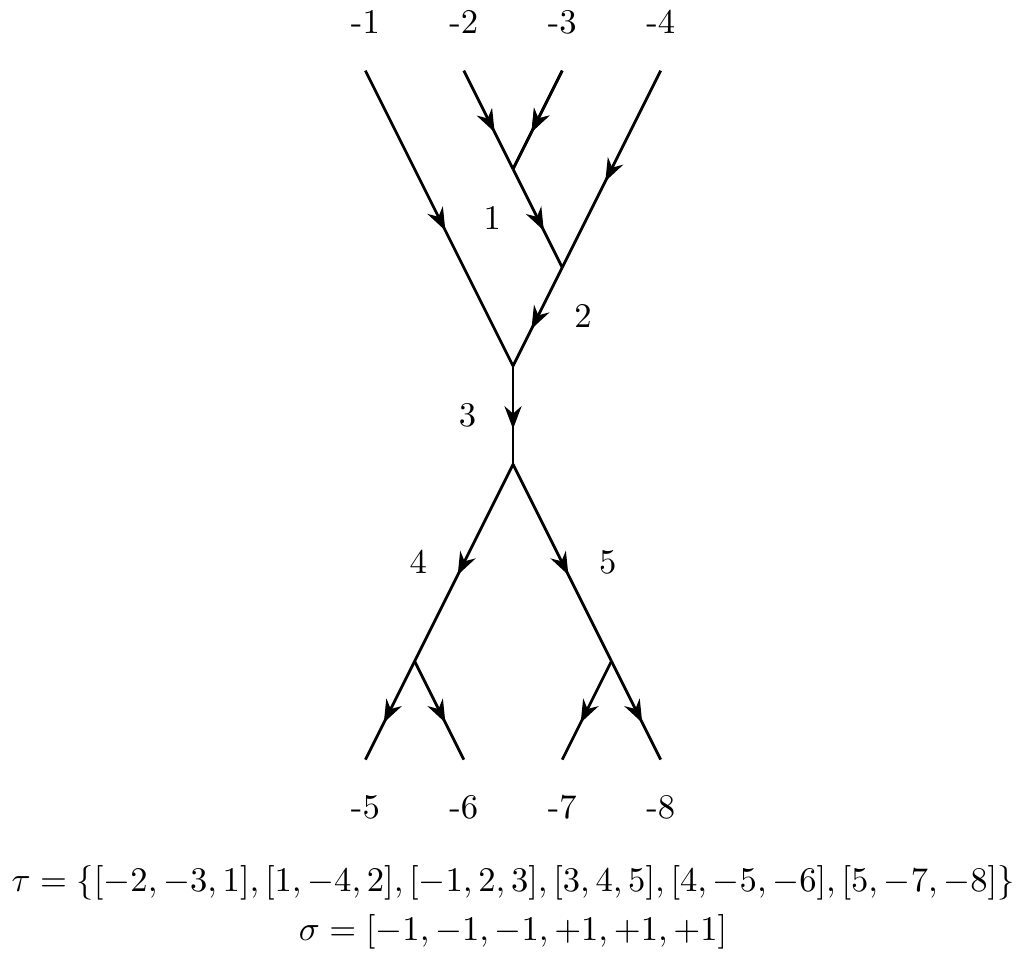}
	\caption{A simple fusion tree for an 8-index symmetric tensor. The incoming edges can be separated from the outgoing edges by cutting only one internal edge.}
	\label{fig:TikZ_Files_FusionTreeDirections_2}
\end{figure}

In this section we want to introduce a classification for fusion trees based on their properties. The provided examples of fusion trees have been either elementary Clebsch-Gordan nodes or fusion trees of the first class, which we call \textit{simple fusion trees}. Simple fusion trees have a clear separation of incoming and outgoing legs, which is either immediate as for the trees in \fref{fig:TikZ_Files_Rank_5_TensorSpinNetwork}, or achievable by cutting only one internal leg. An example for this is provided for a more extended 8-index fusion tree in \fref{fig:TikZ_Files_FusionTreeDirections_2}. Simple fusion trees also have the nice property that all the orientation of the nodes can be uniquely determined by specifying the tree graph and the vector $\vec W$ that sets the directions of the open legs. One could therefore use a deterministic algorithm to determine which node is a fusion and which node is a splitting node based on the provided information. In practice however, it is more useful to specify each tree right from the beginning by the tree graph and the list of node orientations, and manipulate these lists during tensor operations.

\begin{figure}
	\centering
	\includegraphics[width=.6\textwidth]{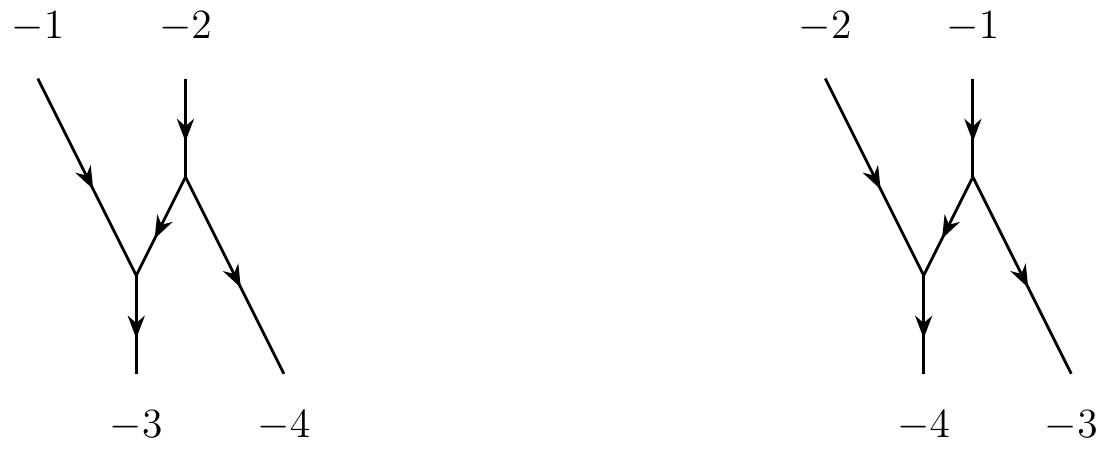}
	\caption{Examples of yoga fusion trees, which correspond to $\tau = \{[-1,1,-3],[-2,1,-4]\}$ but with different lists $\sigma_{\rm l} = [-1,+1]$ and $\sigma_{\rm r} = [+1,-1]$.}
	\label{fig:TikZ_Files_FusionTreeDirections_3}
\end{figure}

The reason for this is the fact that a unique determination of all node orientations fails for specific fusion tree configurations, namely those where an internal edge appears on position 2 of two nodes, see \fref{fig:TikZ_Files_FusionTreeDirections_3}. In this example both trees would correspond to $\tau = \{[-1,1,-3],[-2,1,-4]\}$ and $\vec W = [-1,-1,+1,+1]$, but with different lists $\sigma_{\rm l} = [-1,+1]$ and $\sigma_{\rm r} = [+1,-1]$. Notice that the trees shown in \fref{fig:TikZ_Files_FusionTreeDirections_3} are not simple fusion trees, since one cannot separate the incoming and outgoing open edges by cutting one internal leg. We call these ambiguous trees \emph{yoga trees} or \emph{yoga diagrams}\footnote{The reason for the name is their similarity with some yoga asanas. We used this name informally, but we used it so often that eventually it became our notation. (SS was not involved in devising this terminology, but RO was.)}. An important property of Yoga trees is that they can be transformed into simple trees by so-called $F$-moves, which will be explained in later chapters.

\begin{figure}
	\centering
	\includegraphics[scale = 0.9]{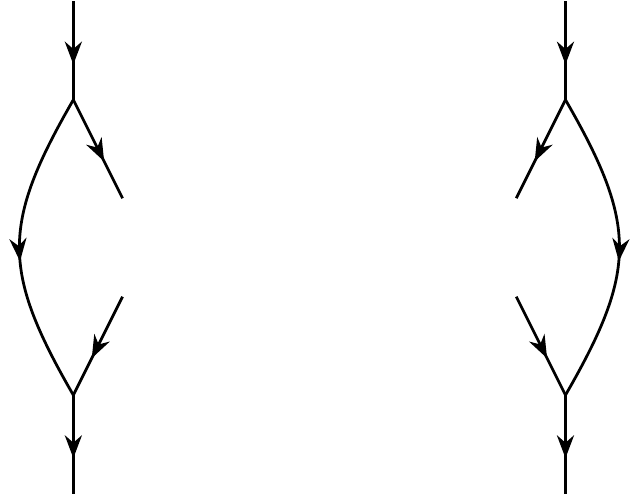}
	\caption{Examples of monster trees. The incoming edges are not separated from the outgoing edges by cutting the internal edge.}
	\label{fig:TikZ_Files_transferMatrix_transferMatrix_1}
\end{figure}

We also note that it is possible to have more general fusion trees, ones in which it is not even possible to separate out the incoming and outgoing open edges by cutting any number of edges. We call such trees \textit{monster fusion trees}\footnote{Storing and manipulating large monster trees can be more challenging to code. {They may even scare some programmers.}}, see e.g. \fref{fig:TikZ_Files_transferMatrix_transferMatrix_1}. Contrary to Yoga trees, Monster trees can no longer be transformed into a simple tree by $F$-moves and are therefore a more robust structure.

It suffices to restrict to simple fusion tree to implement any symmetric tensor network algorithm. In fact, our implementation is based only on simple fusion trees\footnote{Yoga trees may also appear at intermediate steps of algorithms, however they will be transformed back to simple trees.}. However, we remark that the choice of fusion trees may impact the ease and efficiency of a particular implementation of symmetric tensor networks. The use of monster trees, which are the most general type of fusion tree that we can consider, may allow for a more natural and convenient implementation of certain tensor network operations, though requiring a more complex code. The use of monster fusion trees for implementing symmetries will be discussed further in \sref{sec:generalizations_of_fusion_trees_efficiency_and_prospect}.

% subsection simple_yoga_and_monster_fusion_trees (end)

\subsection{Determination of the valid charge sectors} % (fold)
\label{sub:determination_of_the_valid_charge_sectors}

Having introduced fusion trees, let us revisit the decomposition \Eref{eq:gentensor} of a generic symmetric tensor. As described in the previous section, the structural tensors that appear in the decomposition can be specified by means of a fusion tree. As we have seen, the tensor blocks $T_{j_{1} \hdots j_{k}}^{j^\text{int}_1,\hdots,j^\text{int}_l}$ are labeled by the irreps $j_{1} \hdots j_{k}$ carried by the open indices of the tensor and also the internal irreps ${j^\text{int}_1,\hdots,j^\text{int}_l}$. These internal irreps, in fact, correspond to the internal edges of the fusion tree. Given the irreps carried by the indices of the tensor (this data is part of the tensor specification), the possible internal irreps are obtained by pairwise fusing irreps according to the fusion rules, proceeding inwards from the open edges. A necessary task then is to efficiently determine the set of all compatible irrep-decorations of the fusion tree, such that the irreps intersecting at any node obey the fusion rules. We refer to each compatible irrep-decoration of the fusion tree as a \textit{charge sector}; each charge sector labels a block of the symmetric tensor.

Let us now explain how to determine all the charge sectors in practice. We start with the list of irreps (total spins) on internal and open indices according to
\begin{align}
	\text{listOfIrreps}\, = \left\lbrace j^\text{int}_1,j^\text{int}_2,\hdots,j^\text{int}_l\, , j_1,j_2,\hdots,j_k \right\rbrace,
	\label{eq:OrderingOfIrrepsChargeSectors}
\end{align}
where each entry of this list corresponds to a vector with spin configurations for each index. Here we assume that the fusion tree has $l$ internal edges and $k$ open edges. The set of charge sectors is generated by iterating through all the different combinations of irreps on all the edges of the fusion tree, and selecting the ones that satisfy the fusion rules imposed at every node.

As an example, we take the fusion tree in \fref{fig:TikZ_Files_Rank_4_TensorSpinNetwork}(a) and assign irreps $j_i = \lbrack 0,1 \rbrack$ to each open edge. In this case, there is only one internal edge and it can carry irreps $j^\text{int}_1 = \lbrack 0,1,2 \rbrack$. The all possible combinations of irreps are
\beqa
\text{listOfChargeSectors}&=& \{ \lbrack 0,0,0,0,0 \rbrack , \lbrack 0,0,0,1,1 \rbrack , \lbrack 0,1,1,0,0 \rbrack , \lbrack 0,1,1,1,1 \rbrack , \lbrack 1,0,1,0,1 \rbrack , \nonumber \\
&& \lbrack 1,0,1,1,0 \rbrack , \lbrack 1,0,1,1,1 \rbrack , \lbrack 1,1,0,0,1 \rbrack , \lbrack 1,1,0,1,0 \rbrack , \lbrack 1,1,0,1,1 \rbrack , \nonumber \\
&& \lbrack 1,1,1,0,1 \rbrack , \lbrack 1,1,1,1,0 \rbrack , \lbrack 1,1,1,1,1 \rbrack , \lbrack 2,1,1,1,1 \rbrack \},
\eeqa
% \beqa
% \text{tableOfChargeSectors} = \left\lbrack 
% 	\begin{array}{c c c c c}
% 	0 & 0 & 0 & 0 & 0 \\
% 	0 & 0 & 0 & 1 & 1 \\
% 	0 & 1 & 1 & 0 & 0 \\
% 	0 & 1 & 1 & 1 & 1 \\
% 	1 & 0 & 1 & 0 & 1 \\
% 	1 & 0 & 1 & 1 & 0 \\
% 	1 & 0 & 1 & 1 & 1 \\
% 	1 & 1 & 0 & 0 & 1 \\
% 	1 & 1 & 0 & 1 & 0 \\
% 	1 & 1 & 0 & 1 & 1 \\
% 	1 & 1 & 1 & 0 & 1 \\
% 	1 & 1 & 1 & 1 & 0 \\
% 	1 & 1 & 1 & 1 & 1 \\
% 	2 & 1 & 1 & 1 & 1 \\
% 	\end{array} \right\rbrack
% \eeqa
where the irreps in a vector of the list are ordered as in \Eref{eq:OrderingOfIrrepsChargeSectors}, such that internal irreps appear before the open irreps. In order to determine the valid combinations -- that is, the charge sectors -- we check the fusion rules on each node of the fusion tree, beginning at the open edges and traversing the internal edges successively. The testing for set membership according to \Eref{eq:fusionrules} can also be achieved by testing the conditions
\begin{align}
	\left\lbrace \begin{array}{c} \vert j_1 - j_2 \vert\, \le\, j_3 \, \le\, j_1 + j_2\\ j_1 + j_2 + j_3\ \text{is an integer} \end{array}\right.,
\end{align}
which are equivalent. This procedure is performed node by node in the fusion tree following the internal indices successively. The final list of charge sectors is then assembled by concatenating the valid irrep combinations for all the nodes.

\begin{figure}
	\centering
	\includegraphics[width=.7\textwidth]{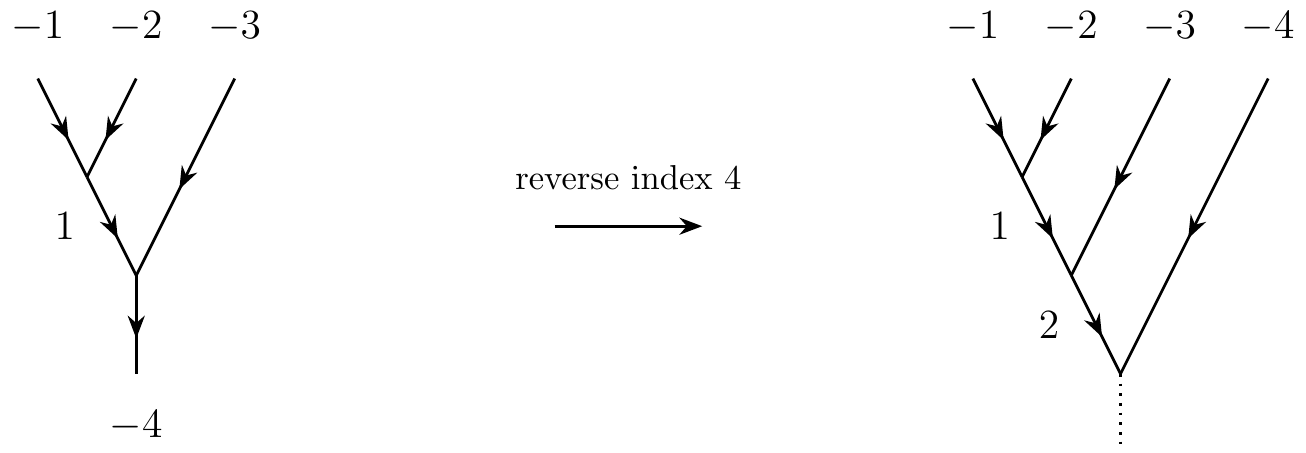}
	\caption{The reversal of the only outgoing index results in a tensor for which all incoming spins fuse to total spin 0, and adds a node in the fusion tree. This is denoted by an explicit ``0'' in the fusion tree, which corresponds to a ``dummy index'' (dotted line) in the extra node.}
	\label{fig:TikZ_Files_Rank_4_FuseToSpinZero}
\end{figure}

A slight subtlety appears when dealing with structural tensors that describe the fusion to total irrep zero or the splitting from total irrep zero, i.e., tensors that have only incoming or only outgoing indices. In this case, the fusion tree is expanded by one node (adding a ``dummy index''), see \fref{fig:TikZ_Files_Rank_4_FuseToSpinZero} for the case of a 4-index tensor. While the fusion tree on the left-hand side of the figure has two nodes, the one on the right-hand side has an extra node at the bottom, accounting for the fusion to total spin zero. In our notation for fusion trees we get
\begin{align}
	\tau^\text{l} &= \lbrace [-1,-2,1],[1,-3,-4] \rbrace\ ,	\\
	\tau^\text{r} &= \lbrace [-1,-2,1],[1,-3,2],[2,-4,{\color{TensorGreen}0}] \rbrace\ ,
\end{align}
where it is clear that in both cases we are dealing with a 4-index tensor. For the second case, this can be identified since the fusion tree is filled with the auxiliary zero, which shows that there is a dummy index even if the tree looks like the one for a 5-index tensor. Once this is clear, the determination for the possible charge sectors for this type of tensors proceeds exactly in the same way as described above. 

% subsection determination_of_the_valid_charge_sectors (end)

\subsection{Building structural tensors from a fusion tree} % (fold)
\label{sub:building_structural_tensors_from_a_fusion_tree}

In our implementation, we do not store and manipulate explicitly the structural tensors, but instead work directly with fusion trees. However, sometimes it may be useful to build the structural tensor. When required, one can build a structural tensor from a fusion tree as follows. Assign an irrep ($j$ label) to each edge and attach the corresponding fusion or splitting Clebsch-Gordan tensor to each fusion or splitting node in the tree respectively. The structural tensor, specified by such an irrep-decorated fusion tree, is obtained by contracting all the fusing and splitting tensors according to the tree graph.

Take for instance the case of a 4-index tensor with the same fusion tree of \fref{fig:TikZ_Files_Rank_4_TensorSpinNetwork}(a) as before. The Clebsch-Gordan tensor for the charge sector $\lbrack 2,1,1,1,1 \rbrack$ of the fusion tree $\tau = \lbrace [-1,-2,1],[1,-3,-4] \rbrace$ is then given by the contraction of $C^\text{fuse}(1,1,2)$ and $C^\text{fuse}(2,1,1)$ over the common internal edge, where the labels on the tensors $C^\text{fuse}$ indicate the corresponding irreps that enter the Clebsch-Gordan coefficients. The resulting structural tensor is a $3 \times 3 \times 3 \times 3$ object with to $m = -1,0,+1$ for all four legs.

% subsection building_structural_tensors_from_a_fusion_tree (end)

\subsection{\texorpdfstring{$F$}{TEXT}-moves} % (fold)
\label{sub:f_moves}

We have already seen that the tree decomposition for $k$-index tensors with $k \ge 4$ is not unique, since the irreps can be pairwise fused to a total irrep in different ways. It turns out that some trees may be better suited for certain tensor operations than others. Imagine for instance that we want to fuse together indices 1 and 2 of the 5-index tensors shown in \fref{fig:TikZ_Files_Rank_5_TensorSpinNetwork}, an operation typically known as \textit{tensor reshaping}. This can be easily implemented for the trees (a) and (b), simply by removing the fusion node and thus exposing the internal edge carrying $j_1^{{\rm int}}$ as the fused or reshaped open edge. However, fusing indices 1 and 2 is not as straightforward for the tree shown in (c), where 1 and 2 are not attached to a single node. On the other hand, for the same reason, the tree shown in (c) is more suitable for fusing together indices 3 and 4. Also, if one wants to fuse indices 1 and 2 and also indices 3 and 4, then the tree shown in (b) is the best choice. Finally, if we wanted to fuse together indices 2 and 3, none of the trees shown in \fref{fig:TikZ_Files_Rank_5_TensorSpinNetwork} are a good option. Thus, when implementing basic tensor operations such as reshape, one may have to transform one fusion tree into another.

\begin{figure}
	\centering
	\begin{minipage}{.425\textwidth}
		\centering
		\includegraphics[width=\textwidth]{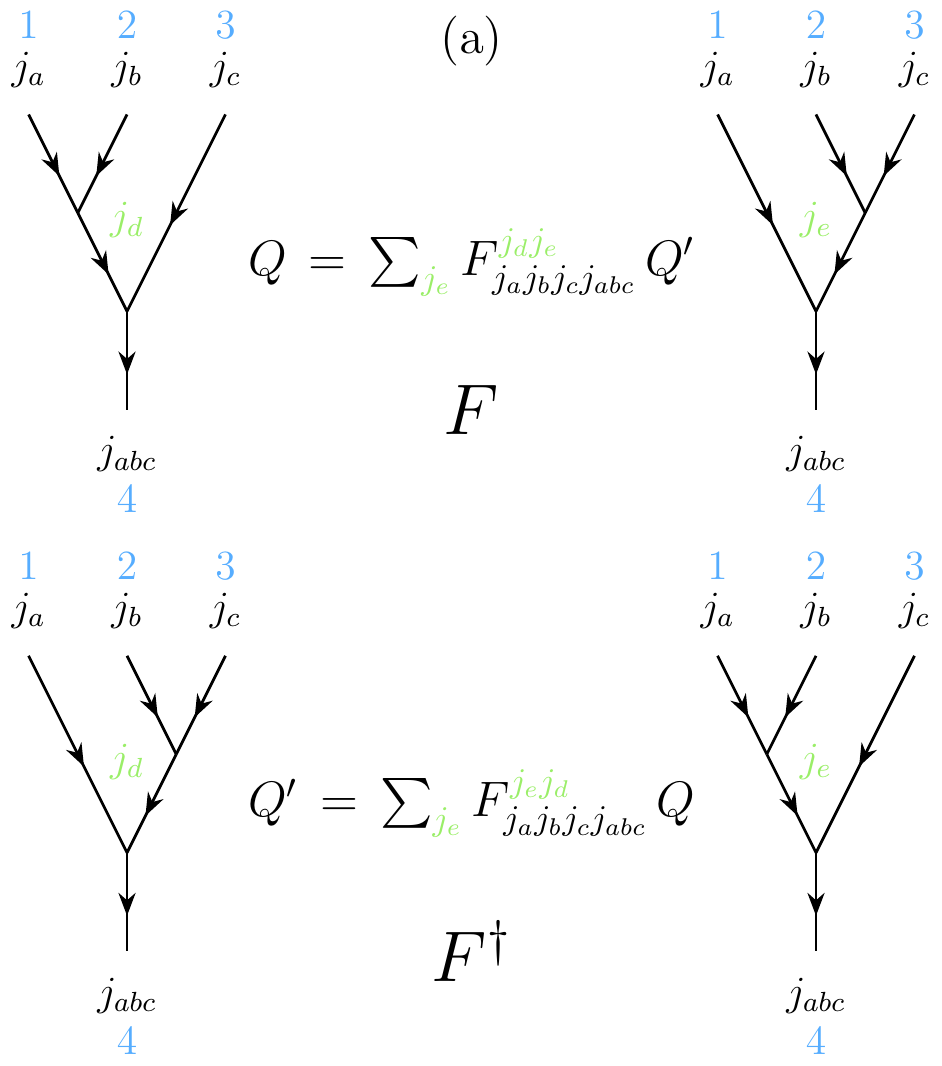}
	\end{minipage}
	\hfill
	\begin{minipage}{.425\textwidth}
		\centering
		\includegraphics[width=\textwidth]{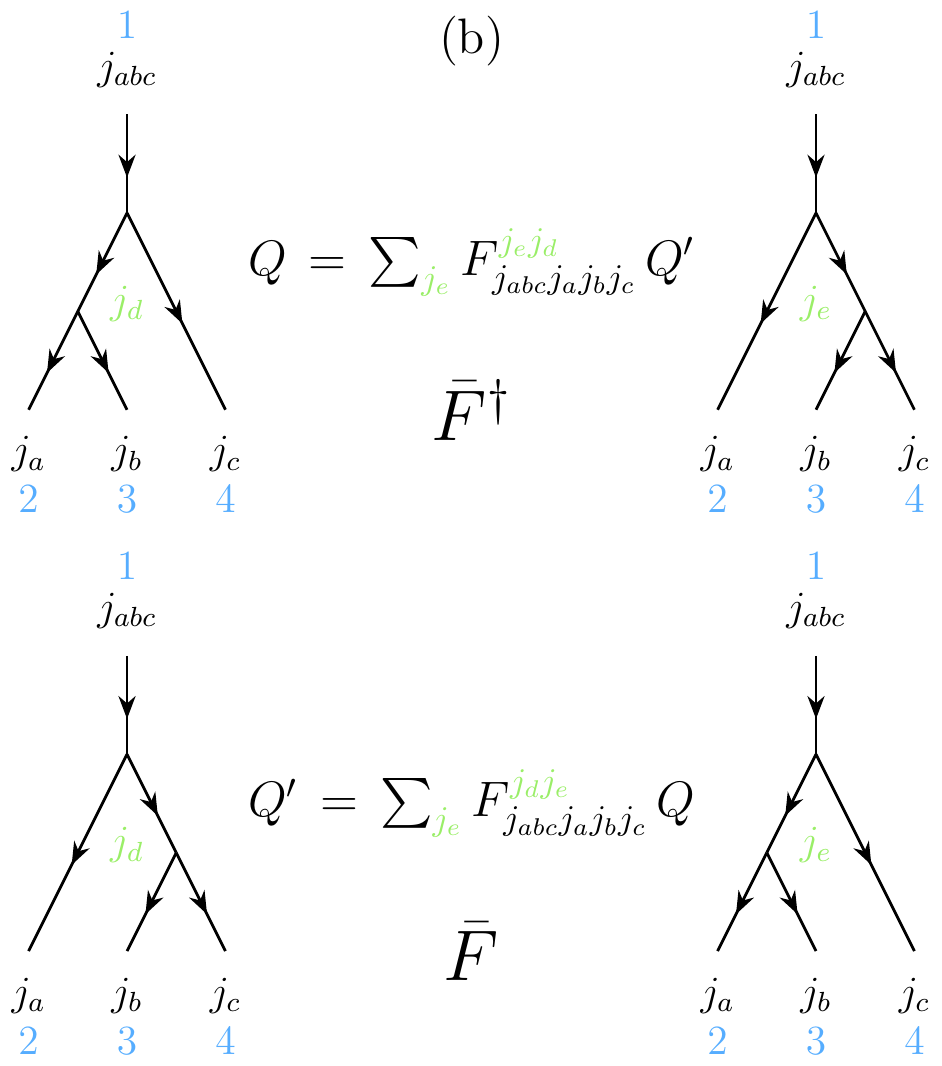}
	\end{minipage}
	\caption{An $F$-move relates different pairwise fusions of three irreps, i.e., it is a change of basis between the corresponding structural tensors $Q$ (each fusion tree corresponds to a $Q$). In (a) three spins fuse into one spin, and in (b) one spin splits into three spins. Note that different $F$ coefficients appear in the four cases shown here. Large symbols ($F$, $F^\dagger$, $\bar F$, $\bar F^\dagger$) label the four different operations. The blue number labels indicate the oriented labeling of the fusion trees (use in the data structure for storing the fusion trees). These labels need not coincide with the order of the open indices in the corresponding symmetric tensor.}
	\label{fig:TikZ_Files_F_Move_3_4}
\end{figure}

 \begin{figure}
	\centering
	\includegraphics[width=0.8\textwidth]{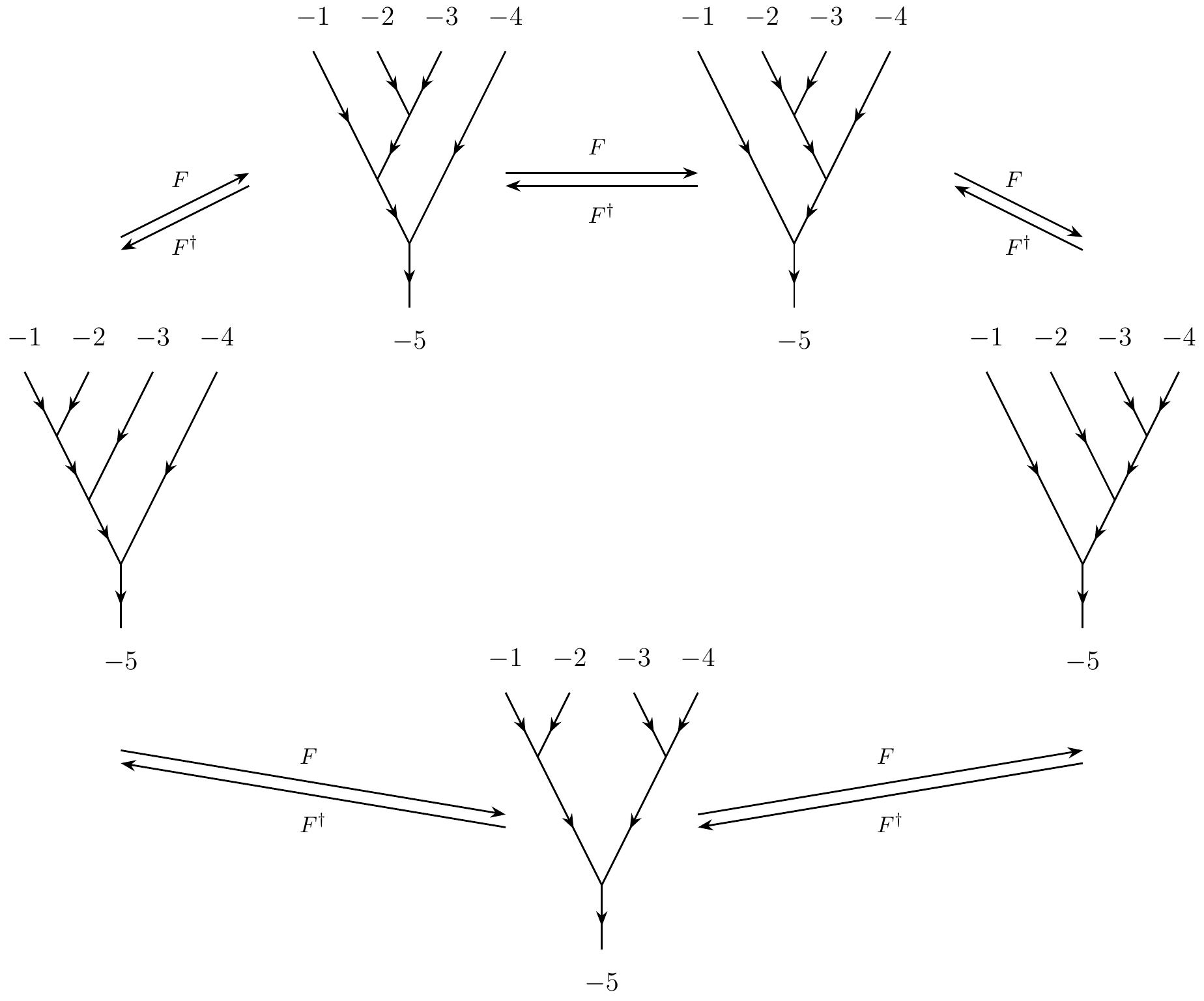}
	\caption{The fusion tree on the far left can be transformed to the fusion tree on the far right by applying either three $F$-moves (the top route) or only two $F$-moves (the bottom route). These two transformations must be equivalent, which leads to a constraint equation for the $F$-symbols called the pentagon equation.}
	\label{fig:TikZ_Files_Pairwise_Mapping_4}
\end{figure}

A given fusion tree can be transformed to another fusion tree by applying a sequence of an elementary transformation known as an $\mathit{F}$\textit{-move}. An $F$-move transforms a fusion tree with four open edges, see \fref{fig:TikZ_Files_F_Move_3_4}. Notice that $F$-moves do neither change the directions of the indices, nor the ordering of the open indices. An $F$-move is essentially a change of basis for the structural tensors that correspond to the fusion trees on both sides of the equation, and the coefficients $F_{j_a j_b j_c j_{abc}}^{j_d j_e}$ that appear on the right-hand side are the \textit{recoupling coefficients} of $SU(2)$. They are closely related to the well-known Wigner 6-$j$ symbols as 
\begin{align}
	F_{j_a j_b j_c j_{abc}}^{j_d j_e} = (-1)^{j_a + j_b + j_c + j_{abc}} \sqrt{(2j_d + 1)(2j_e + 1)} \cdot \left\lbrace \begin{array}{c c c} j_a & j_b & j_d \\ j_c & j_{abc} & j_e \end{array} \right\rbrace\ .
\end{align}
where the $\{\cdots\}$ denotes a 6-j symbol. For large fusion trees, the mappings amongst them can always be reduced to a concatenation of pairwise mappings such as the ones shown in \fref{fig:TikZ_Files_F_Move_3_4} for a {4}-index tensor. {In fact it may be possible to apply different sequences of $F$-moves that transform the fusion tree to a given final fusion tree. From a computational perspective one sequence may be faster to perform than others and is therefore favorable. For example, \fref{fig:TikZ_Files_Pairwise_Mapping_4} shows two sequences of $F$-move that transforms the fusion tree at the far left to the one on the far right. The top sequence is composed of three $F$-move, while the bottom one has only two $F$-moves and is thus more computationally efficient. The fact that both sequences in \fref{fig:TikZ_Files_Pairwise_Mapping_4} lead to the same fusion tree and the same overall change of basis is thanks to the so called \emph{pentagon equation} -- an important consistency constraint that is satisfied by the $F$-moves.

% subsection f_moves (end)

\subsection{Finding the minimal sequence of \texorpdfstring{$F$}{TEXT}-moves to transform one fusion tree to another} % (fold)
\label{sub:finding_the_minimal_sequence_of_f_moves_to_transform_one_fusion_tree_to_another}

\begin{figure}
	\centering
	\includegraphics[width=.5\textwidth]{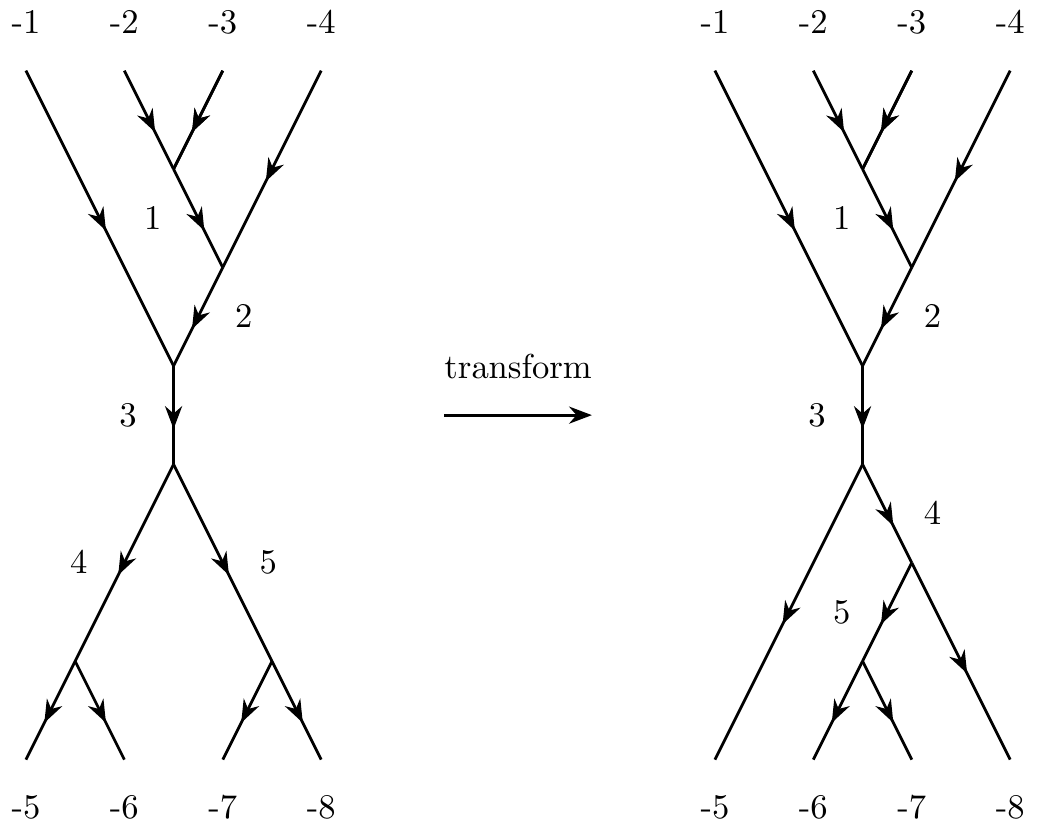}
	\caption{These fusion trees can be transformed to each other by applying two $F$-moves. The trees are defined in \Eref{eq:FMoveTree_1} and \Eref{eq:FMoveTree_2} with the same list $\sigma = [-1,-1,-1,+1,+1,+1]$.}
	\label{fig:TikZ_Files_updateFusionTree_described_Mapping_1}
\end{figure}

Let us consider, for example, the transformation between the following two fusion trees 
\begin{align}
	\tau^\text{i} &= \lbrace [-2,-3,1],[1,-4,2],[-1,2,3],[3,4,5],[4,-5,-6],[5,-7,-8] \rbrace\ ,	\label{eq:FMoveTree_1}\\
	\tau^\text{f} &= \lbrace [-2,-3,1],[1,-4,2],[-1,2,3],[3,-5,4],[4,5,-8],[5,-6,-7] \rbrace\ \label{eq:FMoveTree_2}
\end{align}
illustrated in \fref{fig:TikZ_Files_updateFusionTree_described_Mapping_1}. Our goal is to find the minimal number of $F$-moves that transform the fusion tree $\tau^\text{i}$ to the fusion tree $\tau^\text{f}$. We can start by scanning for the internal edges that differ in the two trees $\tau^\text{i}$ and $\tau^\text{f}$. These are the edges that will need to be re-organized by applying $F$-moves. Note that for each internal edge there is only one possible $F$-move that can be applied to move that edge. Therefore, there is a one to one correspondence between internal edges and the possible $F$-moves that can applied on the tree. It turns out that there exists at least one sequence of $F$-move, and therefore {one sequence} of internal edges that differ in the two trees, that transforms one tree to the other. Once these edges have been identified -- in this example, they correspond to edges 4 and 5 -- we can try every permutation of these indices. Here the only permutations are [$4,5$] and [$5,4$] {and only the sequence [$4,5$] gives the correct transformation}. Corresponding to each sequence, we build the resulting fusion tree and compare it to the target fusion tree. We may find several sequences that lead to the target tree. In this case, we select the cheapest sequence i.e., the one that contains the minimum number of moves. In this way, we determine the sequence of $F$-moves that has to be applied.

A point to remember is that the labeling of the internal indices is not unique. For example, consider the two fusion trees
\begin{align}
	\tau_\text{old} &= \left\lbrace [-2,-3,1],[1,-4,2],[-1,2,3],[3,4,5],[4,-5,-6],[5,-7,-8] \right\rbrace \ , \\
	\tau_\text{new} &= \left\lbrace [-2,-3,1],[1,-4,2],[-1,2,3],[3,5,-8],[4,-5,-6],[5,4,-7] \right\rbrace \  
\end{align}
represented in \fref{fig:TikZ_Files_updateFusionTree_described_Mapping_2}. As easily seen in the figure, both trees differ in exactly two nodes. Since the labeling of the internal labels does not have any strict meaning, one could also choose the following representation for the third tree in the figure
\begin{align}
	\tau_\text{new}^\prime = \left\lbrace [-2,-3,1],[1,-4,2],[-1,2,{\color{TensorGreen}4}],[{\color{TensorGreen}4},5,-8],[{\color{TensorGreen}3},-5,-6],[5,{\color{TensorGreen}3},-7] \right\rbrace \ ,
\end{align}
where internal index 3 and 4 have been exchanged. Both representations $\tau_\text{new}$ and $\tau_\text{new}^\prime$ describe exactly the same tensor. Therefore two trees are equivalent up to a relabeling of their internal indices. {However $\tau_\text{old}$ and $\tau_\text{new}^\prime$ differ in more than two nodes despite being related to one another by a single $F$-move.} This {ambiguity} has to be taken into account when comparing fusion trees in order to determine the required $F$-move sequence.

\begin{figure}
	\centering
	\includegraphics[width=.8\textwidth]{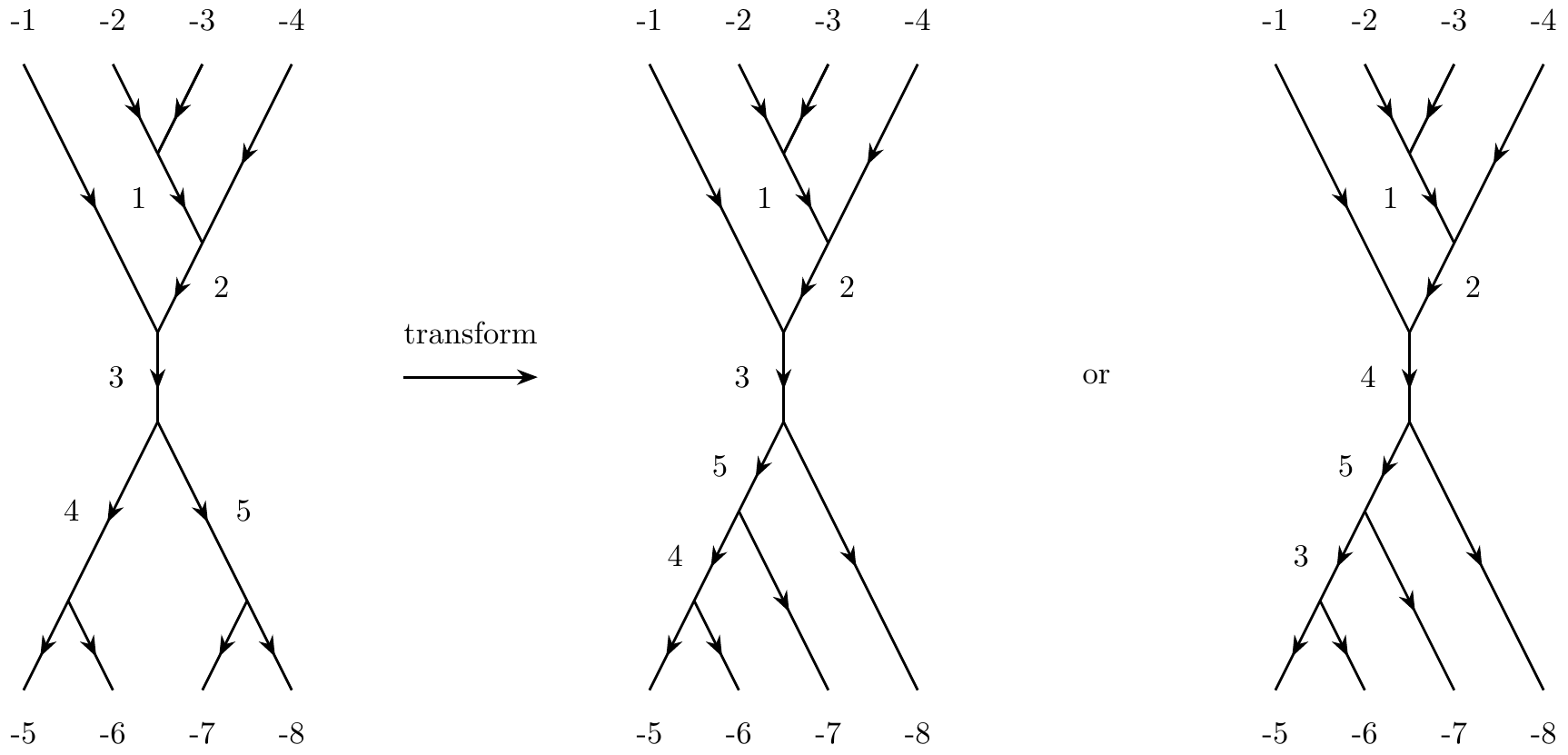}
	\caption{Two fusion trees related by a single $F$-move. Importantly, every fusion tree is unique only up to a relabeling of the internal legs. This ambiguity has to be taken into account to find the minimal number of $F$-moves.}
	\label{fig:TikZ_Files_updateFusionTree_described_Mapping_2}
\end{figure}

The next task is to determine exactly which $F$-move must be applied to each internal edge in the optimal sequence. Note that the above procedure does not specify the ordering of the nodes in the fusion trees. For instance, the two trees
\begin{align}
	\tau &= \left\lbrace[-1,-2,1],[1,-3,-4] \right\rbrace \ , 	\label{eq:ConstructNumericalFusionTree_1}	\\
	\tau^\prime &= \left\lbrace[1,-3,-4],[-1,-2,1] \right\rbrace		\label{eq:ConstructNumericalFusionTree_2}
\end{align}
together with the direction of the nodes describe exactly the same tensor (also see \sref{sec:fusion_trees}). However, it is important to have a convention for the ordering here so that the different $F$-moves can be unambiguously and consistently defined. We adopt the convention shown in \fref{fig:TikZ_Files_F_Move_5}. Each figure consists of two nodes (either both fusion nodes or both splitting nodes) and one internal edge. The numbers in brackets indicate which index of each node corresponds to the internal edge. The bracketed pair of numbers can be used to label and distinguish these different trees on which a particular $F$-move applies. Note that in the figure the two trees in the center share the same structure (and are thus labeled by the same bracketed pair of numbers), whereas the remaining two are different.

\begin{figure}
	\centering
	\includegraphics[width=.75\textwidth]{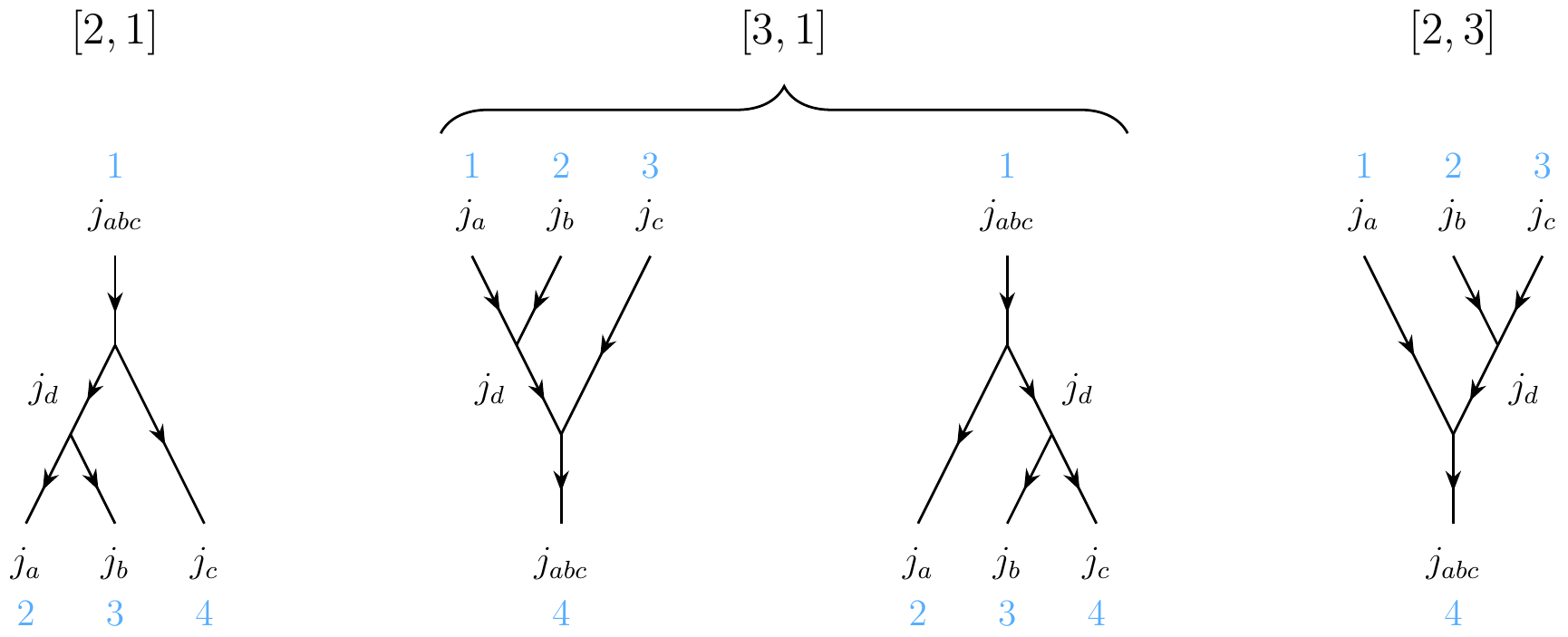}
	\caption{Convention for ordering of nodes in fusion trees, each consisting of two nodes and one internal edge. The pair of numbers in brackets is the position (according to the node orientation convention shown in \fref{fig:TikZ_Files_ClebschGordan_2}) of the internal edge in each of the two nodes.}
	\label{fig:TikZ_Files_F_Move_5}
\end{figure}

With these conventions, the four possible $F$-moves from \fref{fig:TikZ_Files_F_Move_3_4} can be represented as in \tref{tab:FusionTreeMapping} using the array notation (we use $F$ for the ordinary $F$-move, $F^\dagger$ for its inverse, $\bar F$ for the reversed $F$-move and $\bar F^\dagger$ for the reversed inverse).
\begin{table}[ht]
	\centering
	\begin{tabular}{c c c c}
		$F:$					& $\tau = \lbrace [\alpha,{\color{TensorGreen}i},\delta],[\beta,\gamma,{\color{TensorGreen}i}] \rbrace$	& $\xrightarrow{[2,3] \rightarrow [3,1]}$		& $\tau^\prime = \lbrace [\alpha,\beta,{\color{TensorGreen}i}],[{\color{TensorGreen}i},\gamma,\delta] \rbrace$	\\
		$~F^\dagger:$			& $\tau = \lbrace [\alpha,\beta,{\color{TensorGreen}i}],[{\color{TensorGreen}i},\gamma,\delta] \rbrace$		& $\xrightarrow{[3,1] \rightarrow [2,3]}$		& $\tau^\prime = \lbrace [\alpha,{\color{TensorGreen}i},\delta],[\beta,\gamma,{\color{TensorGreen}i}] \rbrace$	\\\\
		$\bar F^\dagger:$	& $\tau = \lbrace [\alpha,\beta,{\color{TensorGreen}i}],[{\color{TensorGreen}i},\gamma,\delta] \rbrace$		& $\xrightarrow{[3,1] \rightarrow [2,1]}$		& $\tau^\prime = \lbrace [\alpha,{\color{TensorGreen}i},\delta],[{\color{TensorGreen}i},\beta,\gamma] \rbrace$	\\
		$\bar F:$			& $\tau = \lbrace [\alpha,{\color{TensorGreen}i},\delta],[{\color{TensorGreen}i},\beta,\gamma] \rbrace$		& $\xrightarrow{[2,1] \rightarrow [3,1]}$		& $\tau^\prime = \lbrace [\alpha,\beta,{\color{TensorGreen}i}],[{\color{TensorGreen}i},\gamma,\delta] \rbrace$
	\end{tabular}
	\caption{Different $F$-moves, in the array notation, following our convention. The numbers in brackets above the arrows indicate the change of trees, as defined in \fref{fig:TikZ_Files_F_Move_5}.}
	\label{tab:FusionTreeMapping}
\end{table}

% subsection finding_the_minimal_sequence_of_f_moves_to_transform_one_fusion_tree_to_another (end)

% section fusion_trees (end)

\section{{Data structure for an \texorpdfstring{$SU(2)$}{TEXT}-symmetric tensor}} % (fold)
\label{sec:blue_data_structure_for_an_su_2_symmetric_tensor}

Let us summarize the discussion so far by listing the data structure that we used in our implementation to store a generic $SU(2)$-symmetric tensor in memory. From the point of view of implementing symmetric tensors, a $k$-index $SU(2)$-symmetric tensor is not just a $k$-dimensional numerical array, but has a rich internal structure. It is specified by the following data.
\begin{itemize}
	\item \textit{numberOfOpenEdges}: integer variable to store the number of open edges in the fusion tree ($k \ge 0$)
	\item \textit{numberOfInternalEdges}: integer variable to store the number of internal edges in the fusion tree ($l \ge 0$)
	\item \textit{numberOfAuxiliaryEdges}: integer variable to store the number of dummy indices in the fusion tree ($m \in [0,1,2,3]$)
	\item \textit{listOfOpenEdges}: list to store the open edges in the fusion tree with the following data for each edge
	\begin{itemize}
		\item \textit{edgeNumber}: number label of the edge ($\in\ [1,\hdots,k]$)
		\item \textit{edgeDirection}: incoming or outgoing edge ($\in\ [-1,+1]$)
		\item \textit{edgeIrreps}: list of irreps carried by the edge $\lbrace j_i,t_{j_i} \rbrace$
		\item \textit{isFused}: boolean variable to indicate whether the edge was obtained by fusing other edges (refer to \sref{sec:reshaping_a_tensor})
		\item \textit{originalIrreps}: list of irreps before fusion (refer to \sref{sec:reshaping_a_tensor})
	\end{itemize}
	\item \textit{listOfInternalEdges}: list to store the internal edges in the fusion tree with the following data for each internal edge
	\begin{itemize}
		\item \textit{edgeNumber}: number label of the edge ($\in\ [1,\hdots,l]$)
		\item \textit{edgeIrreps}: list of irreps carried by the edge $\lbrace j_i,t_{j_i} \rbrace$ (determined by pairwise fusing all the open edges and applying fusion rules at each step) 
	\end{itemize}
	\item \textit{listOfChargeSectors}: list to store all the charge sectors $\lbrace \lbrack j^\text{int}_1,j^\text{int}_2,\hdots,j^\text{int}_l\, , j_1,j_2,\hdots,j_k \rbrack \rbrace$ (refer to \sref{sub:determination_of_the_valid_charge_sectors})
	\item \textit{listOfDegeneracyTensors}: list of $k$-dimensional numerical arrays, each one corresponding to one entry in $\textit{listOfChargeSectors}$ (these are the variational parameters of the total symmetric tensor)
	\item \textit{listOfStructuralTensors}: list of $k$-dimensional numerical arrays, each one corresponding to one entry in $\textit{listOfChargeSectors}$ (do not need to store this if working directly with fusion trees)
	\item \textit{fusionTree}: list of nodes (the $\tau$ vectors described previously in \sref{sec:fusion_trees}) that describes the fusion tree associated with the symmetric tensor 
	\item \textit{fusionTreeDirections}: list of directions for each node in the fusion tree (see \sref{sub:data_structure_for_a_fusion_tree})
\end{itemize}

% section blue_data_structure_for_an_su_2_symmetric_tensor (end)

\section{A general template for symmetric tensor operations} % (fold)
\label{sec:a_general_template_for_symmetric_tensor_operations}

Having described our fundamental object -- the symmetric tensor -- our next goal is to describe all the necessary symmetric tensor operations that can be used to compose any tensor network algorithm.
\begin{enumerate}
\item Index reversal: reversing the direction of any index of a tensor.
\item Permuting a tensor: changing the order of the indices of a tensor.
\item Reshaping a tensor: fusing two indices of a tensor into an effective index, or splitting an index of a tensor into two indices.
\item Contracting two tensors into a single tensor.
\item Decomposing a tensor into a product of tensors.
\end{enumerate}
Our goal here is to implement these operations for a symmetric tensor. Importantly, the idea of any implementation of symmetric tensor networks is to work directly with the degeneracy tensors, which are much smaller than the total symmetric tensor. In order to do this, we have to describe (and implement) how the degeneracy tensors of a symmetric tensor are updated as a result of these operations. Any operation on the input symmetric tensor(s) generally follows these update steps:
\begin{enumerate}
\item Build the fusion tree of the output tensor by manipulating the input fusion tree(s) in some way,
\item Determine the charge sectors $\{c_{\text out}\}$ for the output tensor,
\item Build a table $E$ that determines all the input charge sectors (for each input tensor) that contribute to the update required for each output charge sector,
\item Transform the input degeneracy tensors by performing some operations on them,
\item Build the output degeneracy tensor for each output charge sector $c_{\text out}$ by combining (e.g. adding or concatenating), possibly several, transformed input degeneracy tensors -- those that correspond to the contributing input charge sectors that appear against the entry of $c_{\text out}$ in the table $E$.
\end{enumerate}
In our implementation, each elementary symmetric tensor operation follows the above steps. In the rest of the paper, we describe how to implement these elementary operations in detail.

% section a_general_template_for_symmetric_tensor_operations (end)

\section{Transforming the fusion tree} % (fold)
\label{sec:transforming_the_fusion_tree}

The transformation of fusion trees has been described in detail in \sref{sub:f_moves}. When it comes to the actual implementation of these transformations for a full tensor with degeneracy and structural parts (represented as fusion trees), we see that by following their definition they act as in \fref{fig:TikZ_Files_F_Move_6}. That equation is exact, and clearly accounts for the recoupling of the coefficients in the structural tensors. However, as stated several times throughout the paper, the structural tensors are never explicitly stored, since we only deal with fusion trees and their array representation. Therefore, we do not account for the effect of an $F$-move in the structural tensor, \emph{but in the degeneracy tensor instead}. This trick is needed in order to have the appropriate degeneracy tensors when, e.g., doing a reshape of indices, as we shall see. Let us be more specific. Consider the general decomposition of a 4-index tensor $T$, 
\beq
T = \sum_{j_d} P'_{j_d} Q'_{j_d}, 
\label{firsteq}
\eeq
with degeneracy and structural parts $P'_{j_d}$ and $Q'_{j_d}$ respectively. Here we are assuming that index $j_d$ is an internal index of a given fusion tree, see \fref{fig:TikZ_Files_F_Move_6}. Let us now suppose that, for a given reshape of indices (i.e., index fusion), the fusion tree $Q'_{j_d}$ is not a good choice. Instead, one can rewrite this tensor in terms of (say) an $F$-move and a good fusion tree, i.e., 
\beq
{Q'}_{j_d} = \sum_{j_e} F^{j_d j_e} {Q}_{j_e}.
\eeq
Since the transformation is applied to the structural part of the tensor, the degeneracy tensor $P'_{j_d}$ can be treated as a constant factor. This means that tensor $T$ is given by 
\beq
T = \sum_{j_d, j_e} P'_{j_d} F^{j_d j_e} {Q}_{j_e}, 
\eeq
see \fref{fig:TikZ_Files_F_Move_6}. Using the standard decomposition this in turn can be rewritten as 
\beq
T = \sum_{j_e} P_{j_e} {Q}_{j_e}, 
\eeq
with 
\beq
{P}_{j_e} = \sum_{j_d} F^{j_d j_e} {P'}_{j_d}.
\label{thiseq}
\eeq
The conclusion is that a change of basis (via the $F$-move) in the structural tensor induces also a change of basis in the degeneracy tensor, in order to leave the overall tensor $T$ unchanged. Thus, changing the fusion tree in the representation of the tensor implies a change in the degeneracy tensor as prescribed by \Eref{thiseq}. Notice that the structural part, or rather the fusion tree, cannot be manipulated numerically by means of an $F$-move due to its analytic array form -- it will be therefore simply replaced. The change of basis for the degeneracy tensor is then performed by taking the inverse change of basis to that in \Eref{thiseq}, which is given by 
\beq
{P'}_{j_d} = \sum_{j_e} F^{j_e j_d} {P}_{j_e}.
\label{thateq}
\eeq
Using this equation in combination with \Eref{firsteq} implies that the transformed tensor $T$ can finally be written as 
\beq
T = \sum_{j_d} \left( \sum_{j_e} F^{j_e j_d} {P}_{j_e} \right) Q'_{j_d},
\eeq
where $Q'_{j_d}$ is the new fusion tree that replaces the old one, and the transformation in parenthesis makes sure that the new degeneracy tensors suit the new fusion tree. This relation is quite interesting, since it rewrites the tensor $T$ in terms of a change of basis for the degeneracy tensor, while leaving the fusion tree in the structural part unchanged, see \fref{fig:TikZ_Files_F_Move_7}.

\begin{figure}
	\centering
	\includegraphics[width=.75\textwidth]{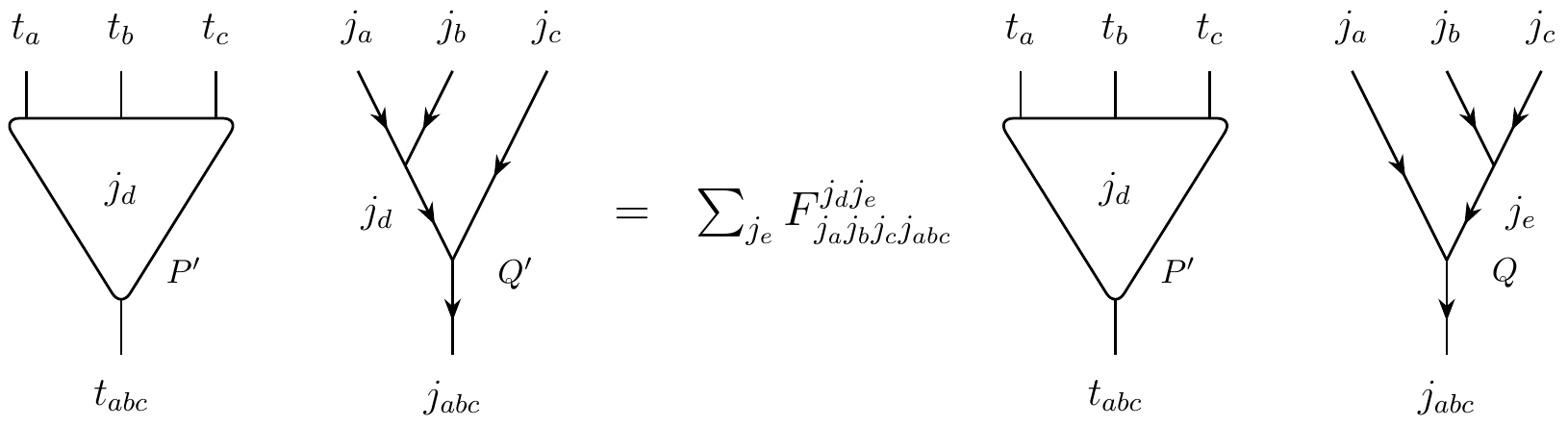}
	\caption{An $F$-move accounting for a change on a structural tensor only as described in \sref{sub:f_moves}.}
	\label{fig:TikZ_Files_F_Move_6}
\end{figure}
\begin{figure}
	\centering
	\includegraphics[width=.75\textwidth]{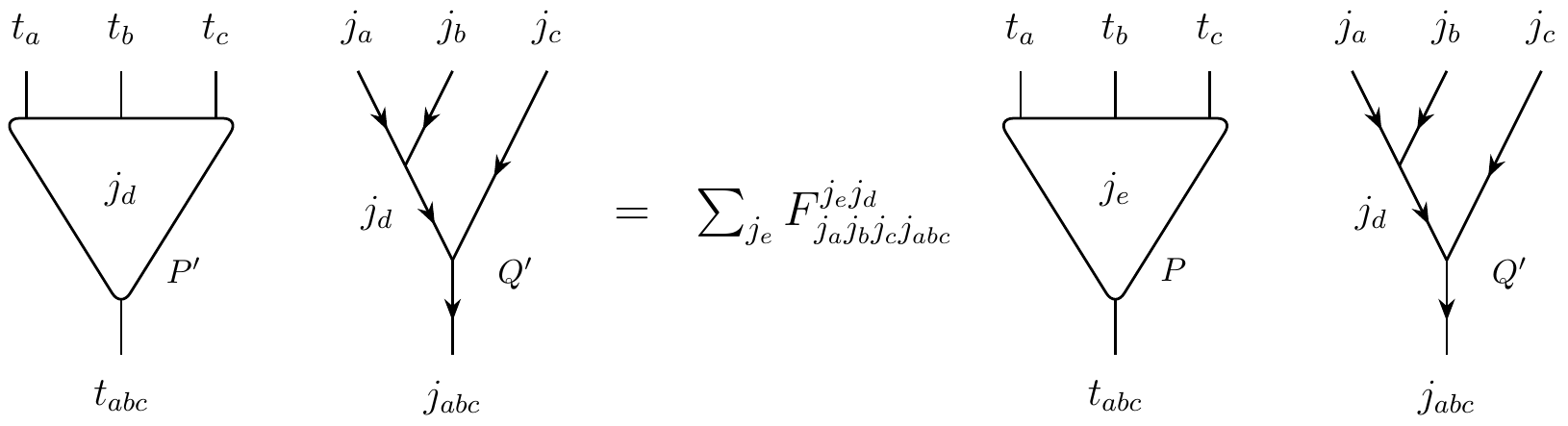}
	\caption{The same $F$-move as in \fref{fig:TikZ_Files_F_Move_6} but now its whole effect is a change in the degeneracy part only.}
	\label{fig:TikZ_Files_F_Move_7}
\end{figure}

The procedure of transforming the fusion tree of an $SU(2)$-symmetric tensor can be summarized in the following steps:
\begin{enumerate}
\item Replace the input fusion tree by the transformed output fusion tree,
\item Determine the charge sectors for the transformed output tensor,
\item Build a table $E$ that lists all contributing input charge sectors for each output charge sector, together with the weight for the input degeneracy tensor given by the numerical $F$-moves,
\item Build the output degeneracy tensor in each output charge sector by taking linear combinations of the input degeneracy tensors that appear in table $E$ with their respective weights.
\end{enumerate}

% section transforming_the_fusion_tree (end)

\section{Reversing an index} % (fold)
\label{sec:reversing_an_index}

Reversing the direction of an index of a regular tensor is a trivial operation. This is not the case for a symmetric tensor though. Reversing indices in a fusion tree corresponds to changing the Clebsch-Gordan coefficients that are associated with one or more nodes, and thus the updating the structural tensors. Furthermore, the resulting index reversed fusion tree may no longer be a simple fusion tree and may have to be restored to a simple tree. (Recall that we have restricted ourselves to consider only simple fusion trees). This restoration may require a sequence of $F$-moves, which in turn requires updating the degeneracy tensors by taking linear combinations of the input degeneracy tensors (refer to \sref{sub:f_moves} about $F$-moves).

To reverse an index, we follow the convenient approach introduced in Ref.~\cite{sukhi} where index reversals are implemented by contracting the symmetric tensor with a simple 2-index tensor, along the index to be reversed. Thus, in this approach, index reversals can be viewed as a very special case of contracting two symmetric tensors, an operation that we will develop in full generality later.

\subsection{CUP and CAP tensors} % (fold)
\label{sub:cup_and_cap_tensors}

Following Ref.~\cite{sukhi}, let us introduce a 2-index symmetric tensor called ``CUP'', which has two incoming indices and will be used to reverse an outgoing index. Likewise, we introduce its inverse -- a 2-index symmetric tensor tensor called ``CAP'' with two outgoing indices that will be used to reverse an incoming index. The CUP and CAP tensors are trivial in their degeneracy part, and read as
\begin{align}
	\Omega^\text{cup} &\equiv \bigoplus_j \left( \mathbb{I}_{d_j} \otimes \Omega_j^\text{cup} \right), 	\\
	\Omega^\text{cap} &\equiv \bigoplus_j \left( \mathbb{I}_{d_j} \otimes \Omega_j^\text{cap} \right),
\end{align}
where $d_j$ is the dimension of the degeneracy subspace. However, the action of CUP and CAP is non-trivial in the structural part. Such structural parts $\Omega_j^\text{cup}$ and $\Omega_j^\text{cap}$ are given by
\begin{align}
	\Omega_j^\text{cup} &= \sqrt{2j+1} \cdot \omega_j^{\phantom\dagger} , 	\\
	\Omega_j^\text{cap} &= \sqrt{2j+1} \cdot \omega_j^{\dagger},
\end{align}
where the coefficients of $\omega$ and $\omega^\dagger$ are the usual Clebsch-Gordan coefficients that describe fusing to or splitting from a total irrep zero, and $\sqrt{2j+1}$ is a normalization factor\footnote{The Clebsch-Gordan coefficients, encoded in $\omega$ and $\omega^\dagger$, that describe the fusion of two copies of irrep $j$ into zero are not normalized. They are given by
\beq
\langle j, m; j, m^\prime | 0,0 \rangle = \frac{(-1)^{j-m}}{\sqrt{2j+1}}\delta_{m,-m^\prime} 
\eeq
and the conjugate transpose thereof.}. This factor ensures that the CUP and CAP tensors fulfill the relations 
\begin{align}
	\Omega^\text{cup}_j \Omega^\text{cap}_j = \Omega^\text{cap}_j \Omega^\text{cup}_j = \mathbb{I}_{2j+1} .
\end{align}
{This relation simply means that the CUP and CAP tensor are the simple matrix inverse of each other.} The graphical representation of the CUP and CAP tensors is shown in 
\fref{fig:TikZ_Files_IndexReversal_0}.

\begin{figure}
	\centering
	\includegraphics[width=.9\textwidth]{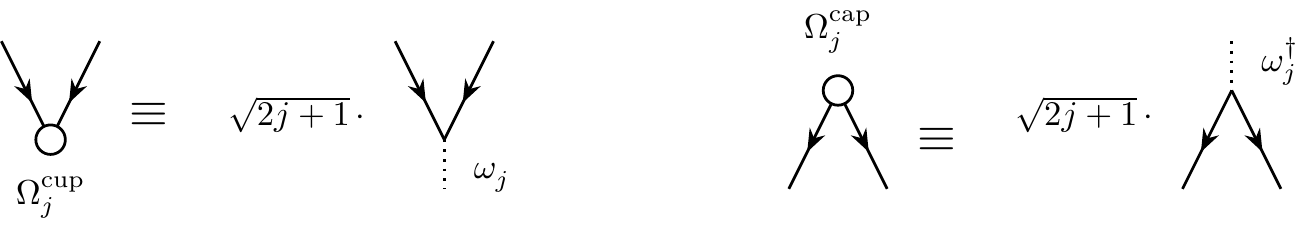}
	\caption{Graphical representation of the CUP and CAP tensors.}
	\label{fig:TikZ_Files_IndexReversal_0}
\end{figure}

% subsection cup_and_cap_tensors (end)

\subsection{Tensors with small number of indices} % (fold)
\label{sub:tensors_with_small_number_of_indices}

Let us first consider the reversal of indices for tensors of with 1 (vectors), 2 (matrices) and 3 indices. For tensors with one index, the label of the single index can only be $(0,t_0)$, since an $SU(2)$-invariant vector can only have total spin $0$ (i.e., be a singlet), also see \fref{fig:TikZ_Files_ClebschGordan_LowRank}. This implies that the structural part is a 1-index intertwiner with spin 0, i.e., it is a scalar and is equal to 1. In other words, the structural part of a 1-index tensor is trivial. Thus, in this case, index reversal has no effect on the structural part, and thus also does not requiring updating the degeneracy part either. For 2-index tensors, their structural part is a set of Clebsch-Gordan tensors where one of the indices are fixed to irrep 0. Thus, reversing an index of a 2-index tensor can be analyzed as a particular case of index reversals for 3-index tensors. 

The case of 3-index tensors requires more attention. However, following an index reversal, the degeneracy tensors for 3-index tensors are only multiplied by a factor, which comes from the corresponding index reversal in the fusion tree, as we will describe below. This is a particularity of the reversal operations and generally, as mentioned previously, the degeneracy tensors are updated by taking linear combinations of the input degeneracy tensors. Therefore, in the following discussion, we will only focus on the index reversal in the fusion tree (which, in this case, is just a single fusion or splitting node) and determine the corresponding factors, which are then multiplied with the degeneracy tensors.

\begin{figure}
	\centering
	\includegraphics[width=1\textwidth]{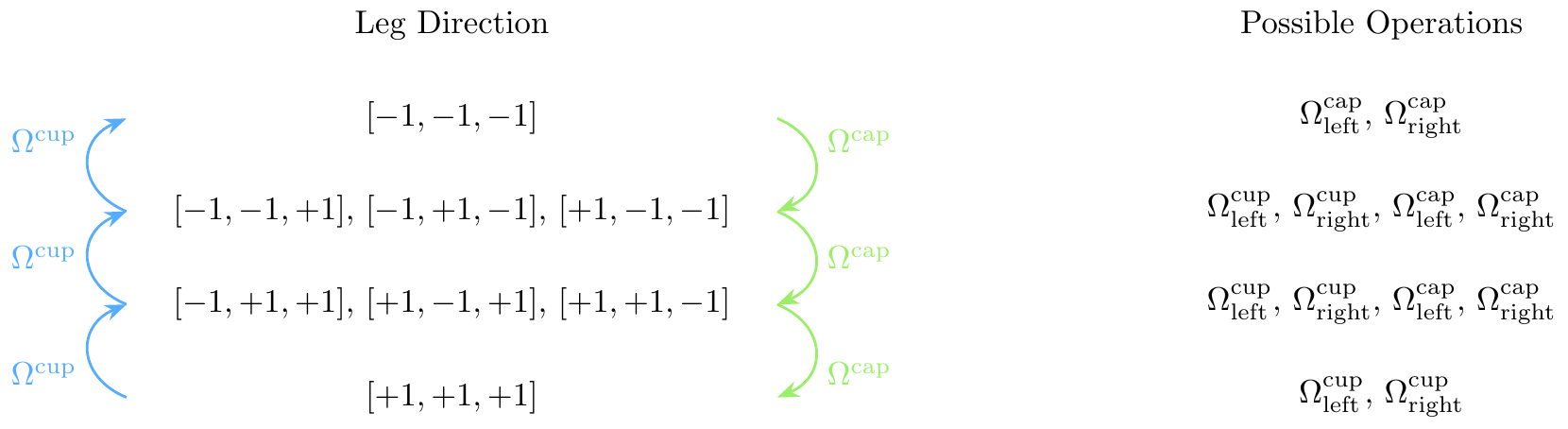}
	\caption{Classification of possible index reversals for a 3-index symmetric tensor.}
	\label{fig:TikZ_Files_Rank_3_Classification}
\end{figure}

In \fref{fig:TikZ_Files_Rank_3_Classification} we list all the possibilities for reversing an index for 3-index tensors. Let us discuss in more detail how to implement such reversals in practice. To begin with, we consider reversing the outgoing index of a fusion tree that is simply a fusion node, see \fref{fig:TikZ_Files_indexReversal_IndexReversal_Rank3_2}. This reversal requires a very simple update (of the data structure) of the symmetric tensor. We simply expand the fusion tree as shown in the figure. The degeneracy tensors are multiplied by the resulting factor. On the other hand, reversing an incoming index of a fusion node does not lead to an expanded fusion tree, but instead replaces it with a splitting node. This also introduces a factor, given as an $F$-symbol, as shown in \fref{fig:TikZ_Files_indexReversal_IndexReversal_Rank3_3}. In order to update the symmetric tensor, we determine the factor for each charge sector and absorb it in to the corresponding degeneracy tensor (resulting in the updated degeneracy tensors). The case of splitting tensors can be dealt with similarly, see \fref{fig:TikZ_Files_indexReversal_IndexReversal_Rank3_4} and \fref{fig:TikZ_Files_indexReversal_IndexReversal_Rank3_5}.

\begin{figure}
	\centering
	\includegraphics[width=.7\textwidth]{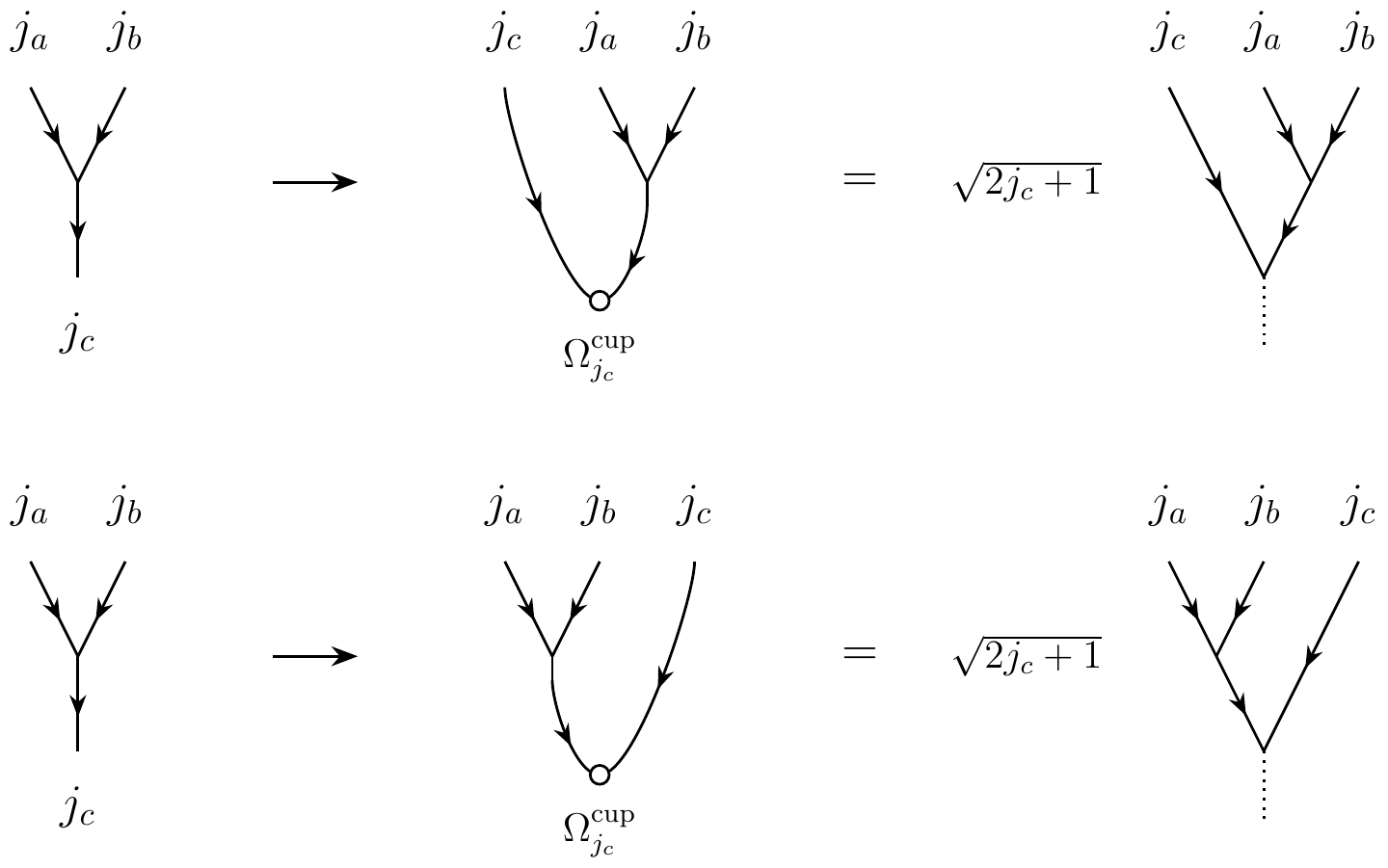}
	\caption{Reversal of the outgoing index on a fusion node using $\Omega^\text{cup}$.}
	\label{fig:TikZ_Files_indexReversal_IndexReversal_Rank3_2}
\end{figure}

\begin{figure}
	\centering
	\includegraphics[width=.9\textwidth]{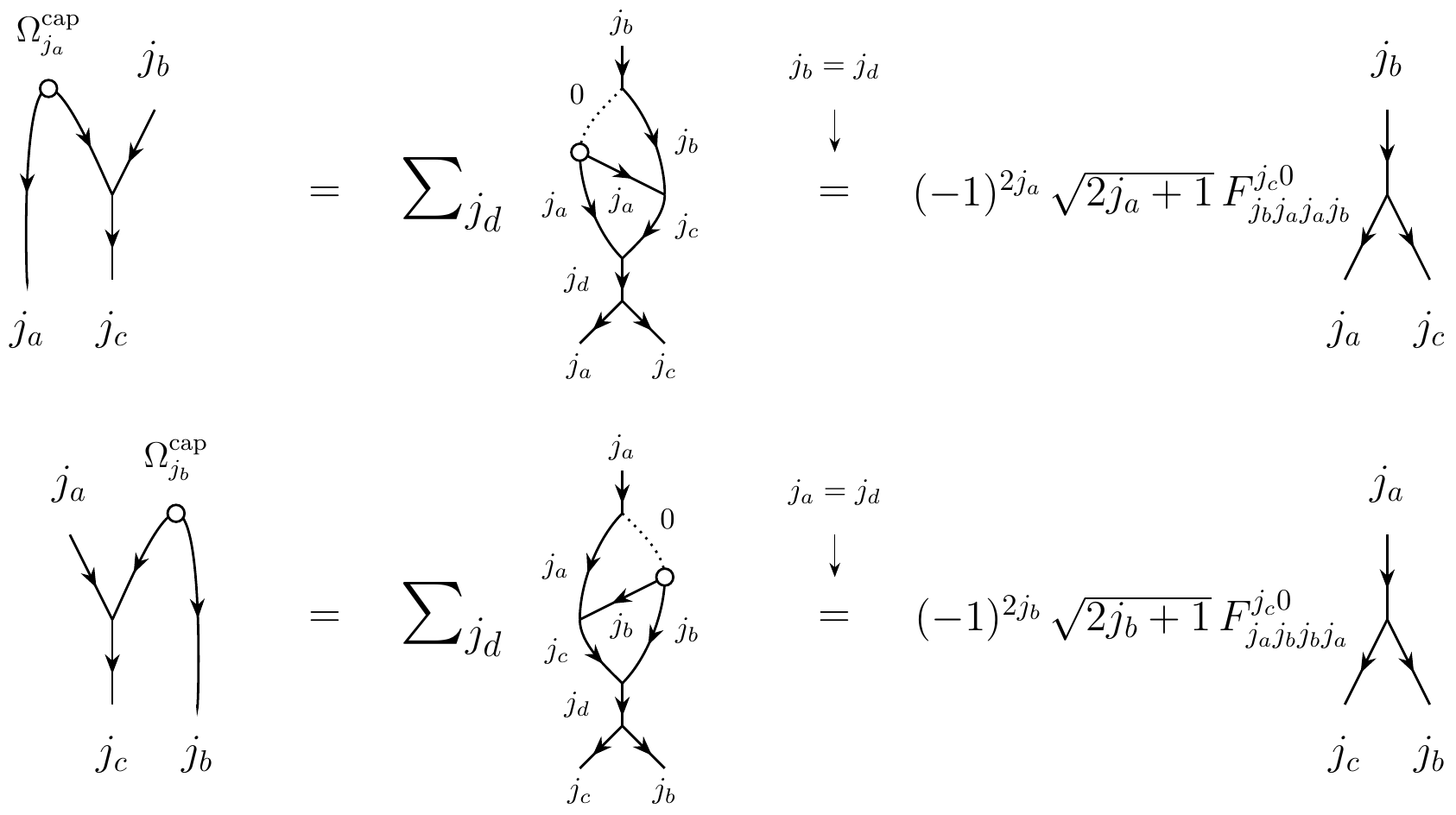}
	\caption{Reversal of incoming indices on a fusion node using $\Omega^\text{cap}$.}
	\label{fig:TikZ_Files_indexReversal_IndexReversal_Rank3_3}
\end{figure}

\begin{figure}
	\centering
	\includegraphics[width=.9\textwidth]{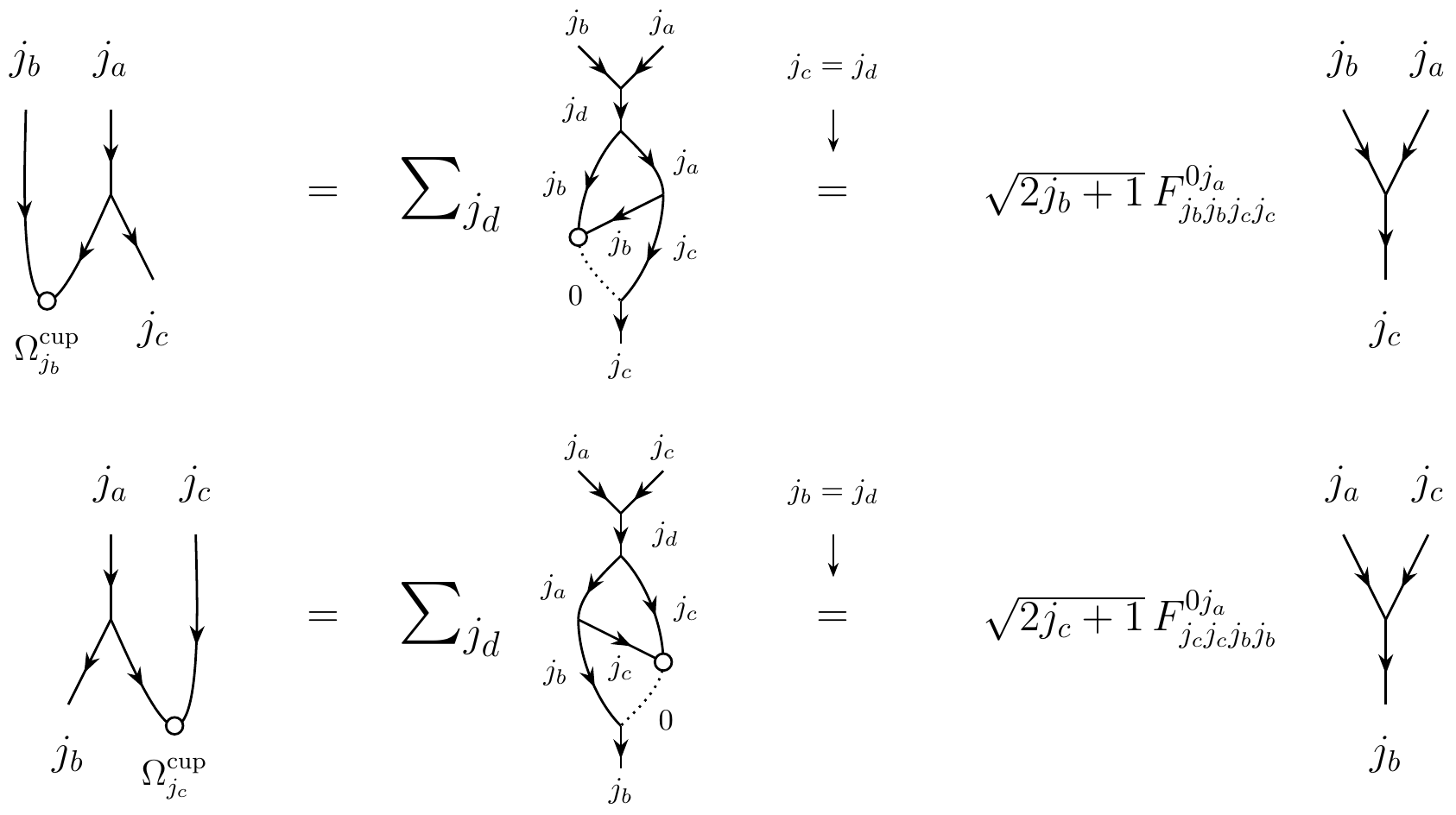}
	\caption{Reversal of outgoing indices on a splitting node using $\Omega^\text{cup}$.}
	\label{fig:TikZ_Files_indexReversal_IndexReversal_Rank3_4}
\end{figure}

\begin{figure}
	\centering
	\includegraphics[width=.7\textwidth]{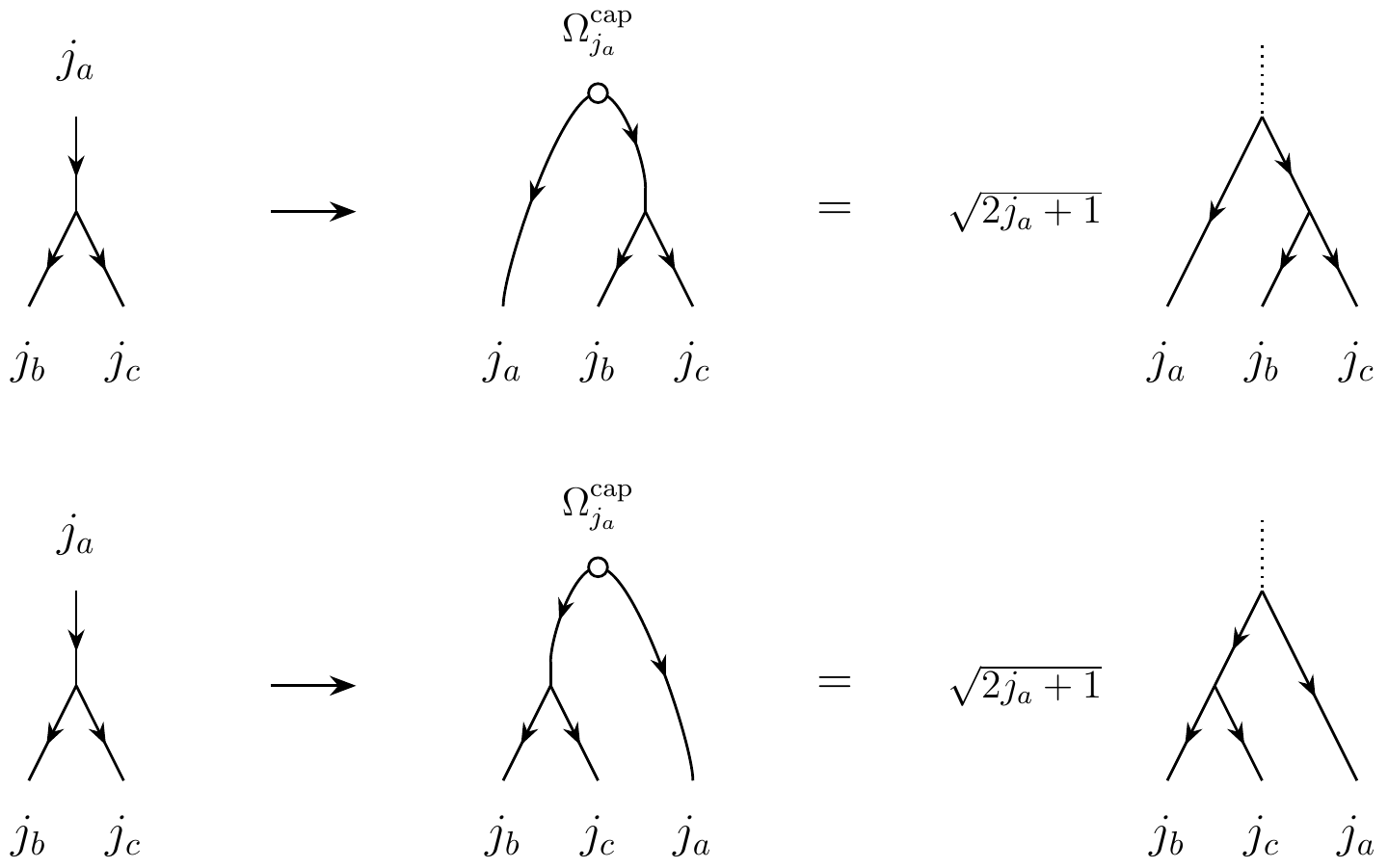}
	\caption{Reversal of the incoming index on a splitting node using $\Omega^\text{cap}$.}
	\label{fig:TikZ_Files_indexReversal_IndexReversal_Rank3_5}
\end{figure}

The $F$-symbols that are appear above can be read off from the figures by making the use the graphical representation of a single $F$-symbol shown in \fref{fig:TikZ_Files_F_Move_2}.

\begin{figure}
	\centering
	\includegraphics[width=.35\textwidth]{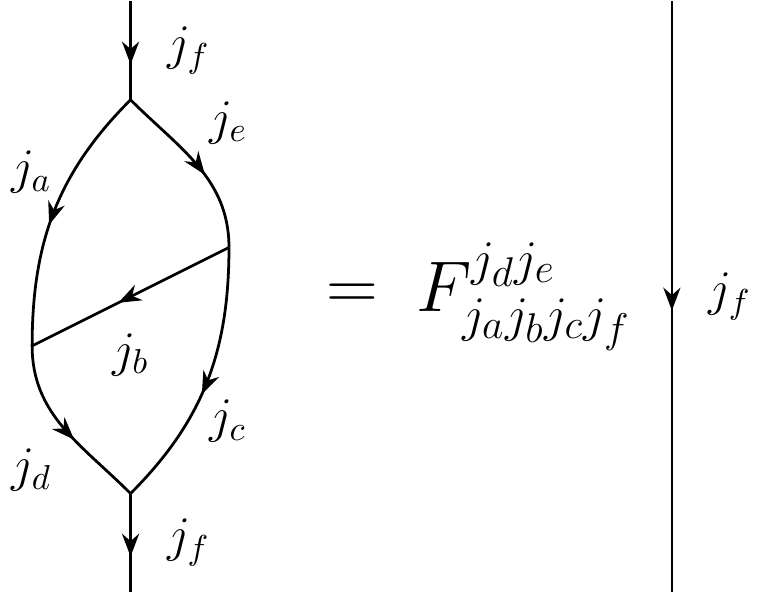}
	\caption{The diagram on the left is equal to the identity times an $F$-symbol.}
	\label{fig:TikZ_Files_F_Move_2}
\end{figure}

% subsection tensors_with_small_number_of_indices (end)

\subsection{Tensors with more than three indices} % (fold)
\label{sub:tensors_with_more_than_three_indices_reversal}

The general idea for implementing the index reversal in tensors with more than three indices reuses factor diagrams shown in the previous section for 3-index tensors, but applied only to a part of the fusion tree. For example, the numerical factors that enter the update for the reversal of index 4 of the fusion tree shown in \fref{fig:TikZ_Files_indexReversal_indexReversal_RelevantPart_1} are determined by the highlighted node only, whereas the remaining nodes play no role here. 
\begin{figure}
	\centering
	\includegraphics[width=.5\textwidth]{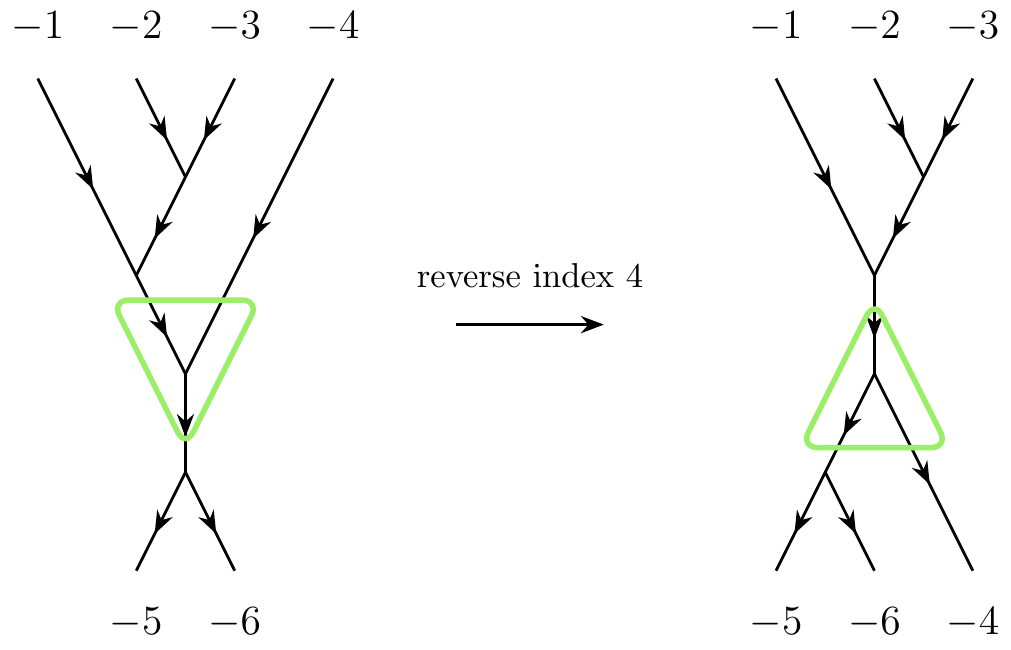}
	\caption{Reversing index 4 in this fusion tree only affects the highlighted node.}
	\label{fig:TikZ_Files_indexReversal_indexReversal_RelevantPart_1}
\end{figure}

\begin{figure}
	\centering
	\includegraphics[width=0.9\textwidth]{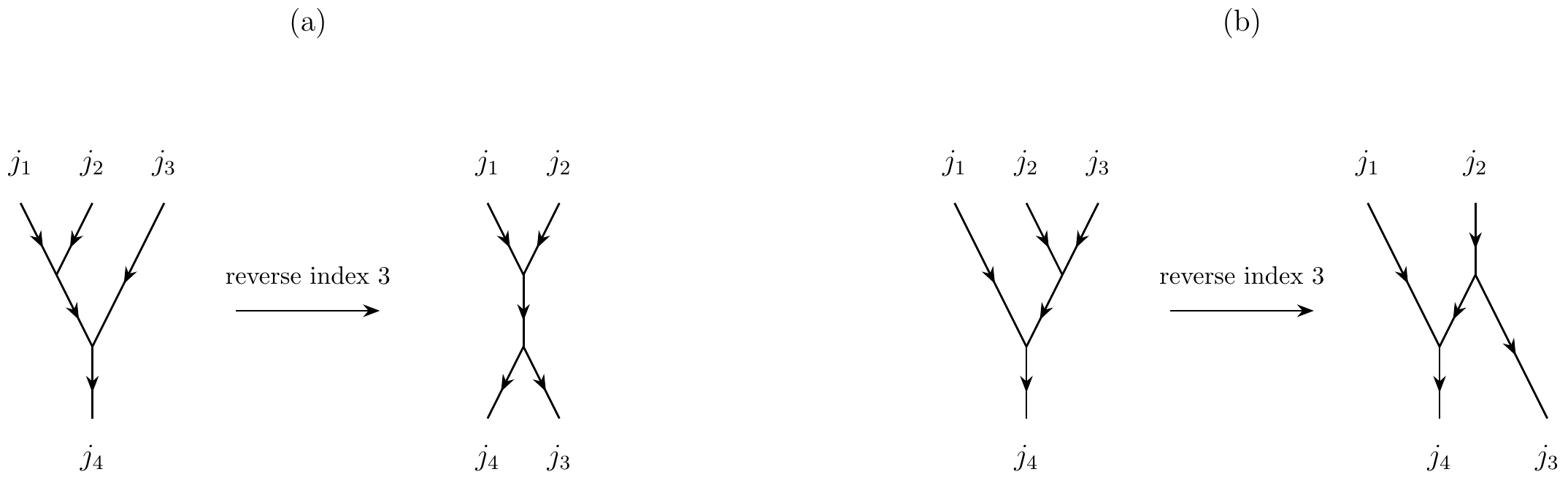}
	\caption{Reversal of the same index for different fusion trees. On the right one can see that the reversal in an unprepared tree results in an unwanted yoga fusion tree.}
	\label{fig:TikZ_Files_Index_Reversal_1}
\end{figure}

However, index reversal is not always as straightforward as in the case shown in \fref{fig:TikZ_Files_indexReversal_indexReversal_RelevantPart_1} or the one shown in \fref{fig:TikZ_Files_Index_Reversal_1}(a). Consider, for instance, the index reversal shown in \fref{fig:TikZ_Files_Index_Reversal_1}(b). In this case, reversing an index leads to an intermediate yoga fusion tree. In order to avoid this, and remain within the set of simple fusion trees, we can first transform the tree (by applying $F$-moves) such that the subsequent index reversal does not lead to a yoga fusion tree, see e.g. \fref{fig:TikZ_Files_SmarterReversal_2}.

Finally, in our implementation we only allow reversals of indices that are located at the edge of the fusion tree. In order to reverse an index that is not at the edge of the fusion tree, an index permutation must be introduced. We explain this next.

\begin{figure}
	\centering
	\includegraphics[width=0.8\textwidth]{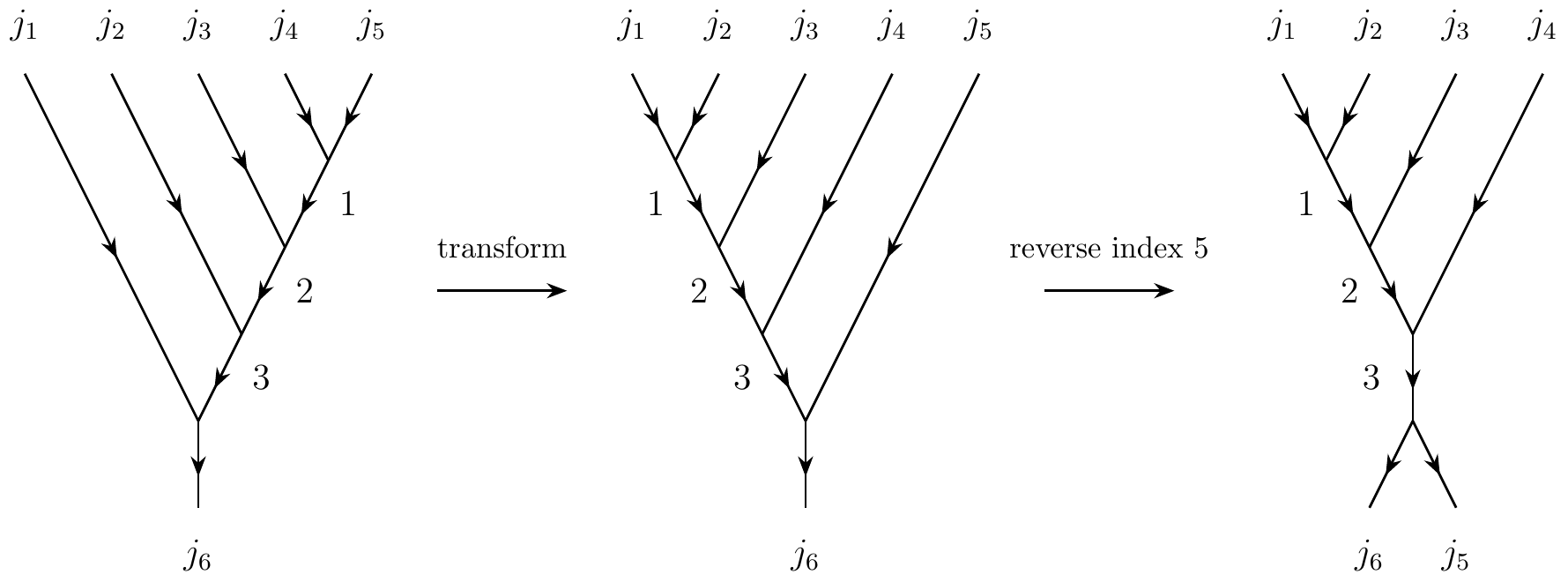}
	\caption{Preparing index 5 to be reversed, so that the reversed index appears either on the same node or in the node closest to its successor index.}
	\label{fig:TikZ_Files_SmarterReversal_2}
\end{figure}

% subsection tensors_with_more_than_three_indices (end)

% section reversing_an_index (end)

\section{Permutation of indices} % (fold)
\label{sec:permutation_of_indices}

Permuting indices of a regular tensor, namely, a numerical array corresponds to simply shuffling the tensor components. However, updating a symmetric tensor after permuting indices is more involved, since once again we focus on updating only the degeneracy tensors.

\subsection{\texorpdfstring{$R$}{TEXT}-symbols} % (fold)
\label{sub:r_symbols}

As also considered in Ref.~\cite{sukhi}, we find it convenient to implement a generic permutation as a sequence of swaps. We will restrict to swaps of indices that (i) have the same direction, (ii) appear at neighbouring locations in the fusion tree, and (iii) belong to the same node. (Two indices with different directions can be swapped by first reversing one of the indices, swapping, and then inverting the reversal.) As we discuss below, such a swap introduces only a factor, which is given by the $\mathit{R}$\textit{-symbol} of $SU(2)$
\begin{align}
	R^\text{swap}_{j_a,j_b \rightarrow j_c} = R^\text{swap}_{j_c \rightarrow j_a , j_b} = (-1)^{j_a + j_b - j_c},
\end{align}
see \fref{fig:TikZ_Files_indexPermutation_Index_Permutation_1}. (We remark that for an anyon model, the swap is replaced by the braiding operator, and clockwise and counterclockwise braiding generally correspond to different $R$-symbols.)

\begin{figure}
	\centering
	\includegraphics[width=1\textwidth]{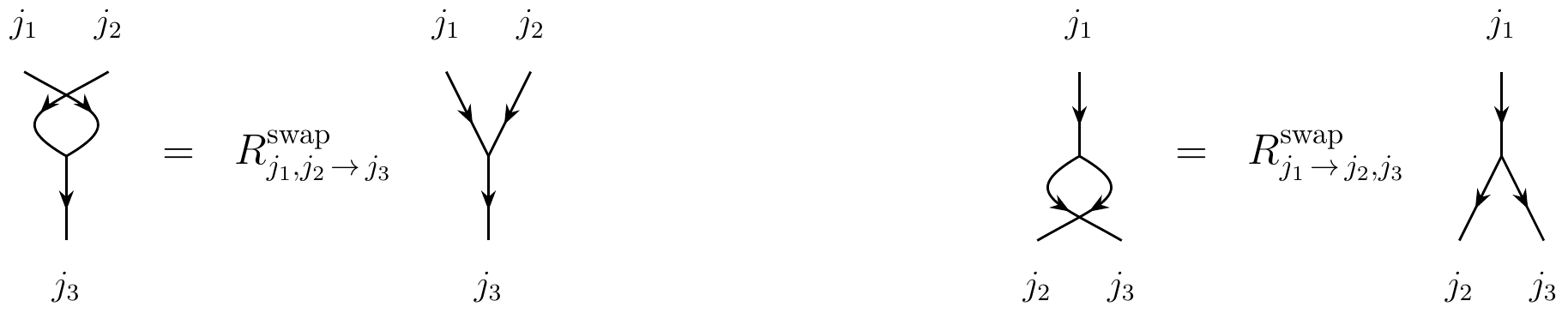}
	\caption{Swap two indices in a fusion (splitting) node is equal to a fusion (splitting) node times a $R$-symbol.}
	\label{fig:TikZ_Files_indexPermutation_Index_Permutation_1}
\end{figure}

% subsection r_symbols (end)

\subsection{Tensors with three indices} % (fold)
\label{sub:tensors_with_three_indices_permutation}

Let us be more specific by considering the example of a 3-index symmetric tensor $T$. We can swap two of its indices (that have the same direction) by swapping indices of the blocks $T_{j_aj_bj_c}$ in each charge sector. We obtain
\begin{align}
	T_{j_aj_bj_c} \hspace{0.25cm} \xrightarrow[\phantom{\text{permutationnn}}]{\text{permutation}} \hspace{0.25cm} T_{j_bj_aj_c} =  P_{j_b j_a j_c}  \otimes  Q_{j_b j_a j_c}.
\end{align}
In the above equation, the swap of indices in the degeneracy tensor $P$ can be done as usual for regular tensors (that is, by simply shuffling components). However, the swapped structural tensors $Q$ are related to the input ones by the $R$-symbols,
\begin{align}
	Q_{j_bj_aj_c}^\prime = R^\text{swap}_{j_a,j_b \rightarrow j_c} Q_{j_aj_bj_c}\ ,
\end{align}
The $R$-factor can be absorbed into the swapped degeneracy tensor, so that we have
\begin{align}\label{eq:updateswap}
	P_{j_bj_aj_c}^\prime = R^\text{swap}_{j_a,j_b \rightarrow j_c} P_{j_aj_bj_c}.
\end{align}
Thus, swapping indices amounts to updating the fusion tree (swapping all the irreps associated with the two indices), and the degeneracy tensors according to \Eref{eq:updateswap}. Note that as for the reversal operations one does not need to take linear combinations of degeneracy tensors here.

Let us explain how the algorithm proceeds more concretely. To this end, let us consider a 3-index tensor with two ingoing and one outgoing index that has irreps and directions as listed in \tref{tab:Rank3Permutation_1}.
\begin{table}[ht]
	\centering
	\begin{tabular}{c c c}
		index		& irreps						& direction	\\\hline
		$1$			& $j = \big\lbrack 0 , \frac 1 2 \big\rbrack$		& $-1$				\\
		$2$			& $j = \big\lbrack 0 , 1 \big\rbrack$				& $-1$				\\
		$3$			& $j = \big\lbrack 0 , \frac 1 2 , 1 , \frac 3 2 \big\rbrack$	& $+1$
	\end{tabular}
	\caption{Index data for a possible 3-index tensor before swap.}
	\label{tab:Rank3Permutation_1}
\end{table}

The list of valid charge sectors can be determined e.g. using the procedure described in \sref{sub:determination_of_the_valid_charge_sectors} and is given by
\begin{align}
	\text{listOfChargeSectors = } \left\lbrace \left\lbrack 0,0,0 \right\rbrack , \left\lbrack 0,1,1 \right\rbrack , \left\lbrack \frac 1 2 , 0 , \frac 1 2 \right\rbrack , \left\lbrack \frac 1 2 , 1 , \frac 1 2 \right\rbrack, \left\lbrack \frac 1 2 , 1 , \frac 3 2 \right\rbrack \right\rbrace .
	\label{eq:ChargeSectors_1}
\end{align}
After swapping indices $1$ and $2$, the updated 3-index tensor will have irreps and directions as listed in \tref{tab:Rank3Permutation_2}.
\begin{table}[ht]
	\centering
	\begin{tabular}{c c c}
		index		& irreps					& direction	\\\hline
		$1$			& $j = \big\lbrack 0 , 1 \big\rbrack$				& $-1$				\\
		$2$			& $j = \big\lbrack 0 , \frac 1 2 \big\rbrack$		& $-1$				\\
		$3$			& $j = \big\lbrack 0 , \frac 1 2 , 1 , \frac 3 2 \big\rbrack$	& $+1$
	\end{tabular}
	\caption{Index data after swap.}
	\label{tab:Rank3Permutation_2}
\end{table}
One then has to determine the valid charge sectors for the swapped tensor (again e.g. using the procedure described in \sref{sub:determination_of_the_valid_charge_sectors}). Following \Eref{eq:ChargeSectors_1}, the updated list of charge sectors is given by
\begin{align}
	\text{listOfChargeSectors = } \left\lbrace \left\lbrack 0,0,0 \right\rbrack , \left\lbrack 0 , \frac 1 2 , \frac 1 2 \right\rbrack , \left\lbrack 1,0,1 \right\rbrack , \left\lbrack 1 , \frac 1 2 , \frac 1 2 \right\rbrack , \left\lbrack 1 , \frac 1 2 , \frac 3 2 \right\rbrack \right\rbrace .
	\label{eq:ChargeSectors_2}
\end{align}
Alternatively, one could determine the charge sectors by simply swapping elements of each vector in \Eref{eq:ChargeSectors_1} as
\begin{align}
	\text{listOfChargeSectors = } \left\lbrace \left\lbrack 0,0,0 \right\rbrack , \left\lbrack 1,0,1 \right\rbrack , \left\lbrack 0 , \frac 1 2 , \frac 1 2 \right\rbrack , \left\lbrack 1 , \frac 1 2 , \frac 1 2 \right\rbrack, \left\lbrack 1 , \frac 1 2 , \frac 3 2 \right\rbrack \right\rbrace .
	\label{eq:ChargeSectors_3}
\end{align}
The two lists above have the same elements, and one can work with either one. However, in our implementation we found it convenient to work with the first option, which always produces a sorted list of charge sectors. Applying the transformation in \Eref{eq:updateswap} the new degeneracy tensors become
\begin{align}
	\begin{split}
		P_{0 0 0}^\prime &= + \text{permute}(P_{0 0 0},[2,1,3])\\
		P_{0 \frac 1 2 \frac 1 2}^\prime &= + \text{permute}(P_{\frac 1 2 0 \frac 1 2},[2,1,3])\\
		P_{1 0 1}^\prime &= + \text{permute}(P_{0 1 1},[2,1,3])\\
		P_{1 \frac 1 2 \frac 1 2}^\prime &= - \text{permute}(P_{\frac 1 2 1 \frac 1 2},[2,1,3])\\
		P_{1 \frac 1 2 \frac 3 2}^\prime &= + \text{permute}(P_{\frac 1 2 1 \frac 3 2},[2,1,3])\ .
	\end{split}
\end{align}

% subsection tensors_with_three_indices_permutation (end)

\subsection{Tensors with more than three indices} % (fold)
\label{sub:tensors_with_more_than_three_indices_permutation}

Since swaps are allowed only for indices that belong to the same node of the fusion tree, swapping indices that belong to the same node in a tensor with more than three indices essentially reduces to the case of
3-index tensors. If, however, we want to swap neighbouring indices that do not belong to the same node, we would first have to transform the fusion tree via $F$-moves to bring the two indices on the same node, see e.g. \fref{fig:TikZ_Files_indexReversal_Rank_6_Permutation_1}.}

\begin{figure}
	\centering
	\includegraphics[width=.8\textwidth]{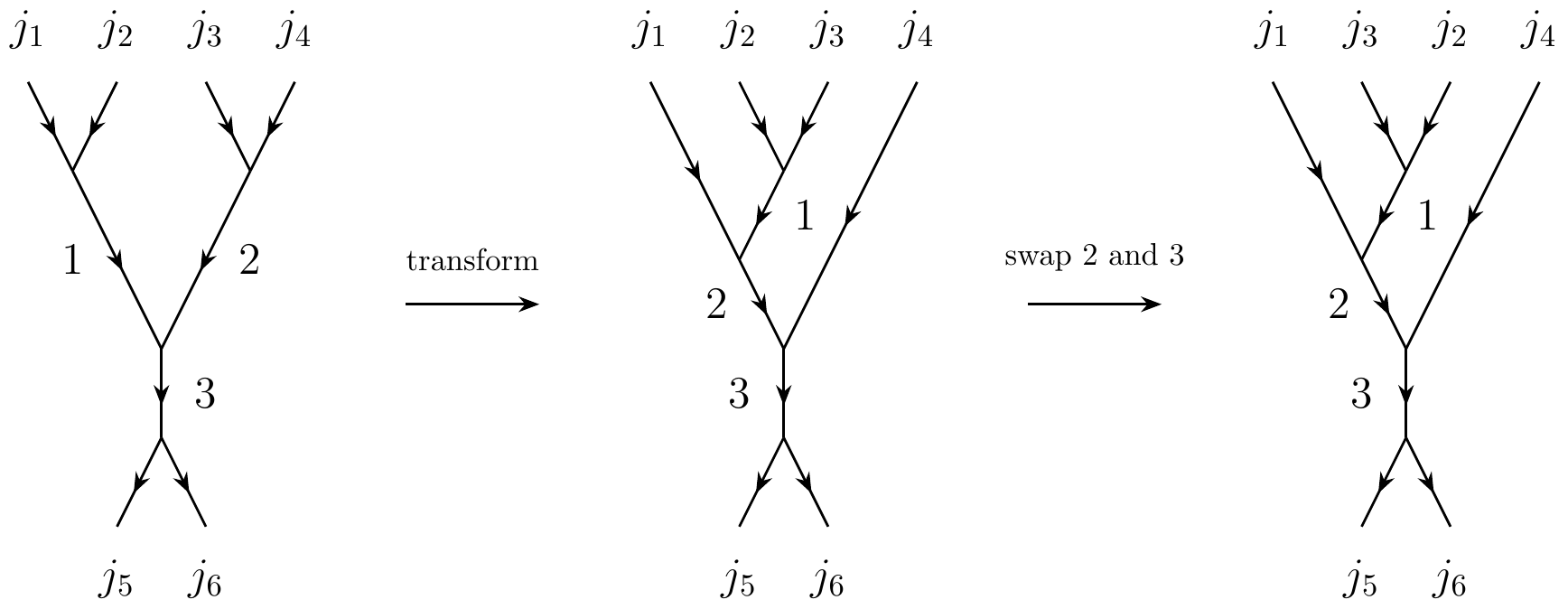}
	\caption{An example of the sequence of operations to swap two indices not belonging to the same node. The final tree could be brought back to its original form while having swapped indices, if required.}
	\label{fig:TikZ_Files_indexReversal_Rank_6_Permutation_1}
\end{figure}

\begin{figure}
	\centering
	\includegraphics[width=.9\textwidth]{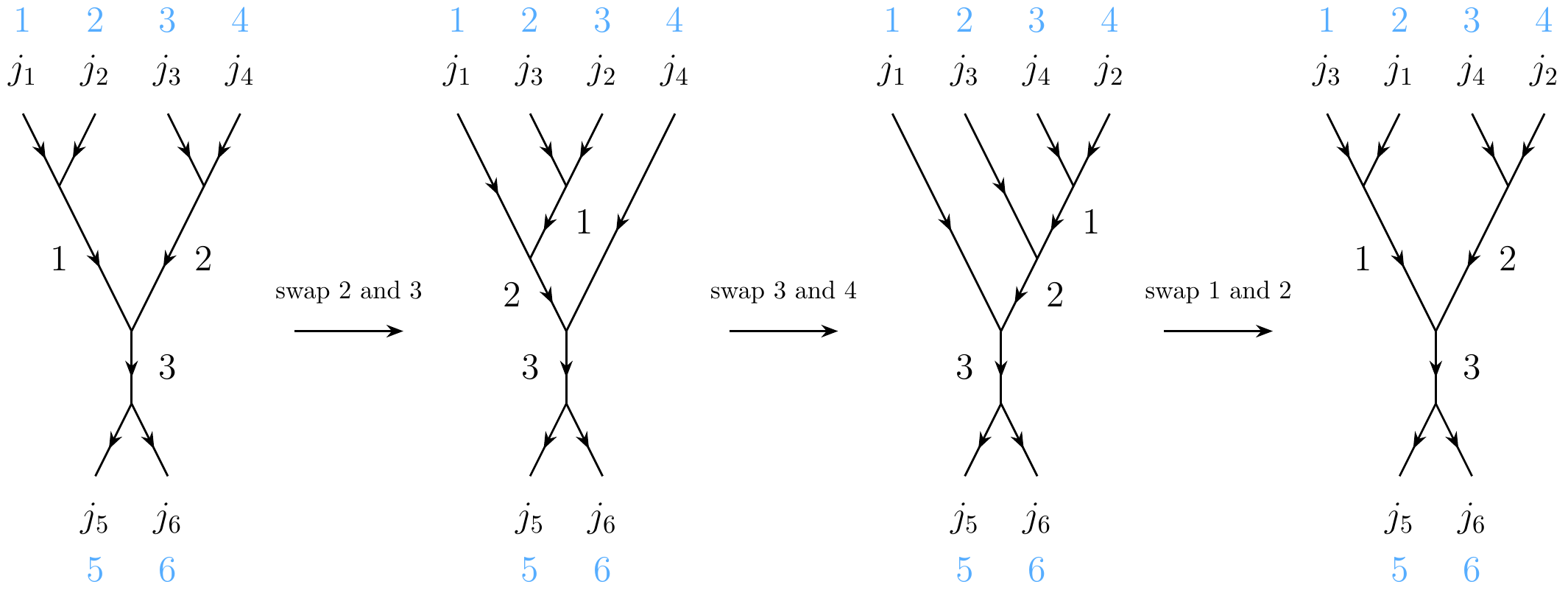}
	\caption{Permuting a 6-index tensor involves applying a sequence of swaps on the fusion tree (left). All the swaps are indicated with respect to the blue number labels for the edges.}
	\label{fig:TikZ_Files_indexPermutation_Rank_6_Permutation_2}
\end{figure}

The sequence of swaps that implement a given index permutation can be determined by means of the bubble sort algorithm~\cite{knuth}, a simple algorithm that is based on repeated pairwise comparisons to sort a list. Consider, for example, the permutation $\text{perm} = [3,1,4,2,5,6]$ of a fictitious 6-index tensor. The algorithm would perform the permutation following three swaps, namely those of positions $[1,2]$, $[3,4]$ and finally $[2,3]$. If we now start from the initial, sorted vector and apply the sequence in reverse order we end up with the desired permutation
\begin{align}
	\text{perm} : \hspace{0.25cm} [1,2,3,4,5,6] \hspace{0.25cm} \longrightarrow \hspace{0.25cm} [1,3,2,4,5,6] \hspace{0.25cm} \longrightarrow \hspace{0.25cm} [1,3,4,2,5,6] \hspace{0.25cm} \longrightarrow \hspace{0.25cm} [3,1,4,2,5,6] . 
\end{align}
The fusion trees for this procedure could look as in \fref{fig:TikZ_Files_indexPermutation_Rank_6_Permutation_2}, where the three swaps are applied in reverse. In this case, each swap is also preceded by a transformation of the fusion tree by applying $F$-moves in order to move indices to the same node in preparation for the swap.

% subsection tensors_with_more_than_three_indices_permutation (end)

% section permutation_of_indices (end)

\section{Reshaping a tensor} % (fold)
\label{sec:reshaping_a_tensor}

Another common operation in tensor network algorithms is ``reshaping'' a tensor. That is, obtaining a new tensor by fusing together indices or splitting back an index into several indices. We will restrict to fusing together only two neighbouring indices into one index or splitting an index into two neighbouring indices. We will also restrict to fusing indices that have the same direction. Similarly, an index with a given direction is split into two indices that have the same direction, such that the resulting two indices can be fused together to recover the original index. More general reshapes can be implemented by composing these basic fusion and splitting operations with index reversals and index permutations that were described previously.

\subsection{Fusion of two indices} % (fold)
\label{sub:fusion_of_two_indices}

Let us illustrate the implementation of fusion with an example of a 3-index symmetric tensor $T_{abc}$ with one incoming index $a$ and two outgoing indices $b$ and $c$. The irreps and degeneracies for each index of $T$ are listed in \tref{tab:Rank3IrrepAssignment}.

\begin{table}[ht]
	\centering
	\begin{tabular}{c c}
		index			& irreps and degeneracies $\lbrace j_i , t_i \rbrace$	\\\hline
		a			& $\left\lbrace \lbrack 0,1,2 \rbrack , \lbrack 1,2,3 \rbrack \right\rbrace$	\\
		b			& $\left\lbrace \lbrack 0,1 \rbrack , \lbrack 1,4 \rbrack \right\rbrace$		\\
		c			& $\left\lbrace \lbrack 0,1 \rbrack , \lbrack 1,6 \rbrack \right\rbrace$
	\end{tabular}
	\caption{Example of irreps and degeneracies for the fusion of a 3-index tensor.}
	\label{tab:Rank3IrrepAssignment}
\end{table}

It is easy to determine all the valid charge sectors, which are listed below.
\beqa
	\text{listOfChargeSectors} & = & \{ \left\lbrack 0,0,0 \right\rbrack , \left\lbrack 0,1,1 \right\rbrack , \left\lbrack 1 , 0 , 1 \right\rbrack , \left\lbrack 1 , 1 , 0 \right\rbrack , \left\lbrack 1 , 1 , 1 \right\rbrack , \left\lbrack 2 , 1 , 1 \right\rbrack \} .
	\label{eq:OldChargeSectors}
\eeqa
These charge sectors label the various degeneracy tensors of $T$, namely,
\begin{align}\label{eq:inputfuse}
\{P_{0,0,0},P_{0,1,1},P_{1,0,1},P_{1,1,0},P_{1,1,1},P_{2,1,1}\}.
\end{align}
Now let us fuse index $b$ and $c$ into a new outgoing index $d$ to obtain a 2-index tensor $T'_{ad}$. The irrep decomposition of the fused index $d$ can be easily obtained by decomposing the tensor product space $\mathbb V^{(b)} \otimes \mathbb V^{(c)}$ as a direct sum of irreps. The resulting irreps and degeneracies that appear on the fused index are listed in \tref{tab:Rank2IrrepAssignment}.

\begin{table}
	\centering
	\begin{tabular}{c c}
		index		& irreps and degeneracies $\lbrace j_i , t_i \rbrace$	\\\hline
		1			& $\left\lbrace \lbrack 0,1,2 \rbrack , \lbrack 1,2,3 \rbrack \right\rbrace$	\\
		2			& $\left\lbrace \lbrack 0,1,2 \rbrack , \lbrack 25,34,24 \rbrack \right\rbrace$		\\
	\end{tabular}
	\caption{Irreps and degeneracies after fusing indices $2$ and $3$ from \tref{tab:Rank3IrrepAssignment}.}
	\label{tab:Rank2IrrepAssignment}
\end{table}

After the fusion the possible charge sectors of $T'$ are given by
\beqa \label{eq:outputfuse}
\text{listOfChargeSectors} & = & \{ \left\lbrack 0,0 \right\rbrack , \left\lbrack 1,1 \right\rbrack , \left\lbrack 2,2 \right\rbrack \},
\eeqa
and denote the corresponding degeneracy tensors as $\{P'_{0,0},P'_{1,1},P'_{2,2}\}$. Our goal is to determine these degeneracy tensors and the fusion tree of $T'$ directly from the degeneracy tensors and fusion tree of $T$ respectively. One begins by identifying which input charge sectors contribute to each of the output charge sectors, in accordance with the fusion rules. The result is shown in \tref{tab:fusemap_in_to_out}.
\begin{table}
	\centering
	\begin{tabular}{c c}
		output charge sector		& contributing input charge sectors	\\\hline
		$\left\lbrack 0,0 \right\rbrack$			& $\{\left\lbrack 0,0,0 \right\rbrack , \left\lbrack 0,1,1 \right\rbrack\}$	\\
		$\left\lbrack 1,1 \right\rbrack$			& $\{\left\lbrack 1 , 0 , 1 \right\rbrack , \left\lbrack 1 , 1 , 0 \right\rbrack , \left\lbrack 1 , 1 , 1 \right\rbrack\}$	\\
		$\left\lbrack 2,2 \right\rbrack$			& $\{\left\lbrack 2 , 1 , 1 \right\rbrack\}$	\\
	\end{tabular}
	\caption{{Map between the charge sectors of the input and the output tensor. Each degeneracy tensor corresponding to one of the output charge sectors is constructed by combining the transformed input degeneracy tensors that correspond to the contributing input charge sectors.}}
	\label{tab:fusemap_in_to_out}
\end{table}
The information in \tref{tab:fusemap_in_to_out} tells us how to compose the $P'$ degeneracy tensors from $P$ degeneracy tensors. For example, the degeneracy tensor $P'_{0,0}$ is obtained by reshaping the corresponding two input degeneracy tensors -- $P_{0,0,0}$ and $P_{0,1,1}$ -- into 2-index tensors by fusing together indices $t_b$ and $t_c$, and then concatenating them. We denote this operation as
\begin{align}
P'_{0,0} = P_{0,(0,0)} \circ P_{0,(1,1)},
\end{align}
where, for example, $P_{0,(0,0)}$ denotes the 2-index tensor obtained by fusing indices $t_b$ and $t_c$ (indicated by bracketing the indices) of the 3-index tensor $P_{0,0,0}$, and $\circ$ denotes the concatenation of arrays along the dimension that has the same size in the two arrays (row-wise in this case). Here $P_{0,(0,0)}$ is a $1 \times 1$ array whereas $P_{0,(1,1)}$ is $1\times 24$ array, and the resulting $P'_{0,0}$ is a $1\times 25$ array. The remaining $P'$ degeneracy tensors can be obtained in a similar way. We have
\begin{align}
P'_{1,1} &= P_{1,(0,1)} \circ P_{1,(1,0)} \circ P_{1,(1,1)}, \\
P'_{2,2} &= P_{2,(1,1)},
\end{align}
where $P'_{1,1}$ is a $2 \times 34$ array that is obtained by concatenating $P_{1,(0,1)}$ (a $2\times 6$ array), $P_{1,(1,0)}$ (a $2\times 4$ array), and $P_{1,(1,1)}$ (a $2\times 24$ array). On the other hand, the degeneracy tensor $P'_{2,2}$ is simply the reshaped tensor $P_{2,(1,1)}$ since there are no other contributing input degeneracy tensors in this case.

Having determined all the degeneracy tensors of $T'$, its fusion tree is obtained from that of $T$ by simply deleting the one node, and introducing a dummy index to represent a fusion vertex that corresponds to the identity.

% subsection fusion_of_two_indices (end)

\subsection{Tensors with more than three indices} % (fold)
\label{sub:tensors_with_more_than_three_indices_reshaping}

Notice that the structural tensors did not play any role in the fusion of the simple case of a 3-index tensor as described in the previous section. This will remain the case for fusion in a tensor with more indices as long as the two indices to be fused belong to the same node and have the same direction. This is the elementary fusion operation that we will restrict ourselves to. Recall that a more general reshape can be obtained by composing this elementary fusion with index reversals and permutations. The fusion procedure for tensors with more than three indices proceeds as for the 3-index case: 
\begin{enumerate}
\item Determine the irreps on the fused index,
\item Determine the charge sectors for the reshaped tensor,
\item Build a table $E$ that lists all contributing input charge sectors for each output charge sector,
\item Reshape each of the input degeneracy tensors by fusing the two indices,
\item Build the output degeneracy tensor in each output charge sector by concatenating the corresponding input degeneracy tensors that appear in table $E$.
\end{enumerate}
The update of the fusion tree is quite straightforward. One only has to remove a node, which is attached to the two indices that are fused, from the fusion tree, and update its data structure accordingly.

\begin{figure}
	\centering
	\includegraphics[width=.9\textwidth]{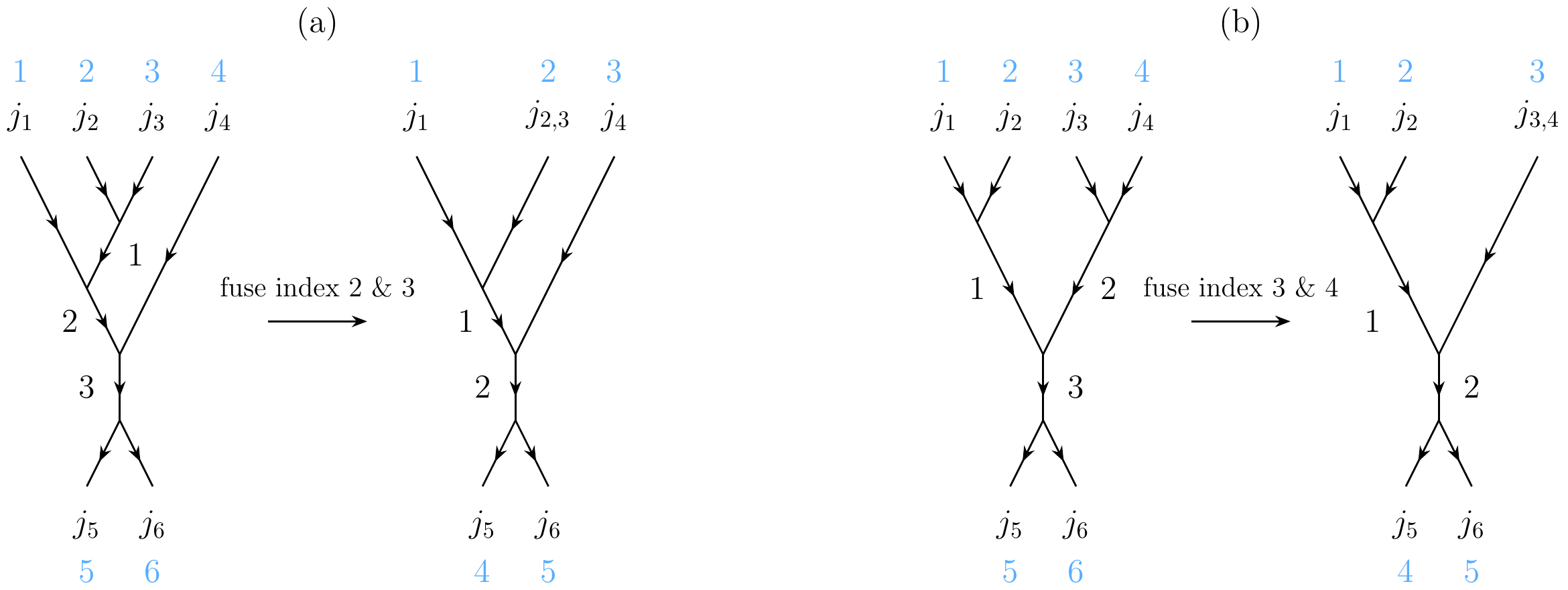}
	\caption{Fusion of two indices for two different possible fusion trees of a 6-index tensor.}
	\label{fig:TikZ_Files_Rank6_Reshaping_1}
\end{figure}

Let us illustrate the update of the fusion tree data structure for the fusions illustrated in \fref{fig:TikZ_Files_Rank6_Reshaping_1}. Figure (a) shows the fusion of indices $[2,3]$ into a new index $2$, and figure (b) on the right shows the fusion of indices $[3,4]$ into a new index $3$.

\begin{figure}
	\centering
	\includegraphics[width=.7\textwidth]{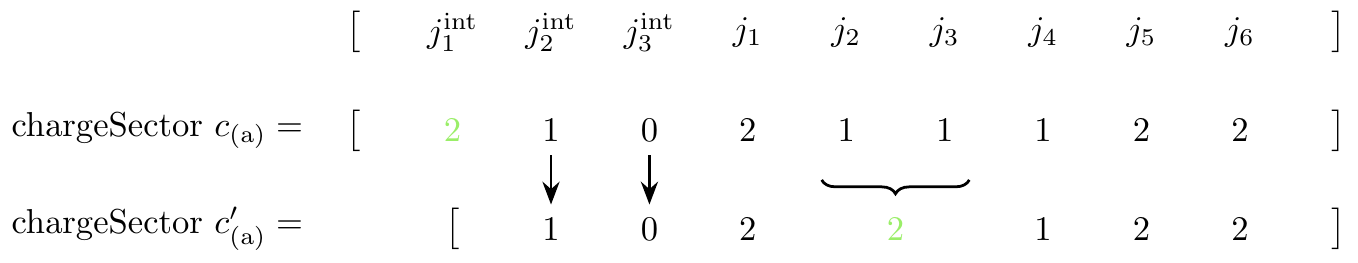}
	\caption{Sorting of possible charge sectors for the fusion indices $2$ and $3$ in \fref{fig:TikZ_Files_Rank6_Reshaping_1}(a). The arrows indicate the internal indices that remain after fusion. The charge (irrep) of the internal index that disappears fixes the value of the charge for the fused index, which we show in green.}
	\label{fig:TikZ_Files_ChargeSector_Reshaping_1}
\end{figure}
\begin{figure}
	\centering
	\includegraphics[width=.7\textwidth]{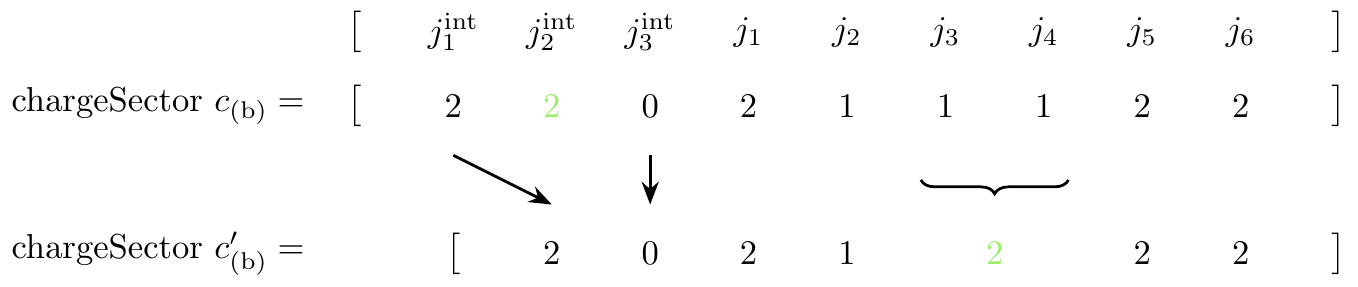}
	\caption{Sorting of possible charge sectors for the fusion indices $3$ and $4$ in \fref{fig:TikZ_Files_Rank6_Reshaping_1}(b). The arrows indicate the internal indices that remain after fusion. The charge of the internal index that disappears fixes the value of the charge for the fused index, which we show in green.}
	\label{fig:TikZ_Files_ChargeSector_Reshaping_2}
\end{figure}

Consider now a possible charge sector. First, for the input tree in \fref{fig:TikZ_Files_Rank6_Reshaping_1}(a) lets choose the charge sector $c_\text{(a)}$ listed in \fref{fig:TikZ_Files_ChargeSector_Reshaping_1}. Recall our convention for labeling charge sectors (introduced in \sref{sub:determination_of_the_valid_charge_sectors}): irreps on the internal indices of a fusion tree appear first, followed by irreps on the open indices. Thus, the first three entries from the left in $c_\text{(a)}$ are the irreps for the three internal indices of the input tree in \fref{fig:TikZ_Files_Rank6_Reshaping_1}(a), and the last six entries are the irreps for the open indices. After fusing the open indices $2$ and $3$, an internal index disappears and we obtain the tree shown on the right-hand side of \fref{fig:TikZ_Files_Rank6_Reshaping_1}(a). The charge sector for the fused tree is listed as $c'_\text{(a)}$ in \fref{fig:TikZ_Files_ChargeSector_Reshaping_1}. Notice that the internal index that disappears becomes the open fused index of the updated tree, as indicated by the green labels in \fref{fig:TikZ_Files_ChargeSector_Reshaping_1}. The analogous update for a different fusion is shown in \fref{fig:TikZ_Files_Rank6_Reshaping_1}(b). A possible charge sector for this tree is listed in \fref{fig:TikZ_Files_ChargeSector_Reshaping_2}. In this case, the second internal index is removed, and becomes the open fused index.

% subsection tensors_with_more_than_three_indices_reshaping (end)

\subsection{Splitting back indices: reversing fusion} % (fold)
\label{sub:splitting_back_indices_reversing_fusion}

So far we have only discussed fusion of indices. We may also want to reverse a fusion, that is, split back a fused index into the constituent indices. In this section, we describe how to reverse a fusion process. In order to reverse a multi-step fusion, we must keep track of how indices are pairwise fused at each step. This information must be created and appended to the data structure of the tensor during fusion.

In our implementation, we introduced a particular internal structure for each index that records the fusion history of the index. To see how it works, let us consider a tensor with five indices, each with the same direction (say all outgoing). Each index can carry multiple irreps and degeneracies,
\begin{align}
	j_i = \lbrace [j_{i1},j_{i2},.\hdots],[t_{j_{i1}},t_{j_{i2}},\hdots] \rbrace .	
\end{align}
Let us imagine that all the five indices of the tensor are fused to obtain a 1-index tensor (a vector) according to the sequence
\begin{align}
	\lbrace j_1,j_2,j_3,j_4,j_5 \rbrace \xrightarrow[]{j_4 \otimes j_5} \lbrace j_1,j_2,j_3,j_4^\prime \rbrace \xrightarrow[]{j_3 \otimes j_4^\prime} \lbrace j_1,j_2,j_3^\prime \rbrace \xrightarrow[]{j_1 \otimes j_2} \lbrace j_1^\prime,j_3^\prime \rbrace \xrightarrow[]{j_1^\prime \otimes j_3^\prime} \lbrace j_1^{\prime\prime} \rbrace\ 
	\label{eq:SuccessiveFusionFiveSpins}
\end{align}
of pairwise fusions. Here we have $j_4^\prime = j_4 \otimes j_5$, $j_3^\prime = j_3 \otimes j_4^\prime$, $j_1^\prime = j_1 \otimes j_2$ and $j_1^{\prime\prime} = j_1^\prime \otimes j_3^\prime$. The lists of irreps that appear on the successively fused indices can be easily computed, and this information needs to be restored to split back the vector into the original 5-index tensor. To achieve this in practice, we store each index as a separate object of a dedicated class, which includes the fusion history of the index (refer to \sref{sec:blue_data_structure_for_an_su_2_symmetric_tensor} for the overview about the stored information in the tensor). The fused index, e.g., the one resulting after the first step in \Eref{eq:SuccessiveFusionFiveSpins}, becomes the new fourth index of the reshaped tensor (after step 1), with irreps and degeneracies $j_4^\prime$. Its direction is the same as the direction of the two indices that were fused. Additionally, we store the fusion history
\begin{align}
	j_4^{\prime,\text{history}} = \left\lbrace j_4^\prime , j_4 , j_5 \right\rbrace .
	\label{eq:CombinedLegPrimaryLegList_1}
\end{align}
The history becomes nested after the next fusion step in which a regular index is combined with an already fused index. This yields
\begin{align}
	j_3^{\prime,\text{history}} = \left\lbrace j_3^\prime , j_3 , \left\lbrace j_4^\prime , j_4 , j_5 \right\rbrace \right\rbrace .
	\label{eq:CombinedLegPrimaryLegList_2}
\end{align}
The history for the remaining steps is
\begin{gather}
		j_1^{\prime,\text{history}} = \left\lbrace j_1^\prime , j_1 , j_2 \right\rbrace , 	\\
		j_1^{\prime\prime,\text{history}} = \left\lbrace j_1^{\prime\prime}, \left\lbrace j_1^\prime , j_1 , j_2 \right\rbrace , \left\lbrace j_3^\prime , j_3 , \left\lbrace j_4^\prime , j_4 , j_5 \right\rbrace \right\rbrace \right\rbrace .
		\label{eq:CombinedLegPrimaryLegList_3}
\end{gather}

Once the fusion history is recorded in this way, one can reverse the fusion following these steps:
\begin{enumerate}
\item Determine the charge sectors for the output tensor,
\item Build a table $E$ that lists all contributing input charge sectors for each output charge sector,
\item Break the degeneracy tensor in each input charge sector into pieces (i.e. reverse the concatenation effected during the fusion). The fusion histories of each index are sorted in the order of the concatenation of the block -- combining this information with the dimensions of the degeneracy tensors of the output contained in the fusion history, the input degeneracy tensors can be appropriately sliced back. Each sliced degeneracy tensor corresponds to an output degeneracy tensor labeled by an output charge sector.
\item Reshape the output degeneracy tensor in each output charge sector by splitting back the fused indices.
\end{enumerate}
The update of the fusion tree requires adding nodes in this case. The necessary information required for this task can also be derived from the fusion histories.

% subsection splitting_back_indices_reversing_fusion (end)

% section reshaping_a_tensor (end)

\section{Tensor contraction} % (fold)
\label{sec:tensor_contraction}

Next, we turn to contraction of a tensor network. Since a tensor network can be contracted by pairwise contracting the tensors in some sequence, we will focus on the contraction of two tensors. A common way to contract two regular tensors is to reshape each of the two tensors to an appropriate matrix (by separating out all indices that are contracted from those that are left open and fusing these two sets of indices into two thicker indices), multiplying the two matrices, and then reshaping the resulting matrix into a tensor by splitting back the fused indices. This contraction can be implemented by composing index reversals, permutations and fusions. (The MATLAB function \textit{ncon}~\cite{ncon} implements tensor contraction in this way.)

In previous sections, we have described how to implement these operations for symmetric tensors. Thus, a possible implementation of a symmetric tensor contraction is to simply implement the regular procedure outlined above by means of symmetric operations. However, in our implementation, we restricted to a subset of all possible tensor contraction scenarios in order to work with only simple fusion trees. Therefore, the input tensors must be prepared in a certain way before the contraction can proceed, as described below.

\bigskip 
{\textbf{\emph{1) Matching of contracted indices.-}} In a regular tensor contraction, the contracted indices must match (be identical) in the two tensors. In our implementation, we require that the indices to be contracted not only match in the two tensors but also that (i) they appear in neighbouring locations in the fusion tree, (ii) all the contracted indices have a common parent node in the fusion tree of both the tensors, and (iii) and these parent fusion trees are the same in the two tensors. This is necessary to ensure that the part of the combined fusion tree that involves the contracted legs can be simplified and removed from the final tree. Generally, this will require preparing the tensors by permuting and transforming the fusion trees of the two tensors. See, for example, the two cases in \fref{fig:TikZ_Files_Contraction_1}. The top fusion tree on the left-hand side needs to be pre-processed before contraction by applying $F$-moves, resulting in the fusion tree shown on the right-hand side, which is now ready for contraction. Another type of preprocessing of one of the input tensors is required if the contraction yields a loop in the combined fusion tree that incorporates one or more open legs. In this case one would need to apply a permutation before contracting the tensors, as demonstrated in \fref{fig:TikZ_Files_Contraction_3}.

\begin{figure}
	\centering
	\includegraphics[width=.8\textwidth]{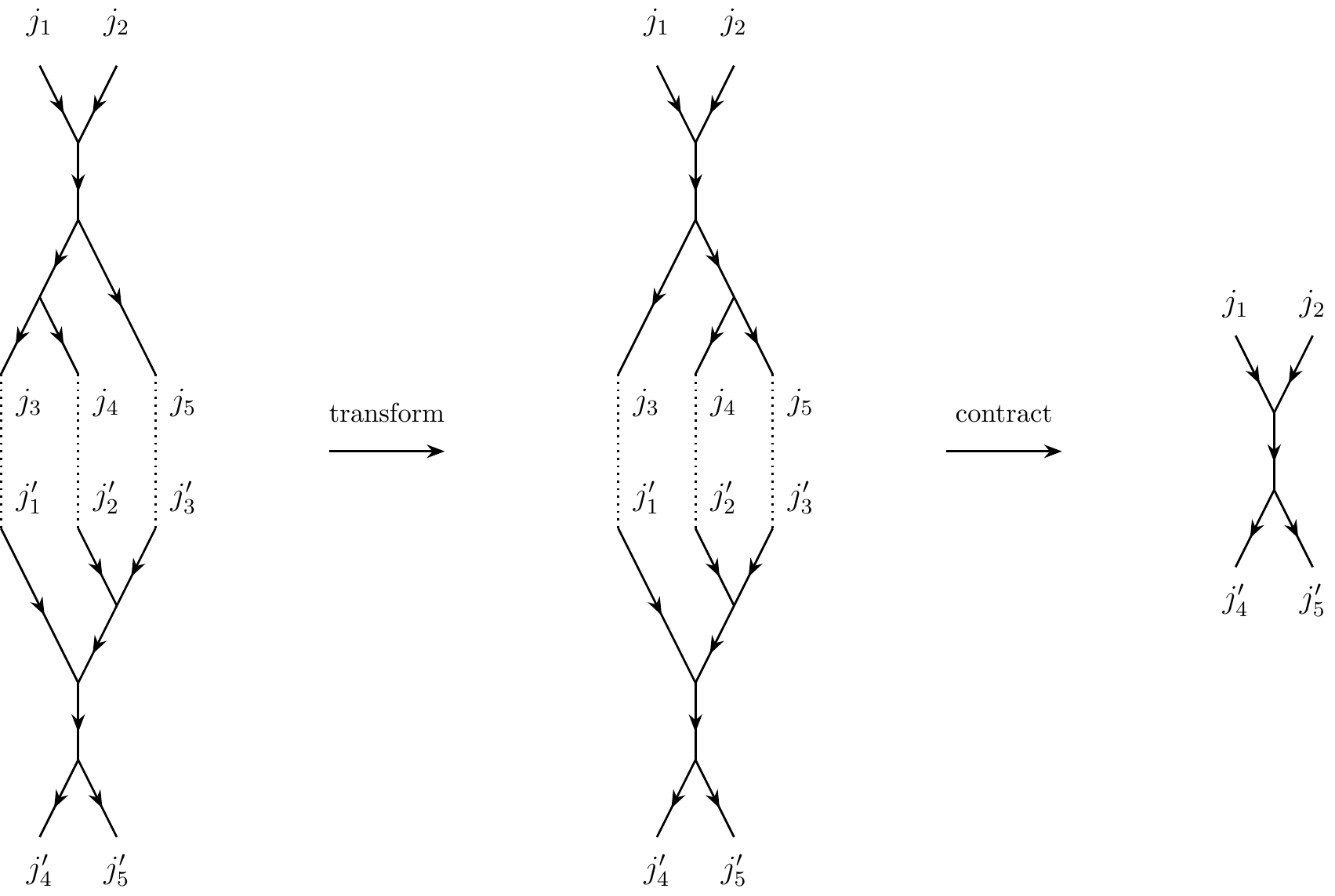}
	\caption{Transforming the top tree on the left as shown simplifies the subsequent contraction of the two fusion trees.}
	\label{fig:TikZ_Files_Contraction_1}
\end{figure}

\bigskip 

{\textbf{2) \emph{The output fusion tree.-}} We will require that the output fusion tree is a simple fusion tree, and most contractions can, in fact, be prepared so that this is the case. But it is possible that the output tree is a yoga as in \fref{fig:TikZ_Files_Contraction_2a} or a monster fusion tree, if the tensors are not prepared carefully. (Notice that the tensor operations that we described in previous sections never resulted in yoga or a monster trees.) Therefore, for completeness we describe how to transform a yoga fusion tree to a simple fusion tree in \sref{sub:transforming_yoga_fusion_trees_to_simple_fusion_trees}. The corresponding transformation for a monster fusion tree also includes index reversals but will not be presented in this guide.

\begin{figure}
	\centering
	\includegraphics[width=.7\textwidth]{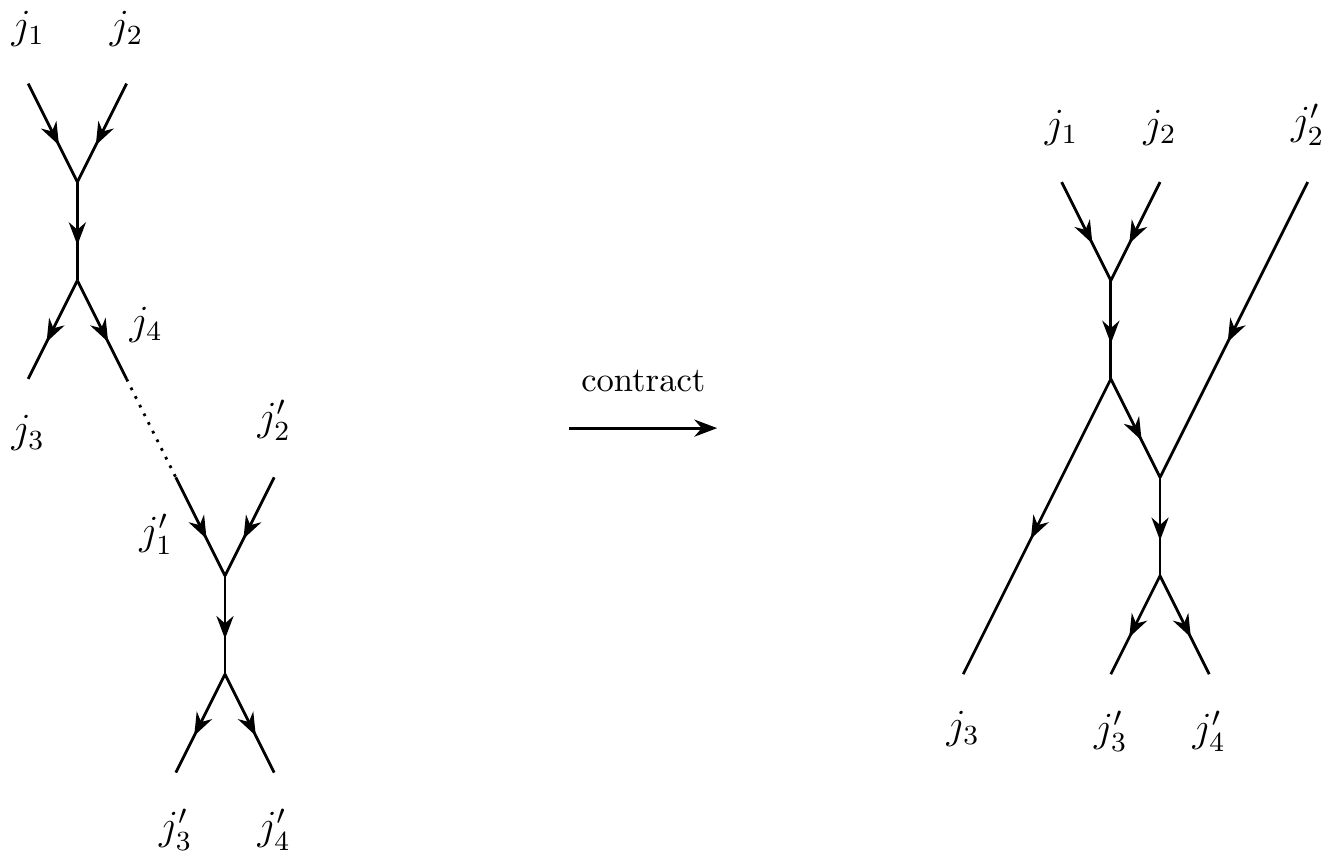}
	\caption{A contraction that leads to a yoga fusion tree.}
	\label{fig:TikZ_Files_Contraction_2a}
\end{figure}

\begin{figure}
	\centering
	\includegraphics[width=.8\textwidth]{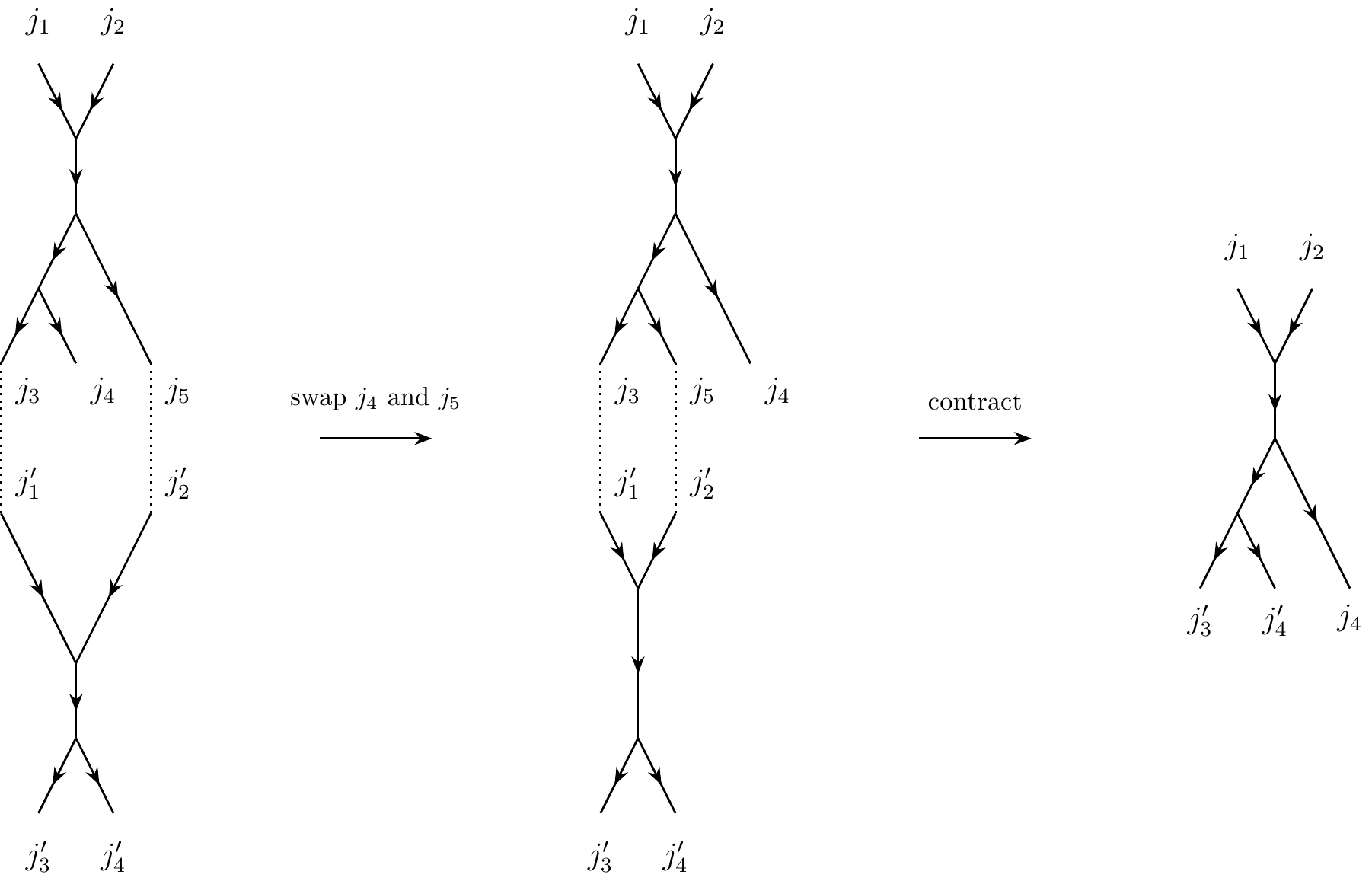}
	\caption{Contractions of fusion trees may involve permutations before contracting.}
	\label{fig:TikZ_Files_Contraction_3}
\end{figure}

\bigskip 

\textbf{\emph{3) Tensor trace.-}} One may also encounter situations where two indices of the same tensor are contracted together, i.e., a partial trace of a tensor is taken. Consider for example the contraction in \fref{fig:TikZ_Files_Contraction_4} for a 5-index tensors. Tensor traces can be implemented by reversing one of the indices, and then contracting with a 2-index identity tensor.

\begin{figure}[ht]
	\centering
	\includegraphics[width=.3\textwidth]{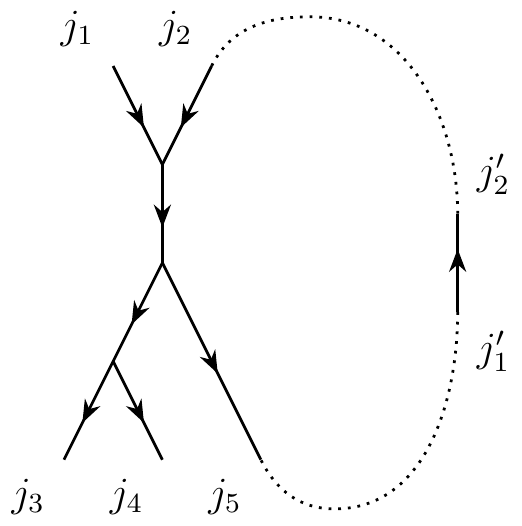}
	\caption{Partial trace as a contraction with the Identity.}
	\label{fig:TikZ_Files_Contraction_4}
\end{figure}

\begin{figure}
	\centering
	\includegraphics[width=.7\textwidth]{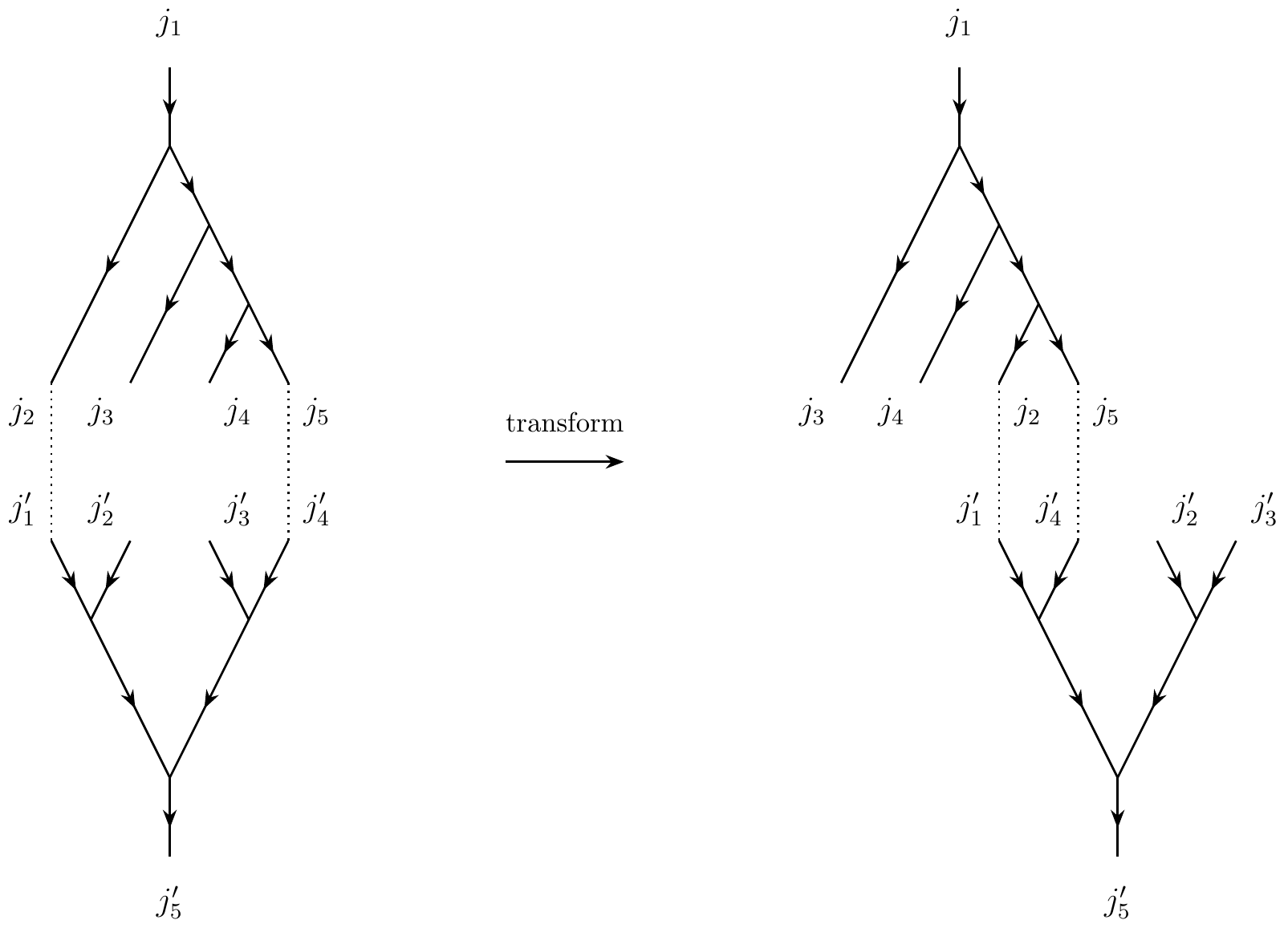}
	\caption{In order to perform the contraction to the left the trees need to be modified in such a way that the loop in the spin network simplifies to the identity.}
	\label{fig:TikZ_Files_Contraction_5}
\end{figure}

\noindent Contraction of two symmetric tensors involves the following three distinct steps:
\begin{enumerate}
\item Merge the fusion trees to obtain the output fusion tree.
\item Build a table $E$ that lists all contributing \textit{pairs} of input charge sectors (corresponding to the two tensors) for each output charge sector.
\item For each output charge sector, the corresponding degeneracy tensor is obtained by contracting together (e.g. using the function \textit{ncon}) and adding all the pairs of input degeneracy tensors that appear in table $E$.
\end{enumerate}

\subsection{Some programming aspects} % (fold)
\label{sub:some_programming_aspects_contractions}

In this section, we discuss some of the programming aspects for contracting two tensors, once they have been prepared as described above. We will focus mostly on how to merge the two input fusion trees, since the remaining two steps of tensor contraction are relatively straightforward to implement.

\begin{figure}
	\centering
	\includegraphics[width=.2\textwidth]{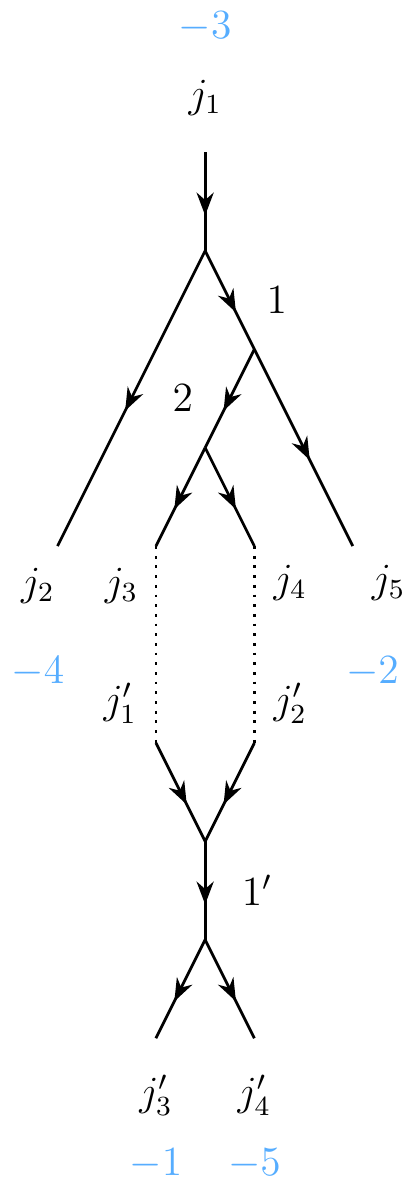}
	\caption{A possible contraction between a 5-index and 4-index symmetric tensor.}
	\label{fig:TikZ_Files_Contraction_6}
\end{figure}

Let us consider a simple example, where we assume the contraction of a 5-index and a 4-index tensor, as shown in \fref{fig:TikZ_Files_Contraction_6}. First of all, we have the following arrays describing the two fusion trees: 
\begin{align}
	\tau_1 &= \lbrace \lbrack -1,-2,1 \rbrack , \lbrack 1,2,-5 \rbrack , \lbrack 2,-3,-4 \rbrack \rbrace,\\
	\tau_2 &= \lbrace \lbrack -1,-2,1 \rbrack , \lbrack 1,-3,-4 \rbrack \rbrace .
\end{align}
Once again we omit the vector of orientations for the nodes because of the visualization of the trees. In order to carry out the contraction, we define an array describing which indices are contracted from each tensor, and which indices are not. For this we follow the index numbering conventions used in the \emph{ncon} MATLAB function~\cite{ncon} for regular tensors: open indices are labeled with negative integers, and contracted indices with positive integers. For this example, we have
\begin{align}
	L = \lbrace \lbrack -3,-4,1,2,-2 \rbrack , \lbrack 1,2,-1,-5 \rbrack \rbrace ,
\end{align}
where the first vector labels the indices of the first tensor, and the second vector those of the second tensor. With this numbering, it is apparent that indices 3 and 4 of the first tensor are contracted with indices 1 and 2 of the second tensor. Note also that the numbering of the open indices indicates the order in which they appear in the resulting tensor (just like in the \textit{ncon} function).

Next, we label the internal indices of both fusion trees with consecutive positive integers, that is, if the first fusion tree has $k$ internal edges and the second one has $k'$ internal edges, we label the internal edges of the first tree with labels in $1$ to $k$, and the internal edges of the second fusion tree with labels from $k+1$ to $k+k'$. With this new labeling, the two fusion trees can be described by the new $\tau$ vectors:
\begin{align}
	\begin{split}
		\tau_1^\prime &= \lbrace \lbrack -1,-2,1 \rbrack , \lbrack 1,2,-5 \rbrack , \lbrack 2,-3,-4 \rbrack \rbrace ,\\
		\tau_2^\prime &= \lbrace \lbrack -1,-2,{\color{TensorGreen}3} \rbrack , \lbrack {\color{TensorGreen}3},-3,-4 \rbrack \rbrace .
	\end{split}
	\label{eq:newconsecutivelabels}
\end{align}
However, for subsequent use we also store a list \textit{internalLegReplacements} of both the old and new labels of each internal index,
\begin{align}
	\text{internalLegReplacements} = \lbrace \lbrace [1,1] , [2,2] \rbrace , \lbrace [1,3] \rbrace \rbrace.
	\label{inter}
\end{align}
Each vector in this list corresponds to an internal index, first entry of the vector is the old label, and the second entry is the new label. This list is required to keep track of and match the charge sectors in the two tensors, as we shall see later. Next, we search array $L$ for the indices in each fusion tree that are contracted, and update the labels in $\tau_1^\prime$ and $\tau_2^\prime$ with the new consecutive labels set in \Eref{eq:newconsecutivelabels}. This gives
\begin{align}
	\tau_1^\prime &= \lbrace \lbrack -1,-2,1 \rbrack , \lbrack 1,2,-5 \rbrack , \lbrack 2,{\color{TensorGreen}4},{\color{TensorGreen}5} \rbrack \rbrace, \\
	\tau_2^\prime &= \lbrace \lbrack {\color{TensorGreen}4},{\color{TensorGreen}5},3 \rbrack , \lbrack 3,-3,-4 \rbrack \rbrace .
\end{align}
Next, we initialize an array \textit{matchIrrepsOnLegs} which stores the pairs of indices that are contracted (to aid in checking for matching conditions later). We also initialize a second list \textit{matchingLegIsOpen} that keeps track of which of these indices correspond to open indices in the output fusion tree (using $0$ to label the contracted indices, and $1$ to label the open indices). In our case, these two lists are
\beqa
		\text{matchIrrepsOnLegs} &=& \lbrace [-3,-1] , [-4,-2] \rbrace, \\
		\text{matchingLegIsOpen} &=& \lbrace [0,0] , [0,0] \rbrace .
\eeqa
We store these two lists for later use. The next step is to build the output fusion tree by merging the two input fusion trees $\tau_1^\prime$ and $\tau_2^\prime$. To this end, first we update the labels of the open indices in $\tau_1^\prime$ and $\tau_2^\prime$ with those specified in the array $L$, which gives
\begin{align}
	\tau_1^{\prime\prime} &= \lbrace \lbrack {\color{TensorGreen}-3},{\color{TensorGreen}-4},1 \rbrack , \lbrack 1,2,{\color{TensorGreen}-2} \rbrack , \lbrack 2,4,5 \rbrack \rbrace,\\
	\tau_2^{\prime\prime} &= \lbrace \lbrack 4,5,3 \rbrack , \lbrack 3,{\color{TensorGreen}-1},{\color{TensorGreen}-5} \rbrack \rbrace .
\end{align}
We also store an array \textit{openLegReplacements} specifying which open indices were updated in the previous step,
\begin{align}
	\text{openLegReplacements} = \lbrace \lbrace [-1,-3] , [-2,-4] , [-5,-2] \rbrace , \lbrace [-3,-1] , [-4,-5] \rbrace \rbrace .
\end{align}
which follows the same convention as \Eref{inter}. As is apparent, the resulting tree contains a closed loop, which should be removed. In order to remove loops, we scan the tree for nodes that have two internal indices on positions $[1,2]$ (fusion node) or positions $[2,3]$ (splitting node). If we find two such nodes whose labels for these internal indices match, then we have found a loop. The only thing left to do is to delete the two nodes from the tree and set the labels of the remaining index of each of these two nodes to the same value, (shown in green below)
\begin{align}
	\tau &= \lbrace [-3,-4,1] , [1,2,-2] , [{\color{TensorGreen}2},4,5] , [4,5,{\color{TensorGreen}3}] , [3,-1,-5] \rbrace , 
\end{align}
where we have chosen the smaller index label for the reset. We obtain
\begin{align}
	\tau^\prime &= \lbrace [-3,-4,1] , [1,{\color{TensorGreen}2},-2] , [{\color{TensorGreen}2},-1,-5] \rbrace .
	\label{eq:ContractionFinalFusionTree}
\end{align}
When a loop is found, we also update the lists \textit{matchIrrepsOnLegs} and \textit{matchingLegIsPhysical} as
\beqa
		\text{matchIrrepsOnLegs} &=& \lbrace [-3,-1] , [-4,-2] , [2,3] \rbrace,\\
		\text{matchingLegIsOpen} &=& \lbrace [0,0] , [0,0] , [0,0] \rbrace. 
\eeqa
After all loops have been removed (in our case we just had the one loop), we have obtained the output fusion tree. Finally, we use the list \textit{internalLegReplacements} to restore the internal indices to their original value from the list \textit{matchIrrepsOnLegs}. We have
\begin{align}
	\text{matchIrrepsOnLegs} = \lbrace [-3,-1] , [-4,-2] , [2,{\color{TensorGreen}1}] \rbrace\ ,
\end{align}
where each element specifies the respective index of the first and the second fusion trees for which the irreps have to match. Since the list \textit{matchingLegIsOpen} is full of zeros, it means that all entries in \textit{matchIrrepsOnLegs} are internal indices of the new fusion tree. The last step of the contraction is to ensure that the irreps of the indices in \textit{matchIrrepsOnLegs} actually match. Notice that, by doing it in this way, we are sure that this list includes \emph{also} those internal indices that end up in the output tree after the simplification of a loop.

\begin{figure}
	\centering
	\includegraphics[width=.2\textwidth]{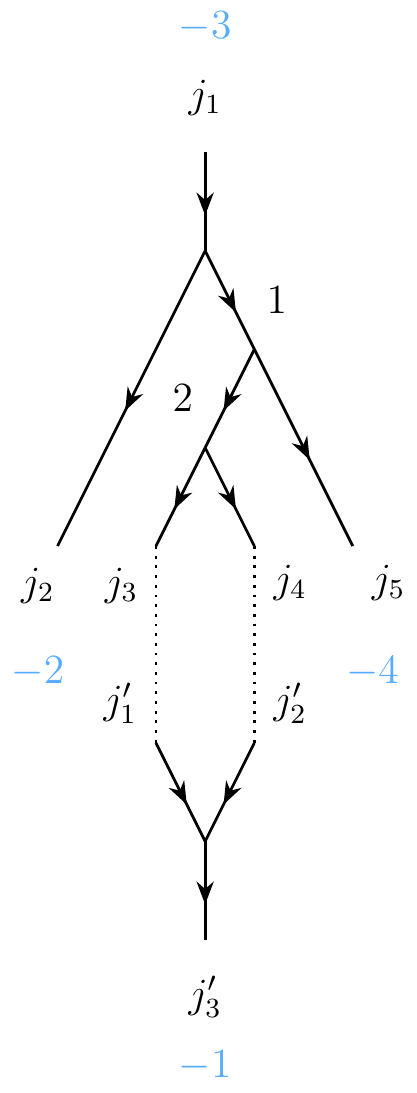}
	\caption{A possible contraction between a 5-index and 3-index symmetric tensor.}
	\label{fig:TikZ_Files_Contraction_7}
\end{figure}

In the case discussed above we only need the list \textit{internalLegReplacements} at the last step, and not the list \textit{openLegReplacements}. However, there are situations where this last list is also needed. Take, for example, the contraction of a 5-index and a 3-index tensor as shown in \fref{fig:TikZ_Files_Contraction_7}. The two fusion trees are
\begin{align}
	\tau_1 &= \lbrace [-1,-2,1] , [1,2,-5] , [2,-3,-4] \rbrace,\\
	\tau_2 &= \lbrace [-1,-2,-3] \rbrace, \
\end{align}
and the final order is chosen arbitrarily. After merging the two trees and removing the loop, we obtain the lists
\beqa
		\text{physicalLegReplacements}& =& \lbrace \lbrace [-1,-3] , [-2,-2] , [-5,-4] \rbrace , \lbrace [-3,-1] \rbrace \rbrace,\\
		\text{matchIrrepsOnLegs} &=& \lbrace [-3,-1] , [-4,-2] , [2,-1] \rbrace,\\
		\text{matchingLegIsOpen} &=& \lbrace [0,0] , [0,0] , [0,1] \rbrace .
\eeqa
Thus, we have a situation where the irreps of an open index in the second fusion tree need to be checked. However, $-1$ is a label of the merged fusion tree and needs to be reverse-updated. The final list is therefore given by
\begin{align}
	\text{matchIrrepsOnLegs} = \lbrace [-3,-1] , [-4,-2] , [2,{\color{TensorGreen}-3}] \rbrace .
\end{align}
For completeness, notice that there are also cases in which \textit{matchingLegIsOpen} shows an entry $[1,1]$, e.g., when contracting a splitting and a fusion node. Such cases can be dealt with similarly. 

\begin{figure}[ht]
	\centering
	\includegraphics[width=.075\textwidth]{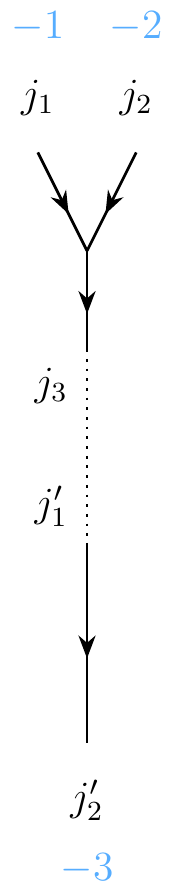}
	\caption{A possible contraction between a 3-index and 2-index symmetric tensor.}
	\label{fig:TikZ_Files_Contraction_8}
\end{figure}

As a last example, we consider the contraction of a tensor with a matrix, since here the fusion trees contain a ``dummy index'' on one of the nodes. The same procedure applies to contractions which involve tensors whose indices all have the same direction. Consider the contraction of a 3-index and a 2-index tensor according to \fref{fig:TikZ_Files_Contraction_8}. Merging the two updated fusion trees yields a valid tree decomposition of the resulting 3-index tensor described by
\begin{align}
	\tau = \lbrace [-1,-2,{\color{TensorGreen}1}] , [{\color{TensorGreen}1},-3,0] \rbrace.
\end{align}
However, since one of the nodes contains a ``dummy index'' (labeled by $0$), the tree can be further simplified. To do this, we scan for nodes with some index label equal to $0$. If such a node is found, then we consider the only internal index in this node and find the other node that is connected by this index. We do this to replace the internal index in the second node by the open index of the first node. Next, we delete the first node. The resulting tree is 
\begin{align}
	\tau' = \lbrace [-1,-2,-3] \rbrace, 
\end{align}
as expected from the figure. 

\begin{figure}
	\centering
	\includegraphics[width=.55\textwidth]{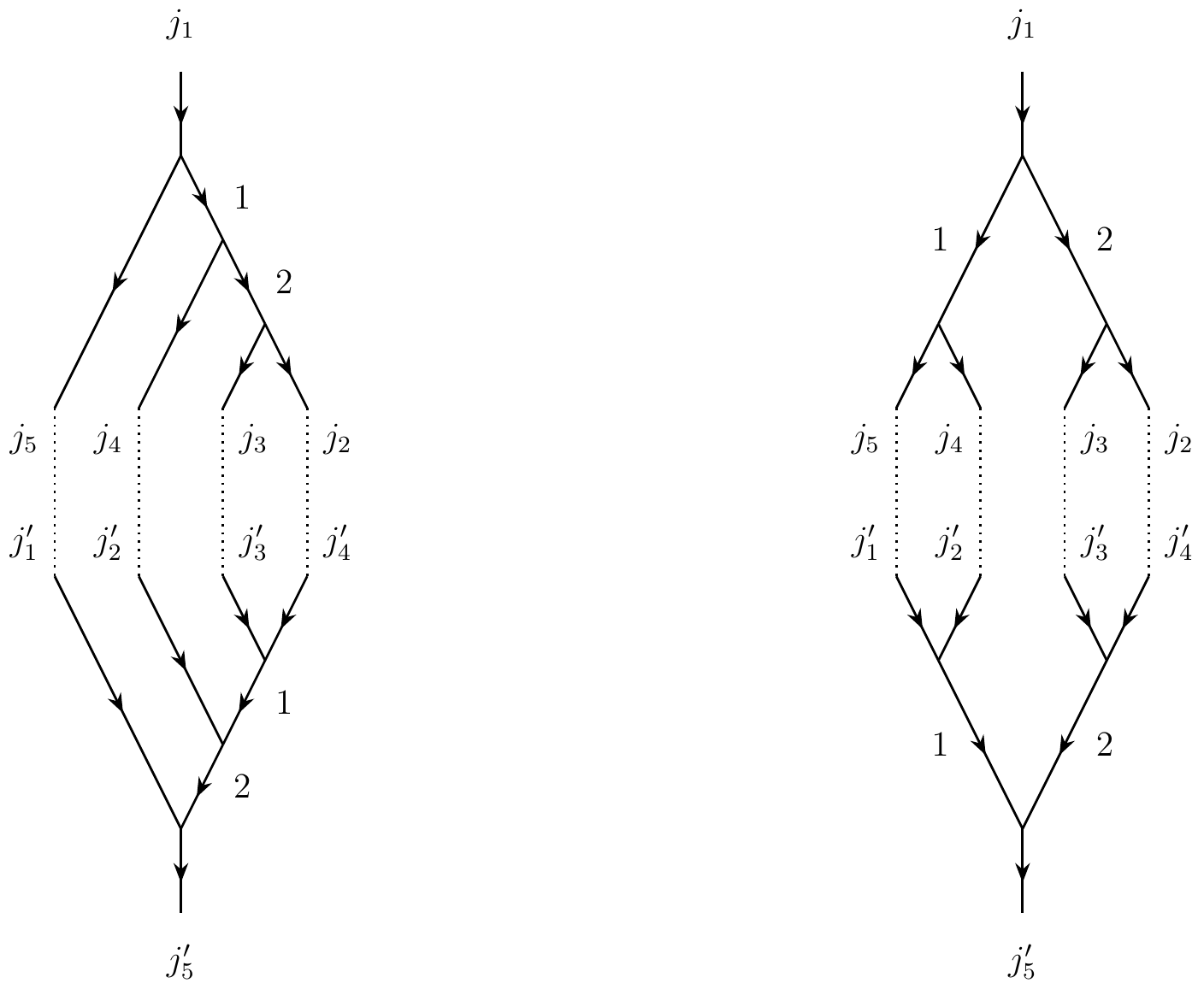}
	\caption{Contraction of two fusion trees involving two nested loops (left) and two parallel loops (right).}
	\label{fig:TikZ_Files_Contraction_9}
\end{figure}

\begin{figure}[ht]
	\centering
	\includegraphics[width=.2\textwidth]{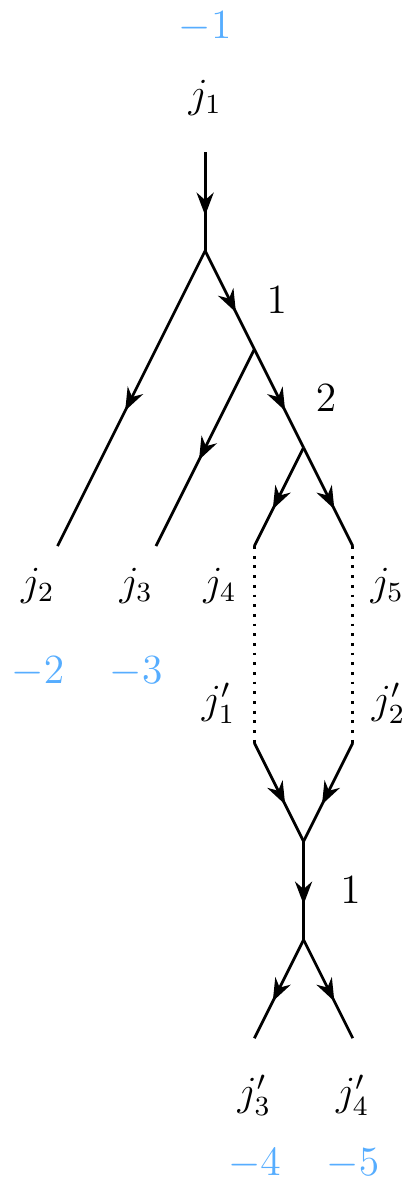}
	\caption{A possible contraction between a 5-index and 4-index symmetric tensor.}
	\label{fig:TikZ_Files_Contraction_10}
\end{figure}

The algorithm described above can be easily applied to merged fusion trees with multiple loops, which can be both nested and parallel, see \fref{fig:TikZ_Files_Contraction_9}). Furthermore, it also keeps track of which index of the input fusion trees corresponds to which indices of the output fusion tree. This is necessary in order to compare the charge sectors of the tensors before and after contraction. Consider the contraction shown in \fref{fig:TikZ_Files_Contraction_10}. For the input tensors the charge sectors are denoted by
\beqa
		\text{chargeSectors}_1 &=& [j_{\text{int},1},j_{\text{int},2},j_1,j_2,j_3,j_4,j_5],	\\
		\text{chargeSectors}_2 &=& [j_{\text{int},1}^{\prime},j_1^{\prime},j_2^{\prime},j_3^{\prime},j_4^{\prime}]. 
\eeqa
All matching configurations in the contraction need to fulfill $j_4 = j_1^\prime$ and $j_5 = j_2^\prime$, but also $j_{\text{int},2} = j_{\text{int},1}^{\prime}$ due to the loop in the fusion tree. The remaining entries of the two input lists of charge sectors are then merged into a single one, which will be compared to the list of charge sectors for the contracted tensor. During this process, we also must ensure that the degeneracies of the irreps also match. 

As an example of matching the degeneracies of irreps, consider again the contraction shown in \fref{fig:TikZ_Files_Contraction_10} with irreps $j = [0,1,2]$ and degeneracies $t = [3,4,5]$ for each index. A possible degeneracy contraction is one for the charge sectors
\beqa
			C_1 &=& [1,{\color{yellow}2},1,0,1,{\color{blue}0},{\color{TensorGreen}2}], \\
			C_2 &=& [{\color{yellow}2},{\color{blue}0},{\color{TensorGreen}2},1,1] ,
\eeqa
where quantum numbers in the same (non-black) color must match. The corresponding degeneracy tensors for these blocks have dimensions
\beqa
		\text{dim}(\text{degeneracyTensor}(C_1)) &=& [4,3,4,{\color{blue}3},{\color{TensorGreen}5}], \\
		\text{dim}(\text{degeneracyTensor}(C_2)) &=& [{\color{blue}3},{\color{TensorGreen}5},4,4].
\eeqa
where again, the colored labels must match. Notice that this ensures that the indices being contracted in the degeneracy tensors are of the same size. The resulting block in the new tensor will then have the charge sector
\begin{align}
	C_\text{final} = [1,2,1,0,1,1,1]
\end{align}
following the labeling convention given in the figure. The two corresponding degeneracy tensors will be contracted (e.g. using the function \textit{ncon}) to yield the output degeneracy tensor with dimensions $\text{dim}(\text{degeneracyTensor}(C_\text{final})) = [4,3,4,4,4]$.

% subsection some_programming_aspects_contractions (end)

\subsection{Transforming yoga fusion trees to simple fusion trees} % (fold)
\label{sub:transforming_yoga_fusion_trees_to_simple_fusion_trees}

Following the contraction procedure described in the previous section, some tensor contractions may result in yoga fusion trees. A typical example is the contraction of two matrix product operators (MPOs), which results in yoga fusion trees of the type illustrated in \fref{fig:TikZ_Files_Resolve_ProblematicStructure_3}.
\begin{figure}[ht]
	\centering
	\includegraphics[width=.6\textwidth]{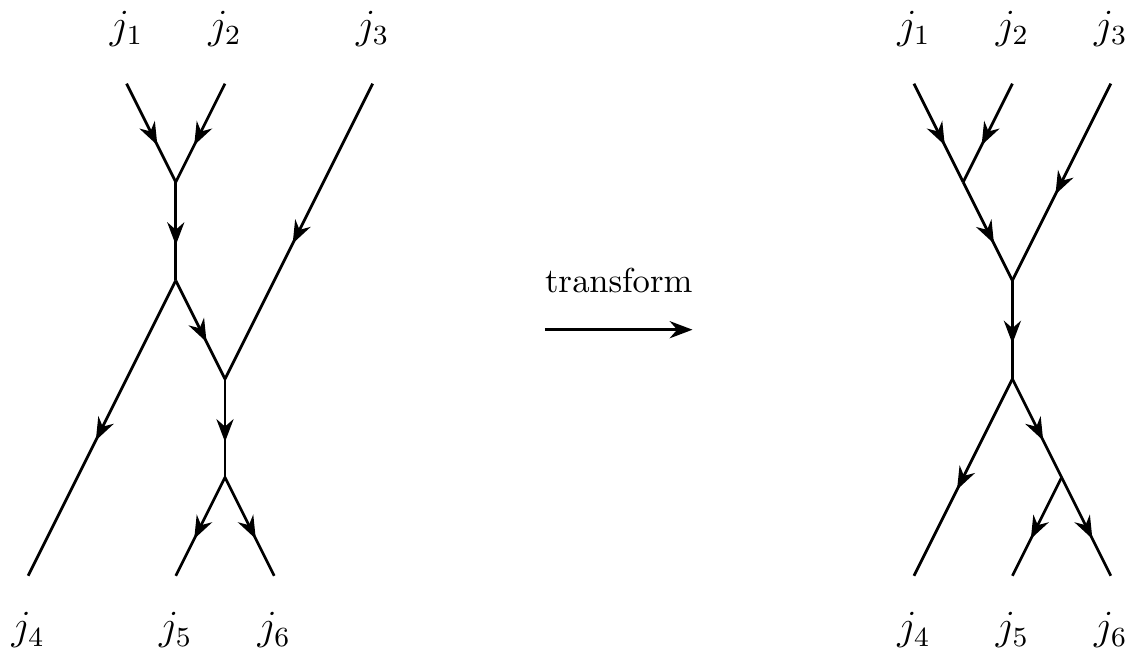}
	\caption{A yoga fusion tree (left) that can result after the contraction of two matrix product operators, but can be transformed to a simple fusion tree.}
	\label{fig:TikZ_Files_Resolve_ProblematicStructure_3}
\end{figure}

Any yoga tree contains one or more elementary yoga subtrees that are shown in \fref{fig:TikZ_Files_Index_Reversal_2}. Fortunately, these elementary yoga trees can be transformed into simple trees, which we prefer in our implementation, by inserting resolutions of the identity (comprised of fusion and splitting) and applying an $F$-move. These transformations are shown in \fref{fig:TikZ_Files_Resolve_ProblematicStructure_1} and \fref{fig:TikZ_Files_Resolve_ProblematicStructure_2}. The degeneracy tensors have to be updated by taking linear combinations of the input degeneracy tensors with some weights that result from transforming the corresponding fusion trees.

However, these weights are not the ones that are depicted in the \fref{fig:TikZ_Files_Resolve_ProblematicStructure_1} and \fref{fig:TikZ_Files_Resolve_ProblematicStructure_2}. The transformations depicted in these figures express each \textit{input} degeneracy tensor (in any given charge sector) as a linear combination of the \textit{updated} degeneracy tensors. Thus, to update the degeneracy tensors we have to invert this transformation.

\begin{figure}
	\centering
	\includegraphics[width=.5\textwidth]{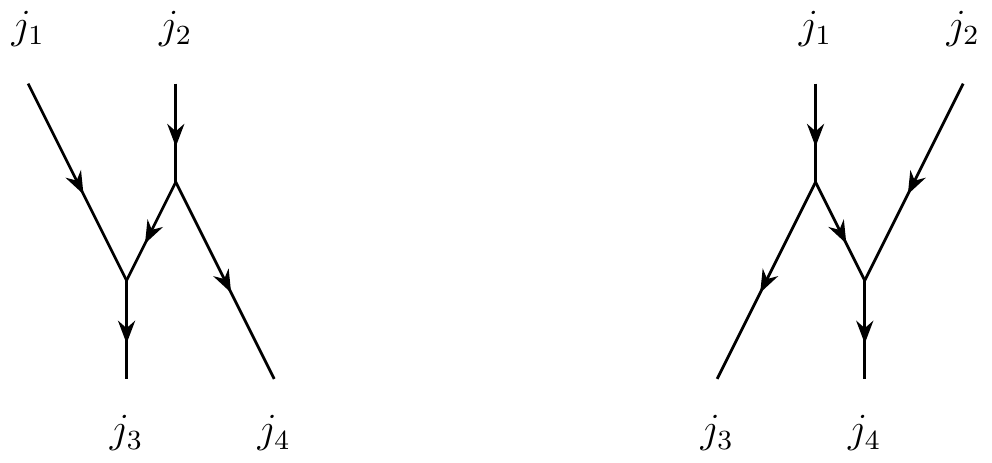}
	\caption{Elementary yoga trees.}
	\label{fig:TikZ_Files_Index_Reversal_2}
\end{figure}
\begin{figure}
	\centering
	\includegraphics[width=.7\textwidth]{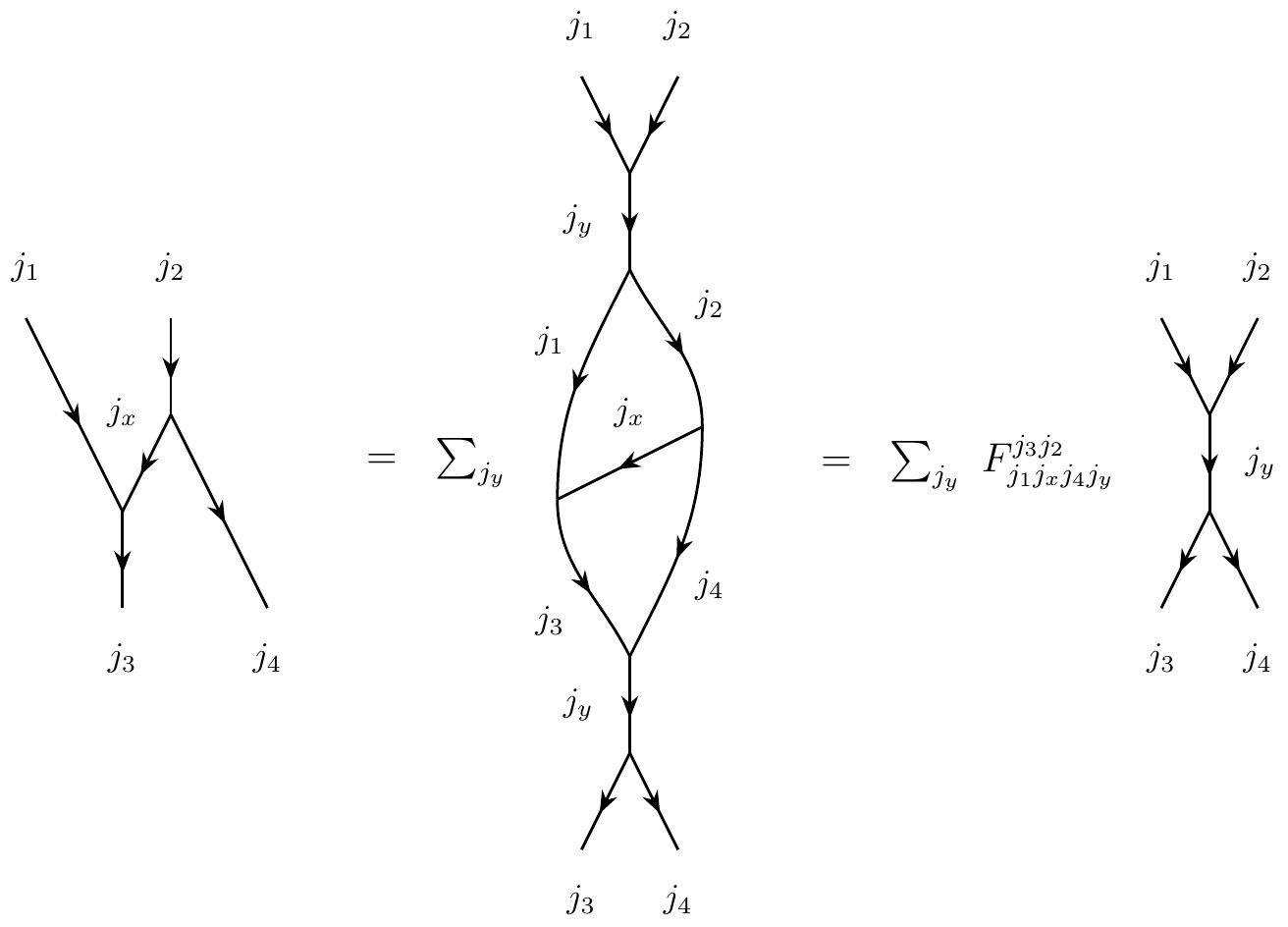}
	\caption{Transforming an elementary yoga tree to a simple fusion tree involves introducing two resolutions of identity by means of fusion and splitting tensors (the first equality) and then applying an $F$-move (the second equality).}
	\label{fig:TikZ_Files_Resolve_ProblematicStructure_1}
\end{figure}
\begin{figure}
	\centering
	\includegraphics[width=.7\textwidth]{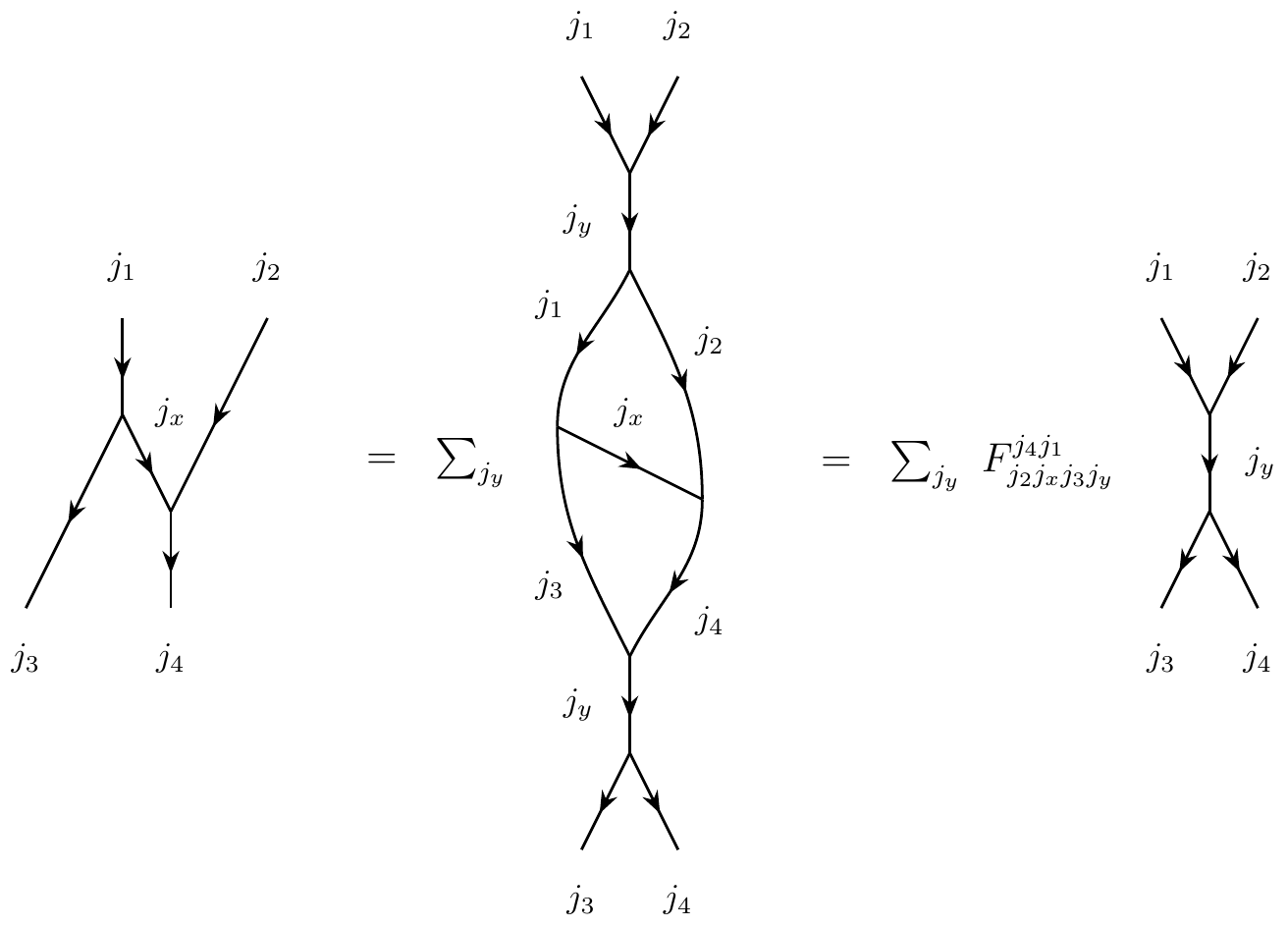}
	\caption{The other elementary yoga tree can also be transformed to a simple fusion tree similar to \fref{fig:TikZ_Files_Resolve_ProblematicStructure_1}.}
	\label{fig:TikZ_Files_Resolve_ProblematicStructure_2}
\end{figure}

To this end, for every charge sector of the yoga fusion tree we first find the corresponding charge sectors of the resulting simple fusion tree that appear in the sums depicted in \fref{fig:TikZ_Files_Resolve_ProblematicStructure_1} and \fref{fig:TikZ_Files_Resolve_ProblematicStructure_2}. For each such pair of charge sectors we determine the corresponding numerical factor (given by the $F$-move) and store it in a matrix $T$, whose rows label the charge sectors for the yoga fusion tree and the columns label the charge sectors for the simple fusion tree:
\begin{align}
	T = 
	\begin{blockarray}{c c c c c c c c c c c c}
		& & \multicolumn{10}{c}{\text{new chargeSectors}}	\\
		& & 1 & 2 & 3 & 4 & 5 & 6 & 7 & 8 & 9 & \hdots\\
		\begin{block}{c c (c c c c c c c c c c)}
			\multirow{5}{*}{\rotatebox[origin=c]{90}{old chargeSectors}}
  			& 1 & F_{11} & 0 & 0 & 0 & 0 & 0 & 0 & 0 & 0 & \\
  			& 2 & F_{21} & 0 & 0 & F_{24} & 0 & 0 & 0 & F_{28} & 0 & \hdots\\
  			& 3 & 0 & F_{32} & 0 & 0 & 0 & F_{36} & 0 & 0 & 0 & \\
  			& 4 & 0 & 0 & F_{43} & 0 & 0 & F_{46} & F_{47} & 0 & 0 & \hdots\\
  			& 5 & 0 & 0 & 0 & 0 & F_{55} & 0 & 0 & 0 & 0 & \\
  			& \vdots & & & \vdots & & \vdots & & \vdots & & & \ddots\\
		\end{block}
	\end{blockarray}
\end{align}

The matrix $T$ corresponds to the transformations shown in the figures, and encodes the basis change from the yoga to the simple fusion tree. But here we want the basis change from the simple fusion tree to the yoga fusion tree (so that we can take linear combinations of the yoga degeneracy tensors with appropriate weights to determine the updated degeneracy tensors). This is given by the inverse transformation $T^\dagger$. The inverse transformations could have been constructed directly by e.g. starting with the simple fusion tree, reversing indices 2 and 3, introducing resolutions of identity (by means of fusing indices 1 and 3 (and 2 and 4), and splitting the fused index back to indices 1 and 3 (2 and 4), and finally reversing back indices 2 and 3. The factors resulting from this transformation are (and must be) the same as those in the matrix $T^\dagger$.

% subsection transforming_yoga_fusion_trees_to_simple_fusion_trees (end)

% section tensor_contraction (end)

\section{{Symmetric Singular Value Decomposition}} % (fold)
\label{sec:symmetric_singular_value_decomposition}

Any rectangular matrix $M \in \mathbb C^{m \times n}$ of arbitrary size can be decomposed into
\begin{align}\label{eq:svd}
	M = U \Lambda V^\dagger\ ,
\end{align}
where $U \in \mathbb C^{m \times m}$ and $V^\dagger \in \mathbb C^{n \times n}$ are unitary matrices and $\Lambda \in \mathbb R^{m \times n}$ is a matrix whose only non-zero components are along the diagonal. The diagonal entries are usually arranged in descending order $\lambda_1 \ge \lambda_2 \ge \hdots \ge \lambda_r$, where $r = \text{min}(m,n)$, and are called the singular values of $M$. The decomposition \Eref{eq:svd} is called the singular value decomposition (SVD) of $M$. This decomposition often appears in tensor network algorithms as a means to discard irrelevant degrees of freedom in the TN~\cite{simple}. It is also closely related to the Schmidt decomposition of a bipartite quantum state.

The singular values $\Lambda$ are unique, whereas the matrices $U$ and $V^\dagger$ are generally not due to possible degeneracies in the $\lambda$'s. The so called ``economic version'' of SVD is obtained by discarding all the zero (or very small) singular values in $\Lambda$ and also the columns and rows corresponding to the zero singular values in matrices $U$ and $V^\dagger$ respectively. The economic SVD is depicted in \fref{fig:TikZ_Files_SVD_Truncation_SVD_BlockStructure_Econ}.

\begin{figure}
	\centering
	\includegraphics[width=.8\textwidth]{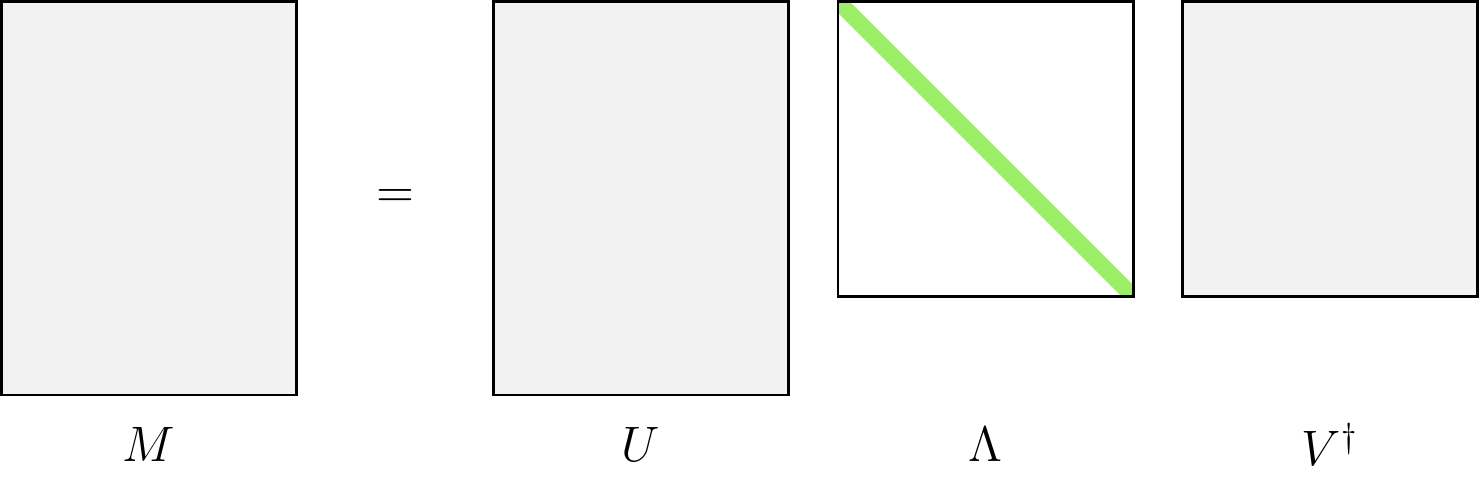}
	\caption[Graphical Representation of the economic \acs{SVD}]{Graphical representation of the economic SVD of an $m \times n$ matrix.}
	\label{fig:TikZ_Files_SVD_Truncation_SVD_BlockStructure_Econ}
\end{figure}

Let us now consider the SVD of a symmetric matrix $M$ -- a symmetric tensor with one incoming and one outgoing index. In the irrep basis, $M$ decomposes as
\begin{align}\label{eq:symMat}
M = \bigoplus_j (M_j \otimes I_j).
\end{align}
Here only a single irrep $j$ appears, since fusion rules require that components of $M$ vanish when different irreps are fixed on the two indices. That is, the matrix $M$ is block diagonal in the irrep labels, see \fref{fig:TikZ_Files_SVD_Truncation_Tensor_BlockStructure}. Each block shown in the figure factorizes as $M_j \otimes I_j$, where $M_j$ the degeneracy matrix corresponding to fixing irrep $j$ on both the indices and $I_j$ is the $2j+1 \times 2j+1$ identity matrix, which acts on the irrep space of $j$. Notice that the blocks (and hence the matrix) do not need to be square. The SVD of a symmetric matrix $M$ can be carried out block-wise, by performing SVD of only the degeneracy matrices $M_j$ for each block. See \fref{fig:TikZ_Files_SVD_Truncation_Tensor_BlockStructure_SVD}.

\begin{figure}
	\centering
	\includegraphics[scale = 0.6]{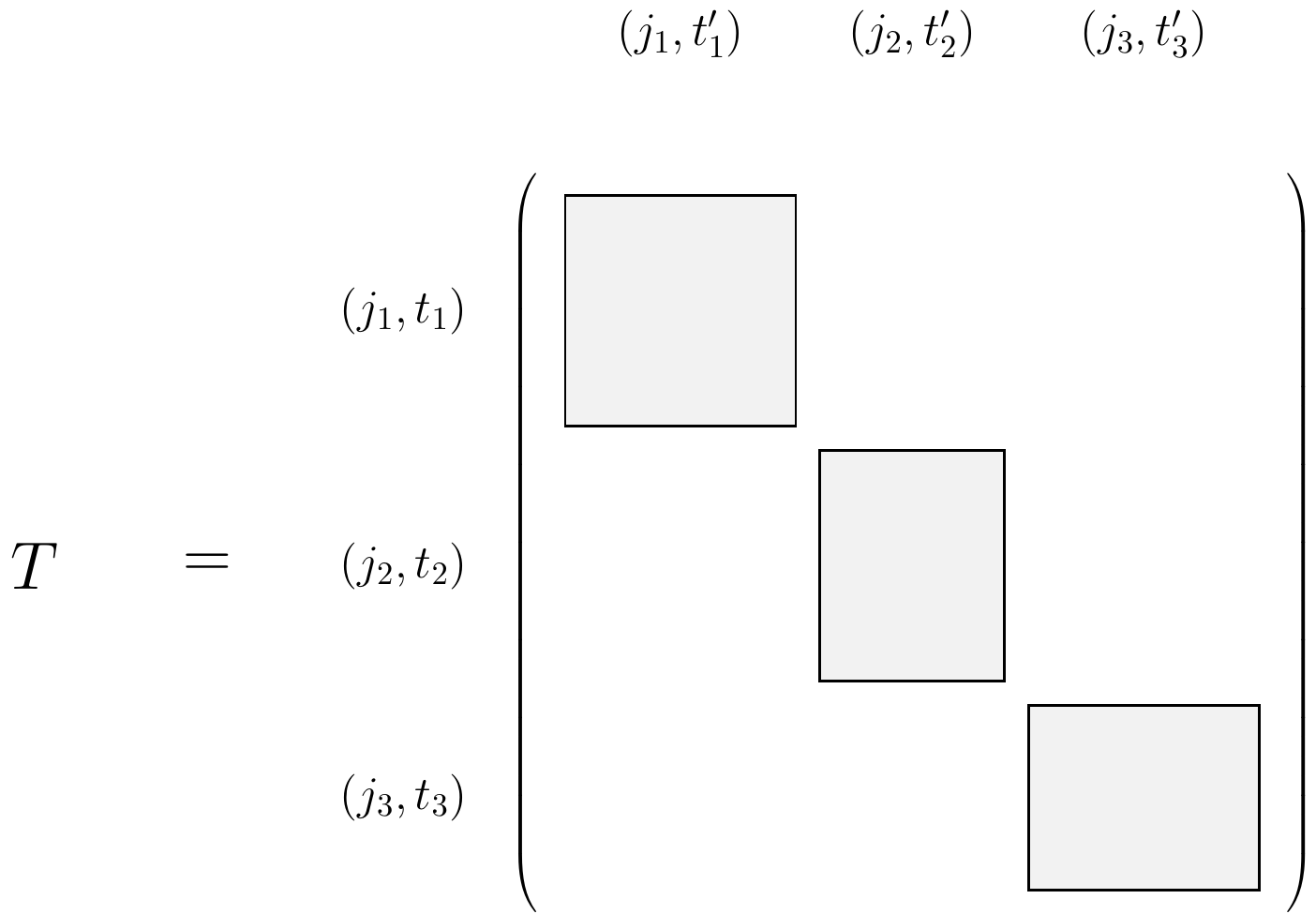}
	\caption{A symmetric matrix is block diagonal in the irrep ($j$) labels. The dimension of each degeneracy blocks is $t_i \times t_i^\prime$.}
	\label{fig:TikZ_Files_SVD_Truncation_Tensor_BlockStructure}
\end{figure}

\begin{figure}
	\centering
	\includegraphics[scale = 0.58]{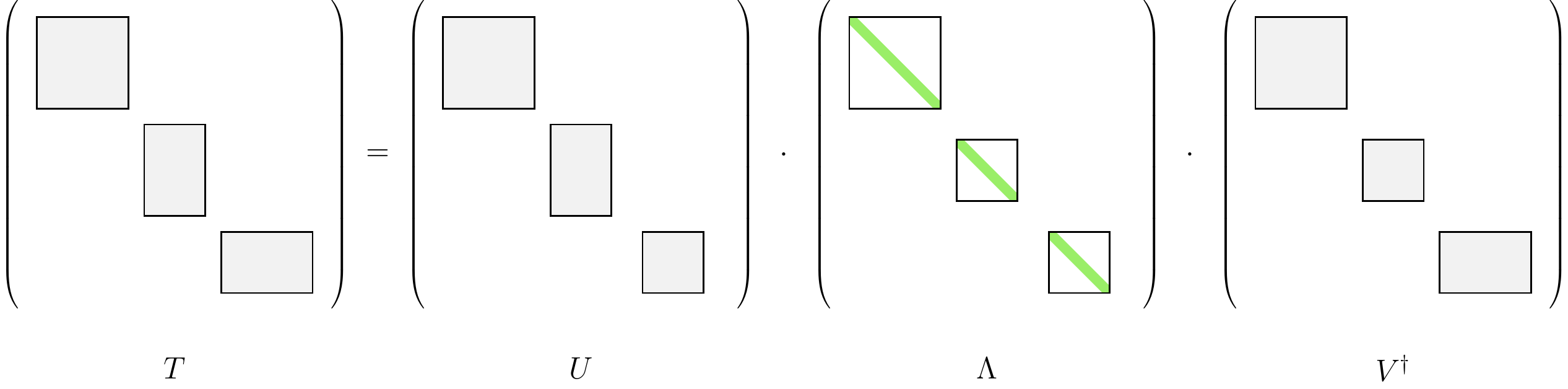}
	\caption{SVD of a symmetric matrix is performed block-wise. The green line indicates the singular values in the diagonal matrices $\Lambda$.}
	\label{fig:TikZ_Files_SVD_Truncation_Tensor_BlockStructure_SVD}
\end{figure}

\subsection{SVD based truncation} % (fold)
\label{sub:svd_based_truncation}

As mentioned previously, SVD often appears in tensor algorithms in the context of approximating one matrix by another one (with a smaller number of singular values). One simply discards singular values that are smaller than a given error $\epsilon$ and truncates the matrices $U$ and $V^\dagger$ accordingly. Such a truncation is depicted in \fref{fig:TikZ_Files_SVD_Truncation_SVD_Truncation_1} and \fref{fig:TikZ_Files_SVD_Truncation_SVD_Truncation_2}. The matrix $M' = U^{[\rm{trunc}]}\Lambda^{[\rm{trunc}]}(V^\dagger)^{[\rm{trunc}]}$ is a good approximation to the input matrix $M$ if $\epsilon$ is small. The accuracy of the approximation is given, e.g.,  by the normalized sum of the discarded squared singular values.

\begin{figure}
	\centering
	\includegraphics[scale = 0.8]{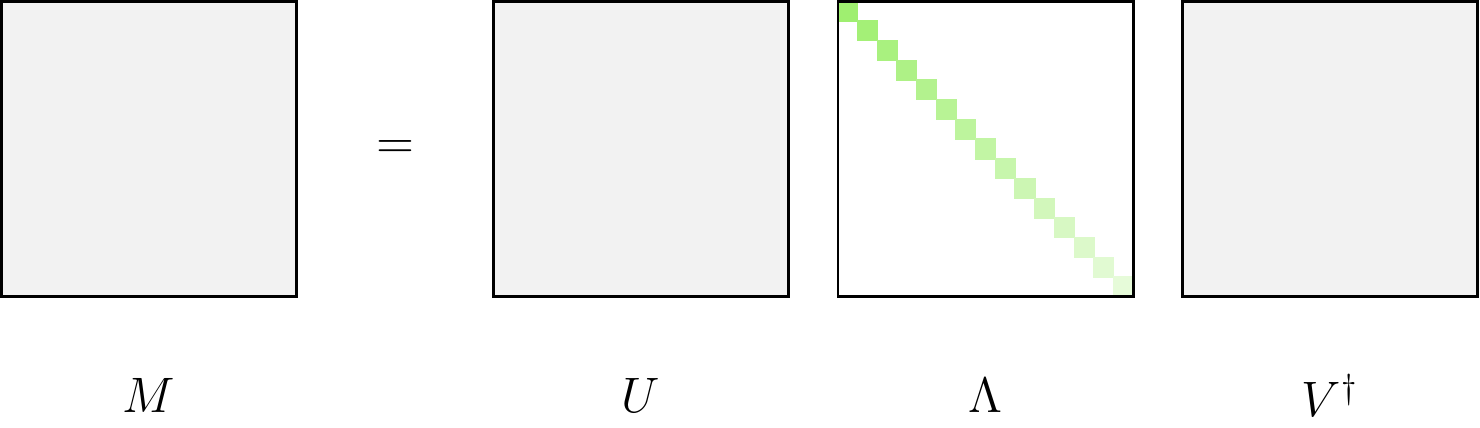}
	\caption{Singular value decomposition of a matrix, where the singular values are ordered from the largest at the top to the smallest at the bottom. (The magnitude of the singular values is indicated by the intensity gradient of green.)}
	\label{fig:TikZ_Files_SVD_Truncation_SVD_Truncation_1}
\end{figure}

\begin{figure}
	\centering
	\includegraphics[scale = 0.8]{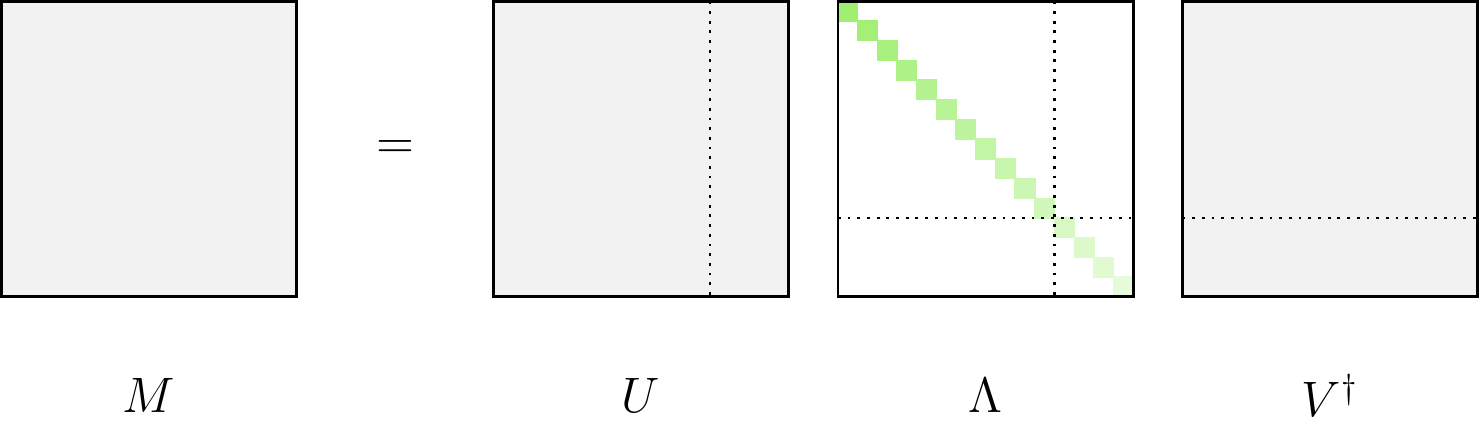}
	\caption{Truncation of the matrices $U$, $\Lambda$ and $V^\dagger$ by retaining only the largest singular values. The rows and columns appearing on the bottom and right of the dotted lines respectively, are deleted.}
	\label{fig:TikZ_Files_SVD_Truncation_SVD_Truncation_2}
\end{figure}

A symmetric truncation is performed by performing blockwise SVD and directly truncating the degeneracy matrices. However, in this case, one must remember that the goal of the truncation is to minimize the trace loss of the total matrix $M$. {We will describe how one must truncate each degeneracy tensor to achieve this. Let $\lambda_{j,t}$ denote the singular values of $M$ in the various blocks labeled by irrep $j$. According to the irrep decomposition \Eref{eq:symMat}, each singular value $\lambda_{j,t}$ appears $2j+1$ times in the singular value spectrum of the total matrix $M$. The trace of $M^\dagger M$ is therefore
\begin{align}
\text{trace}(M^\dagger M) = \sum_{j,t} (2j+1)\lambda^2_{j,t}.
\end{align}
The approximation then proceeds by truncating in the $\lambda$'s. In order to discard a fixed number of singular values, the weight $2j+1$ of each singular value must be taken in account. For example, a small singular value with a large weight may contribute more to the trace than a larger singular value with a smaller weight. Therefore, an overall truncation may require truncating the different blocks by a different amount; there might be blocks for which there is no truncation at all, and blocks for which all the singular values are truncated. In the latter case, it is useful to delete the corresponding irrep from the index data of the matrices $U,\Lambda$ and $V^\dagger$.

As a general comment, let us remark that the discussion of this section also applies to other matrix factorizations, such as eigenvalue decomposition, polar decomposition, and others.

\begin{figure}
 	\centering
	\includegraphics[scale = 0.58]{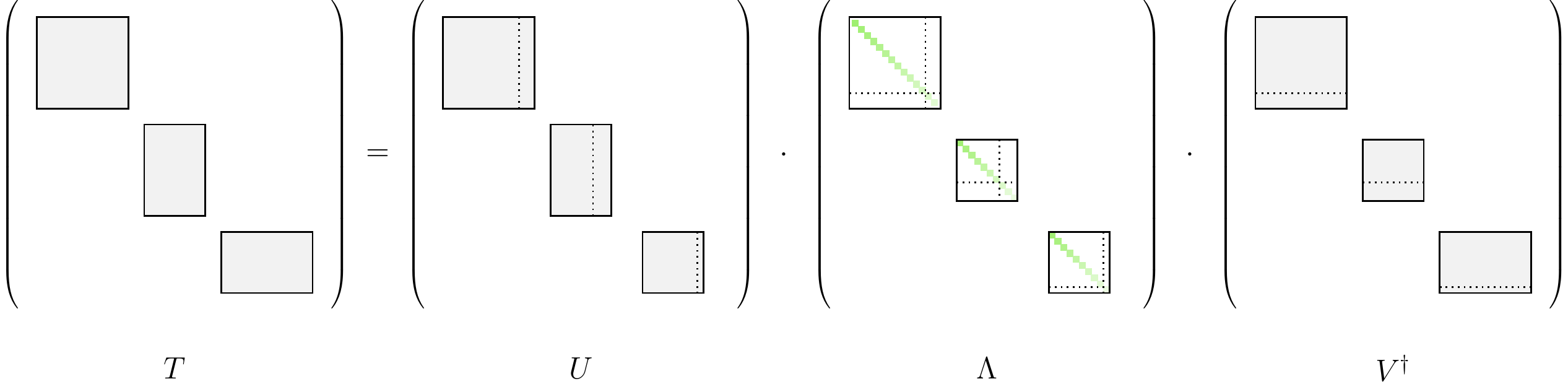}
	\caption{{Truncation of a symmetric matrix by keeping only the largest singular values involves carefully selecting singular values in each degeneracy block.}}
	\label{fig:TikZ_Files_SVD_Truncation_Tensor_BlockStructure_SVD_Truncation_1}
\end{figure}

% subsection svd_based_truncation (end)

% section symmetric_singular_value_decomposition (end)

\section{Generalization and efficiency} % (fold)
\label{sec:generalizations_of_fusion_trees_efficiency_and_prospect}

Our implementation has been restricted to simple fusion trees. This was mostly because simple fusion trees are easier to store and manipulate. Furthermore, most tensor operations can be performed within the domain of simple fusion trees (with a few caveats e.g. as discussed in \sref{sub:transforming_yoga_fusion_trees_to_simple_fusion_trees}). On the other hand, as is perhaps apparent, maintaining a tensor in a simple fusion tree basis (as opposed to a more general fusion tree basis) may require additional manipulations, and generally more restrictive data structures. It is possible to devise an implementation of symmetric tensors that is based more general fusion trees, e.g. which includes yoga fusion trees and even monster fusion trees. We expect that working with more general fusion trees will increase the complexity of the code, but at the same time may perhaps also result in a more efficient implementation. We leave this as a prospect for future implementations of symmetric tensors.

Another avenue for optimizing efficiency is by employing \textit{precomputation}~\cite{sukhi}, which can be integrated in any implementation scheme based on fusion trees. Many tensor network algorithms are iterative. For example, in a variational optimization that approximates the ground state of a given Hamiltonian within a given tensor network ansatz, the tensor network is iteratively updated by sweeping through its tensors, and a {predetermined} sequence of tensor operations is carried out in each iteration. In the context of symmetric tensor network algorithms, the irreps that appear on all the indices of the tensor network can be (and usually are) fixed during such an optimization. Thus, only the degeneracy tensors (and possibly their size) get updated in each iteration.

Recall the general template of a symmetric tensor operation outlined in \sref{sec:a_general_template_for_symmetric_tensor_operations}. Each operation can be regarded as consisting of two logically distinct set of manipulations: ``book-keeping'' manipulations that involve mostly fusion tree manipulations and building the table $E$, and the ``actual'' update that generates the output degeneracy tensors by manipulating the input degeneracy tensors. In an iterative algorithm, the same book-keeping manipulations are unnecessarily repeated over and over again, since these depend only on the fusion trees and irreps (and not on the degeneracy tensors). In practice, we find that the computational cost of doing many book-keeping manipulations (e.g. when dealing with large complicated tensor networks) can become significant. In such a case, one can \textit{precompute} and store the results of all the book-keeping manipulations, say in the first iteration, and reuse this data for subsequent iterations. For example, one can record all the $E$ tables that are generated in the first iteration and reuse them in subsequent iterations to directly update the degeneracy tensors. In practice, we find that partial precomputation leads already to a significant computational speedup in simple cases, such as infinite-DMRG simulations of the spin-1/2 Heisenberg quantum spin chain, see \fref{fig:timing_precomputation} for a plausible comparison. We expect such a speedup to be even more dramatic for more complex tensor network methods, especially those for higher-dimensional systems.

\begin{figure}
 	\centering
	\includegraphics[width = 0.65\textwidth]{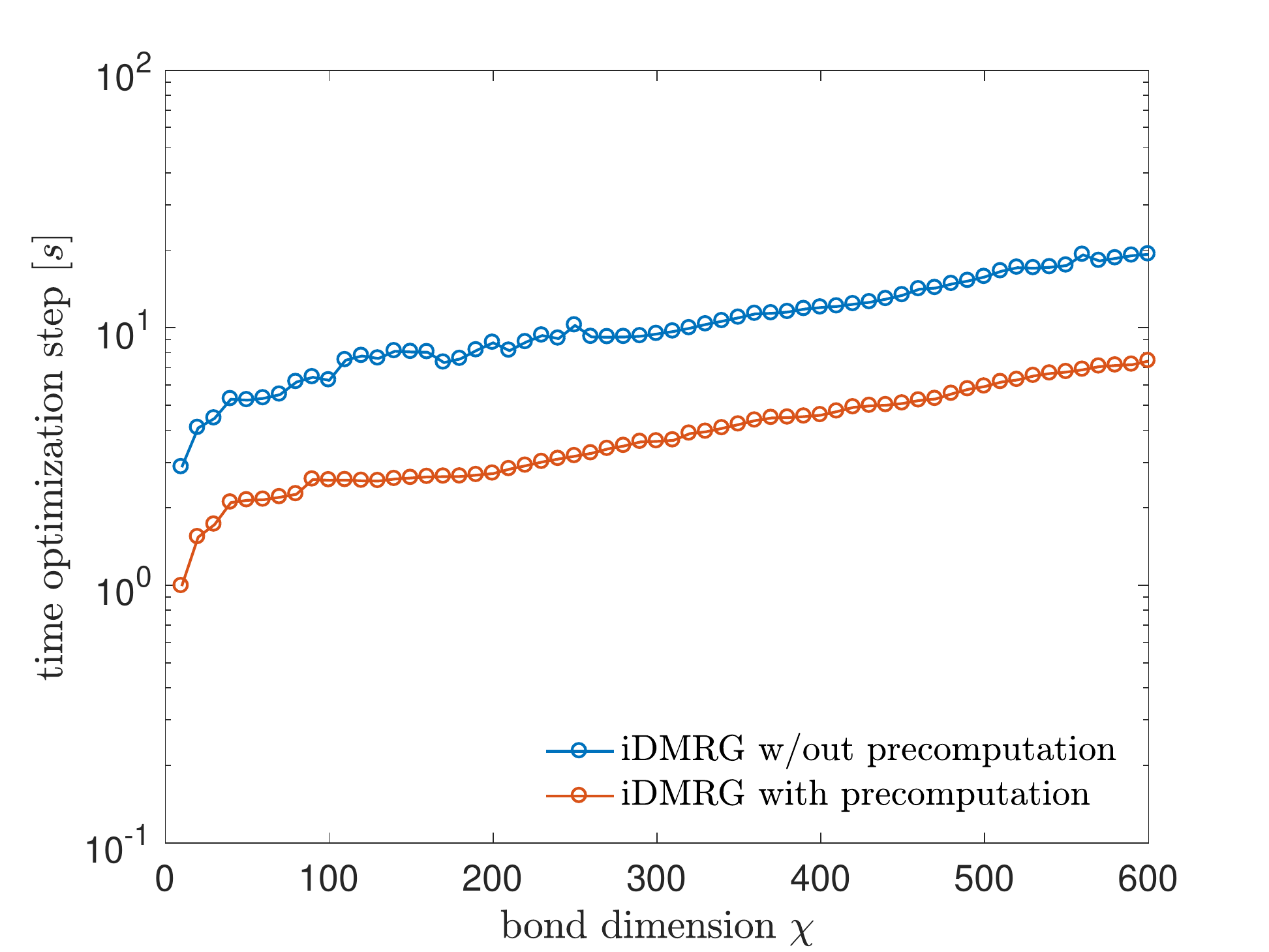}
	\caption{Time (in seconds) required to perform one optimization step in an implementation of the two-site infinite-DMRG algorithm with SU(2) symmetry, for the spin-1/2 Heisenberg quantum spin chain, with partial and without precomputation.} 
	\label{fig:timing_precomputation}
\end{figure}

As a final comparison that gives insight in the simulation overhead due to the manipulation of symmetric tensors as opposed to non-symmetric ones, we show the profiling of the $SU(2)$-symmetric iDMRG simulations for the Heisenberg model presented at the beginning of the paper. Here we break down the simulation time per optimization step in time spent on actual contractions of degeneracy tensors, and time that is spent on manipulations of the fusion tree. The result is shown in \fref{fig:DMRG_Profiling_1}, where the numbers affiliated to the data points show the percentage of the total time.
\begin{figure}
 	\centering
	\includegraphics[scale = 0.58]{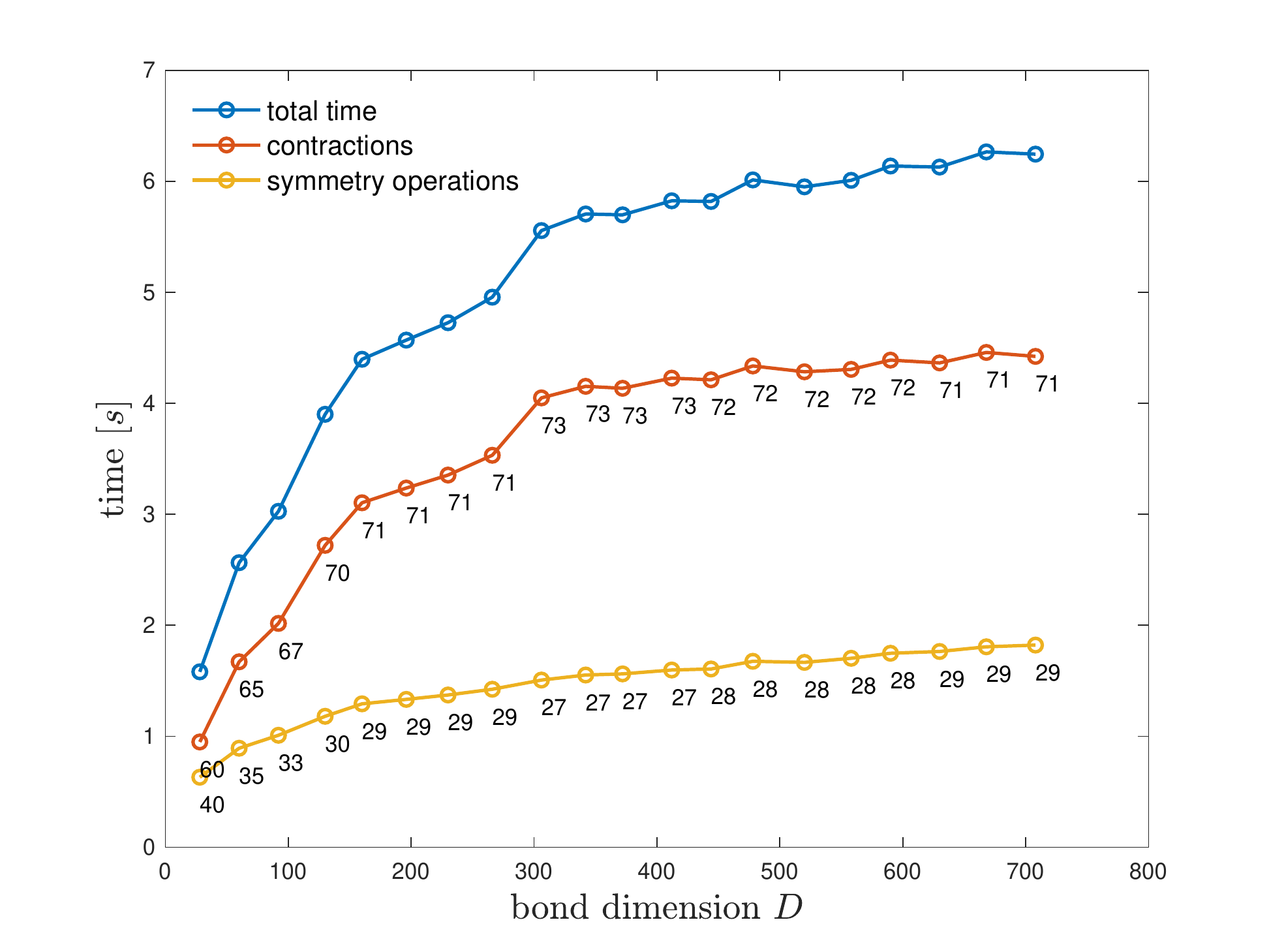}
	\caption{{Profiling of the $SU(2)$-symmetric iDMRG simulation for the spin-1/2 Heisenberg model. The numbers attached to the data points show the percentage of the total time. The total time is given for one iDMRG optimization step.}}
	\label{fig:DMRG_Profiling_1}
\end{figure}
For small bond dimensions the ratio between actual contractions and symmetry-related operations is lower due to an always existing fusion tree overhead. This overhead grows slower than the time spent on contractions, resulting in a $\sim 70\,\%$-$30\,\%$ ratio for reasonable bond dimensions.

\subsection{Simulation of anyonic systems} % (fold)
\label{sub:simulation_of_anyonic_systems}

One of the main advantages of the fusion tree approach is that it is generally suitable to be used in anyonic simulations. Here we want to provide an example of the simulation of a chain of Fibonacci anyons (described by $SU(2)_3$) with nearest neighbour antiferromagnetic interaction. A full description of anyonic MPS simulations will not be provided here, it can be found in e.g.~\cite{anyonicMPS}. Based on the $SU(2)$ framework and the existing iDMRG algorithm anyonic models can be incorporated by specifying the following set of data
\begin{itemize}
	\item (1) the allowed anyon types or topological charges $a,b,c,\hdots$ (which replace the irreps---$0,\frac{1}{2},1,\frac{3}{2},\ldots,$ of SU(2)),
	\item (2) the quantum dimensions $d_i$ associated with each anyon $i$ (which replace the dimensions of SU(2) irreps),
	\item (3) the fusion rules of the anyonic theory described by $a \times b \rightarrow \sum_c N_{ab}^{c} c$ (which replace the SU(2) irrep coupling rules e.g. 
	$0 \times \frac{1}{2} \rightarrow \frac{1}{2}$, $\frac{1}{2} \times \frac{1}{2} \rightarrow 0+1,\ldots,$),
	\item (4) the braiding tensor $R_{c}^{ab}$ that describes the effect of particle exchange (which replace the factor of -1 that results in the fusion basis of two half-integer irreps of SU(2) after the two irreps are swapped) and
	\item (5) the $F$-move $(F^{abc}_d)_{ef}$ for the fusion of three anyons (which replace the Wigner 6-j symbols of $SU(2))$.
\end{itemize}
For Fibonacci anyons there are only two charges denoted by $\boldsymbol 1$ and $\tau$, where $\boldsymbol 1$ is the vacuum and $\tau$ the only non-trivial particle. The fusion rules are given by
\begin{align}
	\begin{gathered}
		\boldsymbol 1 \otimes \tau = \tau \\
		\tau \otimes \boldsymbol 1 = \tau \\
		\tau \otimes \tau = \boldsymbol 1 \oplus \tau\ .
	\end{gathered}
\end{align}
This can be encoded into the three-index fusion tensor $N_{ab}^{c}$, whose components are zero except for $N_{1 1}^{1} = N_{1 \tau}^{\tau} = N_{\tau 1}^{\tau} = N_{\tau \tau}^{1} = N_{\tau \tau}^{\tau} = 1$. The fusion rules can be easily incorporated by representing the particle $\boldsymbol 1$ by spin 0 and the $\tau$ particle by spin 1. The quantum dimensions for the two topological charges are given by
\begin{align}
	d_1 = 1 \hspace{2.0cm} d_\tau = \left( 1 + \sqrt{5} \right)/2 \ .
\end{align}
The exchange of anyonic particles, the so called braiding operation, is omitted since it is does not occur in the iDMRG implementation and we refer to Ref.~\cite{anyonicMPS} for the entries of the tensor $R^{ab}_c$. Relating different ways to fuse three anyons into a single one is described by the $F$-move, whose non-trivial components are given by
\begin{align}
	(F_{\tau}^{\tau \tau \tau})_{ef} = \begin{pmatrix} \phi^{-1} & \phi^{-1/2} \\ \phi^{-1/2} & -\phi^{-1} \end{pmatrix}
\end{align}
where $\phi = (1 + \sqrt 5)/2$ is the golden ratio and $e,f \, \in \, \{\boldsymbol 1,\tau \}$. All other entries are given by $(F_d^{abc})_{ef} = N_{ab}^{e} N_{bc}^{f} N_{ec}^{d} N_{af}^{d}$.

We performed iDMRG simulations for a chain of Fibonacci anyons with one $\tau$ particle per site. The Hamiltonian favors the fusion to the vacuum for two neighbouring particles, similarly to the spin-coupling in the antiferromagnetic Heisenberg model. The convergence of the ground state energy is shown in \fref{fig:groundstateEnergy_FibonacciAnyons_SU2_3}.
\begin{figure}
 	\centering
	\includegraphics[scale = 0.58]{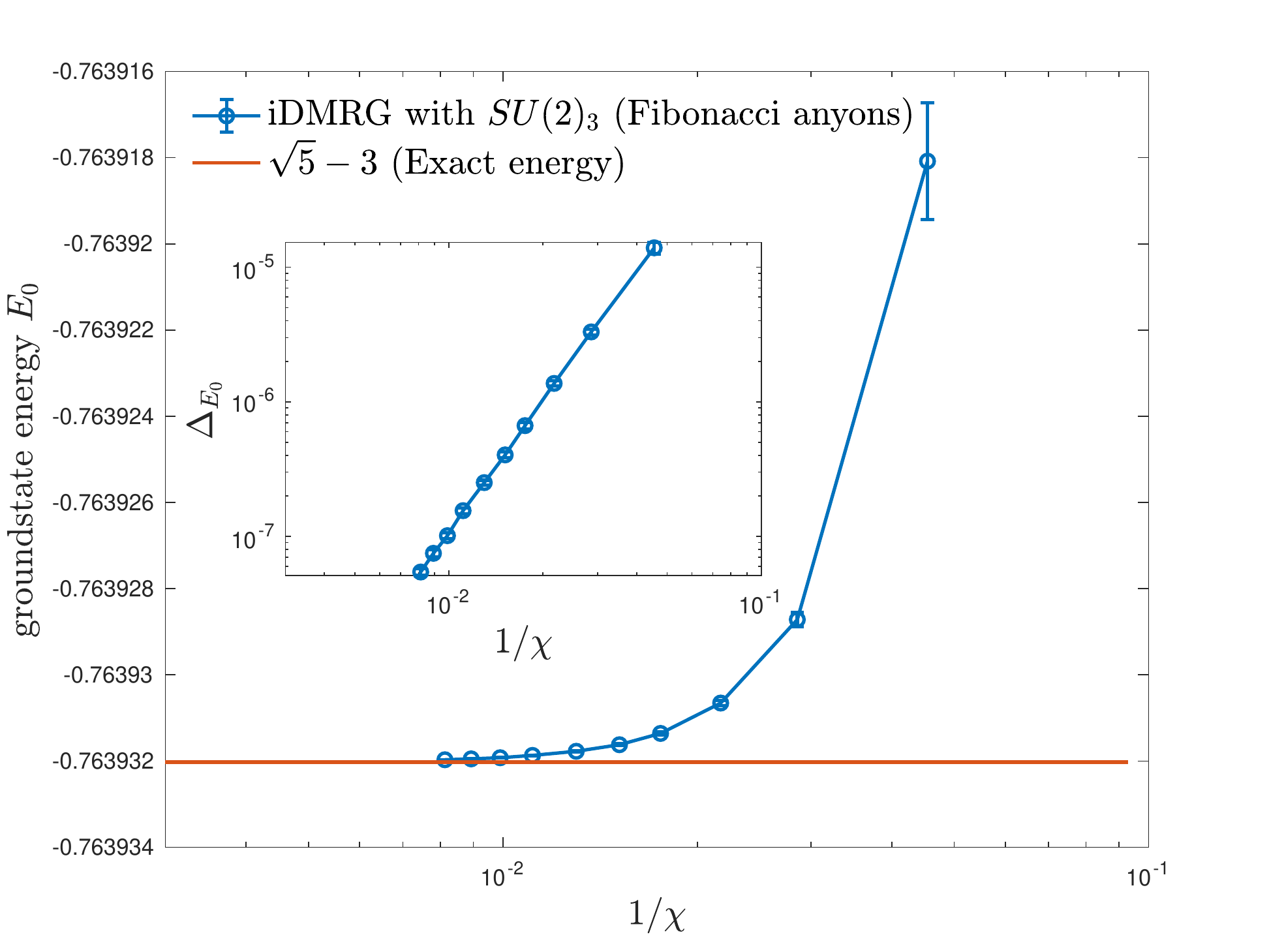}
	\caption{{Convergence of the ground state energy for a chain of Fibonacci anyons. The algorithm converges algebraically to the exact ground state energy which is given by $E_0 = 2(\phi - 2) = \sqrt 5 - 3$ (red solid line) as shown in the inset. \ps{We use the discarded weight (a measure for the accuracy) as an estimate of the relative error for the ground state energy in each simulation.}}}
	\label{fig:groundstateEnergy_FibonacciAnyons_SU2_3}
\end{figure}
The energy converges algebraically to the exact ground state energy of $E_0 = 2(\phi - 2) = \sqrt 5 - 3$. More complicated anyonic theories, e.g. theories with more than two topological charges and fusion multiplicities ($N_{ab}^{c} > 1$) can be implemented as an extension of the presented framework.

% subsection simulation_of_anyonic_systems (end)

\section{Conclusions} % (fold)
\label{sec:conclusions}

In this paper we have described programming tips and tricks to implement $SU(2)$ symmetry in tensor network algorithms. Our implementation is based on the formalism of fusion trees, and proceeds in such a way that structural tensors are never explicitly stored in the computer. Instead, fusion trees impose exact symmetry constraints in the manipulation of degeneracy tensors in the code. This procedure is very accurate, and scales well for tensors with many legs, as required for tensor network simulations of higher-dimensional systems. \ps{Additionally, our fusion tree based formalism also enables simulations of anyonic systems, which we demonstrated for a chain of Fibonacci anyons.}
% Moreover, it has the advantage of being directly applicable to the simulation of anyonic systems. 

The formalism presented in this paper has been successfully applied to study chiral ladder Hamiltonians~\cite{ourLadderPaper}. We also envisage many other applications of this formalism, like the application to ground state calculations with two-dimensional PEPS. We believe that our formalism is well suited in two spatial dimensions, and should allow us to reach large PEPS bond dimension in the study of frustrated quantum antiferromagnets. We leave such a study for future works.

% We envisage many applications of this formalism. In particular, we are currently working on the application of this formalism to study chiral ladder Hamiltonians~\cite{ourLadderPaper}, as well as to ground-state calculations with two-dimensional PEPS. We believe that our formalism is particularly efficient in two spatial dimensions, and should therefore allow us to reach large PEPS bond dimension in the study of frustrated quantum antiferromagnets. We leave such a study for future works. 

\bigskip 

{\bf Acknowledgments.-} We acknowledge fruitful discussions with A. Haller, C. Hubig, I. P. McCulloch, S. Montangero, M. Rakov, P. Silvi and B. Vanhecke. We also acknowledge DFG funding through project GZ OR 381/3-1, as well as the MAINZ Graduate School of Excellence. SS acknowledges the support provided by the Alexander von Humboldt Foundation and the Federal Ministry for Education and Research through the Sofja Kovalevskaja Award.

% section conclusions (end)

\end{document}